%% file: planet_araa.tex
\documentclass{ar-1col}

\usepackage[comma]{natbib}
\usepackage{url}
\usepackage{amsmath,amssymb}
\usepackage{aas_macros}
\usepackage{graphicx}
\graphicspath{{figures/}}

\jname{Xxxx. Xxx. Xxx. Xxx.}
\jvol{AA}
\jyear{YYYY}
\doi{10.1146/((please add article doi))}

\begin{document}

\markboth{Zhu \& Dong}{Exoplanet Statistics and Theoretical Implications}

\title{Exoplanet Statistics and Theoretical Implications}

\author{Wei Zhu$^{1,\,2}$ and Subo Dong$^{3,*}$\footnote{*Corresponding author}
\affil{$^1$Department of Astronomy, Tsinghua University, Beijing, China, Beijing 100084; email: weizhu@tsinghua.edu.cn}
\affil{$^2$Canadian Institute for Theoretical Astrophysics, University of Toronto, Toronto, Canada, ON M5S 3H8}
\affil{$^3$Kavli Institute for Astronomy and Astrophysics, Peking University, Beijing, China, Beijing 100871; email: dongsubo@pku.edu.cn}
}

\begin{abstract}
In the last few years, significant advances have been made in understanding the distributions of exoplanet populations and the architecture of planetary systems. We review the recent progress of planet statistics, with a focus on the inner $\lesssim1\,$AU region of the planetary system that has been fairly thoroughly surveyed by the \emph{Kepler} mission. We also discuss the theoretical implications of these statistical results for planet formation and dynamical evolution.
\end{abstract}

\maketitle

\tableofcontents

\input{sec1.tex}
\input{sec2.tex}

\input{sec3.tex}
\input{sec4.tex}
\input{sec5.tex}

\section*{DISCLOSURE STATEMENT}
The authors are not aware of any affiliations, memberships, funding, or financial holdings that might be perceived as affecting the objectivity of this review. 

\section*{ACKNOWLEDGMENTS}
We thank 
Scott Gaudi,
Andy Gould,
Kento Masuda,
Shude Mao,
Chris Ormel,
Scott Tremaine,
and Josh Winn 
for comments and suggestions on the manuscript.
SD was supported by the National Key R\&D Program of China No.\ 2019YFA0405100.
WZ was supported by the Natural Sciences and Engineering Research Council of Canada (NSERC) under the funding reference \#CITA 490888-16.

\bibliography{review_bib}{}
\bibliographystyle{ar-style2}

\end{document}

%% file: sec1.tex
\section{INTRODUCTION}

\begin{figure}
\includegraphics[width=\typewidth]{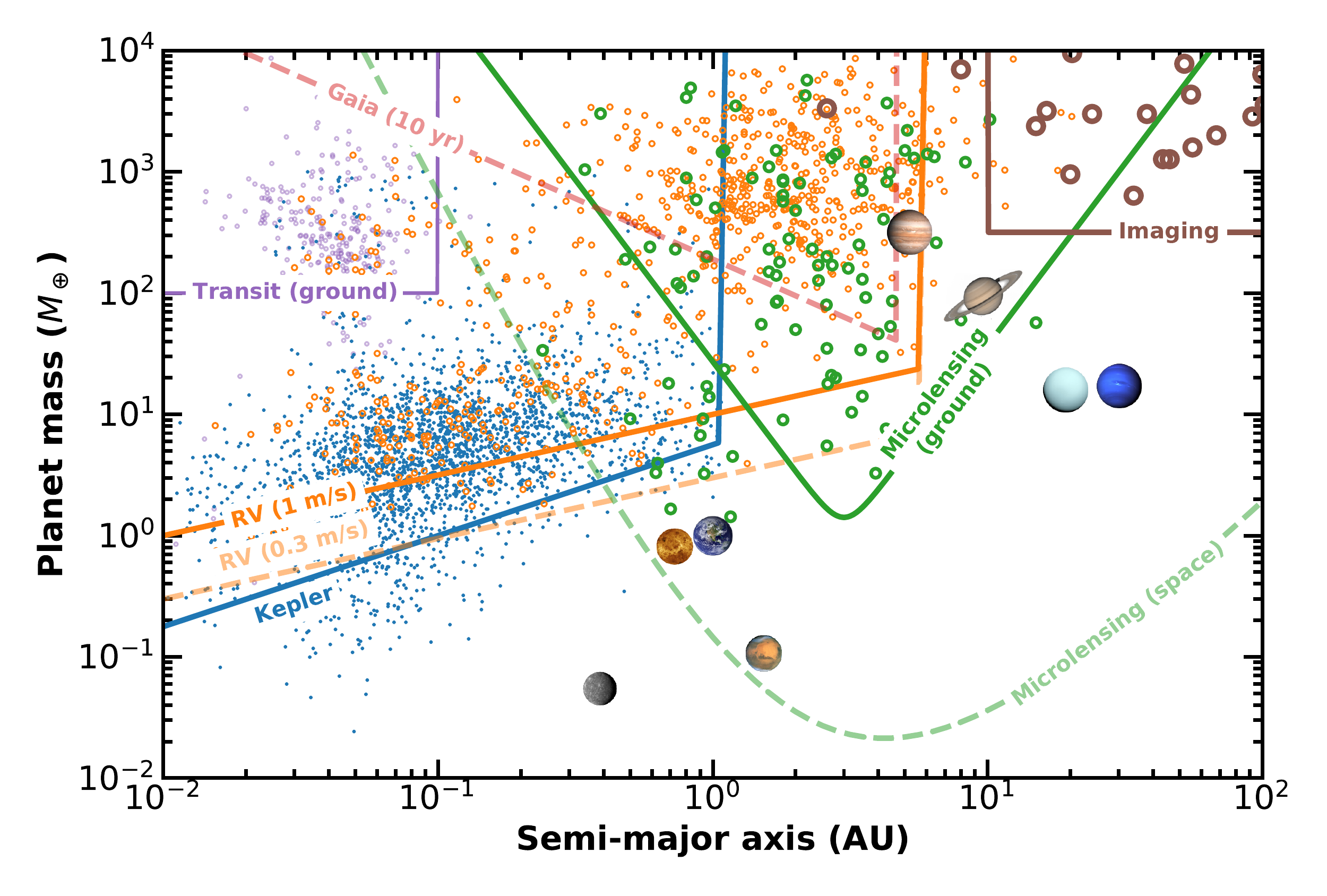}
\caption{Mass versus semi-major axis of known planets, based on the ``Confirmed Planets'' list from NASA Exoplanet Archive \citep[][acquired in September of 2020]{Akeson:2013} and the reliable \emph{Kepler} planet candidates (see Section~\ref{sec:inner_planets} for more details). We differentiate with different colors planet detections as well as the approximate sensitivity curves from ground-based transit (purple), \emph{Kepler} survey (blue), RV surveys (orange), microlensing (green), and direct imaging (brown). 
The masses of the \emph{Kepler} detections are estimated from the measured radii according to the \citet{ChenKipping:2017} mass--radius relation. The sensitivity curve of \emph{Kepler} is also converted in a similar way from that measured in the radius--period plane (see \textbf{Figure~\ref{fig:kepler_planets}}).
The sensitivity curve for the 10-yr \emph{Gaia} astrometry survey is also shown in red, for which we have assumed a Sun-like host at 20$\,$pc and required a 3-$\sigma$ detection over the expected precision. For space-based microlensing, we adopt the sensitivity curve of the microlensing survey that will be performed by the \emph{Nancy Grace Roman Space Telescope} \citep[formerly known as WFIRST,][]{Penny:2019}.
Images of the solar system planets from NASA are shown at their corresponding locations.
\label{fig:overview}}
\end{figure}

``Who ordered that?'' said the theorist I. Rabi when learning about the unexpected discovery of muons in 1936. Little did particle physicists know that it would only be the beginning of uncovering a puzzling ``particle zoo'' filled with diverse particles in the next three decades, until revolutionary theoretical insights were developed to classify the elementary particles. Now nearly three decades since the astonishing discovery of a hot Jupiter \citep{Mayor:1995}, the ``exoplanet zoo''  is ever growing -- whenever the detection territories grow in breadth or depth, nature appears to be teeming with new species. Theorists working on planet formation and evolution face distinctly different sets of challenges from particle physicists: in the popular paradigm, forming planets from dust grains is a daunting march spanning tens of orders of magnitudes in mass and involves many physical processes that are too complex for first-principle calculations. In hindsight, it should probably be of little surprise that a theory involving such complicated physics, which was anchored by the sole sample of our solar system, would have limited predictive success. 

We review the recent progress of planet statistics and identify patterns emerging from the known thousands of exoplanets that cover a broad region of the parameter space (see \textbf{Figure~\ref{fig:overview}}). 
Robustly identifying patterns in the intrinsic distributions of planets can stimulate and test theories. Conversely, theoretical advances may also beam the searchlight on fresh observational ground, as exemplified by the development of the photoevaporation theory leading to the recent discovery of a ``radius valley'' (see Section~\ref{sec:valley}). Since the last \textit{Annual Reviews} article on exoplanet populations \citep{WinnFabrycky:2015}, the field of planet statistics has made significant progress. In particular, the large and homogeneous planet sample from the NASA \emph{Kepler} mission \citep{Borucki:2010} has provided the best source for statistical studies, but a major shortcoming of the \emph{Kepler} data was the initial lack of accurate stellar parameters for both the planet hosts and the target stars (i.e., the parent sample). In the last few years, substantial efforts have been dedicated to systematically characterize the \emph{Kepler} sample and thus unleash its potential for statistical studies. These include asteroseismology \citep[e.g.,][]{Chaplin:2013,VanEylen:2015}, the \emph{Gaia} data releases \citep{Gaia,Gaia_dr2}, follow-up spectroscopic programs such as the LAMOST-\emph{Kepler} survey \citep[e.g.,][]{Dong:2014a,DeCat:2015,Zong:2018} and the California-\emph{Kepler} Survey \cite[CKS,][]{Petigura:2017,Johnson:2017}, as well as many projects of the \emph{Kepler} Follow-up Observation Program \citep[KFOP,][]{Furlan:2017}. Moreover, substantial works to understand the \emph{Kepler} pipeline detection efficiency and vetting false positives have much improved the reliability of \emph{Kepler} statistical inference \citep[e.g.,][]{Christiansen:2015, Morton:2016}. Last but not least, in-depth developments have been recently made to disentangle the intricate observational biases of multi-planet systems. These efforts have made it possible to offer new insights into planet distributions and architectures. 
 
In this review, we first clarify in Sections~\ref{sec:occurrence_rate} and \ref{sec:idem} several common confusions in exoplanet statistical studies. Then we discuss planet distributions in the inner ($\lesssim1\,$AU) and the outer ($\sim1$--$10\,$AU) regions in Sections~\ref{sec:inner_planets} and \ref{sec:outer_planets}, respectively. The former is focused on results from the \emph{Kepler} mission, and the latter includes updated results from radial velocity (RV) and gravitational microlensing. A brief discussion of the free-floating planets (FFPs) from microlensing is also provided. We focus on planets around $\gtrsim$Gyr-old stars, while planets orbiting young stars found by direct imaging are not discussed (see the review by \citealt{Bowler:2016}). The implications to theories of planet formation and evolution are discussed in Section~\ref{sec:theory}. Finally in Section~\ref{sec:summary}, we summarize and outline the promising directions for future developments.

\subsection{On defining and interpreting planet ``occurrence rate''} \label{sec:occurrence_rate}

Many statistical studies focus on deriving the intrinsic ``occurrence rate'' (or the often interchangeably used term ``frequency'') of planets. But from one study to another, the same term can carry different meanings. In the following we clarify these different definitions to avoid further misinterpretations.

In most studies, the derived occurrence rate is the average number of planets per star, and we denote it as $\bar{n}_{\rm p}$, which is defined as
\begin{equation} \label{eqn:np_bar}
\bar{n}_{\rm p} \equiv \frac{\rm Total~\#~of~planets}{\rm Total~\#~of~stars} .
\end{equation}
Here \emph{a planet} is restricted to lie within a predefined parameter space, often in the period--radius plane (for the transit method) or the period--mass (or minimum mass $m_{\rm p}\sin{i}$) plane (for the RV method). Similarly, \emph{a star} is restricted to a star-like target of predefined properties. Since a large fraction of such stars may actually have unresolved stellar companions, the correction for the impact of the stellar binarity can be important for the inference of the planet formation efficiency (see Section~\ref{sec:binary}).

Another important quantity sometimes referred to as occurrence rate is the fraction of stars with planets $F_{\rm p}$
\begin{equation} \label{eqn:fp}
F_{\rm p} \equiv \frac{\rm Total~\#~of~planetary~systems}{\rm Total~\#~of~stars} .
\end{equation}
Here a planetary system has \emph{at least one} planet existing in a predefined parameter space. By definition $F_{\rm p}\le1$, so it is usually reported as a percentage. However,
an occurrence rate reported as a percentage (i.e., ``X\% of stars have planets'') does not necessarily mean that it is the fraction of stars that are hosts of planets, since $\bar{n}_{\rm p}$ is also frequently reported as a percentage.

\begin{marginnote}[]
\entry{Frequency of planets}{the average number of planets per star $\bar{n}_{\rm p}$ (Equation~\ref{eqn:np_bar}).}
\entry{Frequency of planetary systems}{the fraction of stars with planets $F_{\rm p}$ (Equation~\ref{eqn:fp}).}
\entry{Average planet multiplicity}{the average number of planets per planetary system (Equation~\ref{eqn:mp_bar}).}
\end{marginnote}

To distinguish between the two definitions, we refer to $\bar{n}_{\rm p}$ as the frequency of planets and $F_{\rm p}$ as the frequency of planetary systems. The ratio of the two measures the average number of planets per planetary system (within a predefined parameter space), which we call \emph{average planet multiplicity} and denote $\bar{m}_{\rm p}$
\begin{equation} \label{eqn:mp_bar}
\bar{m}_{\rm p} \equiv \frac{\bar{n}_{\rm p}}{F_{\rm p}} = \frac{\rm Total~\#~of~planets}{\rm Total~\#~of~planetary~systems} .
\end{equation}
{\it Kepler} data suggest that multi-planet systems are common, so usually  $\bar{m}_{\rm p}$ is larger than unity, and consequently $\bar{n}_{\rm p}$ and $F_{\rm p}$ substantially differ from each other. They only become similar when the average planet multiplicity $\bar{m}_{\rm p} \rightarrow 1$, which can happen when either a) a category of planets with low intrinsic multiplicity (e.g., short-period giant planets) is concerned, or b) the parameter space of interest is small enough that systems with more than one such planet are rare.

The three quantities, $\bar{n}_{\rm p}$, $F_{\rm p}$, and $\bar{m}_{\rm p}$, are all important for testing theories. To provide a simple example, with only $\bar{n}_{\rm p}$ measured to be unity, it is possible that all stars have one planet ($F_{\rm p}=100\%$ and $\bar{m}_{\rm p}=1$) or that half of the stars have two planets ($F_{\rm p}=50\%$ and $\bar{m}_{\rm p}=2$). These two cases obviously demand different theoretical explanations. 

Observationally, the derivations of $\bar{n}_{\rm p}$ and $F_{\rm p}$ have rather different requirements and follow different procedures. It is generally more straightforward to derive $\bar{n}_{\rm p}$, since correcting the detectability of individual planets concerns observables directly measurable from surveys (e.g., planet size and orbital period for transit, assuming that the properties of the stars are known). In contrast, the detectability of a planetary system usually concerns the intrinsic architecture of the system, including the planet multiplicity and distributions of the orbital and physical parameters, many of which may not be directly observable, so the derivation of $F_{\rm p}$ can rely on assumptions of these unknowns.
This is especially an issue in transit surveys: the derivation of $F_{\rm p}$ requires assumptions about the mutual inclinations between planets, and different assumptions can lead to fairly different values of $F_{\rm p}$ (see Section~\ref{sec:multiplicity}). 

In deriving the two frequencies, statistical studies involving multi-planet systems usually treat the planet occurrence as a Poisson process. 
This may be a reasonable assumption in the derivation of the planet frequency $\bar{n}_{\rm p}$, but it can lead to unreliable results in the derivation of the planetary system frequency $F_{\rm p}$.
This Poisson process assumption implies that the presences of individual planets in the same system are independent and that their physical and orbital properties are independent of the properties of other planets or of the host star. As discussed later in this review, such an assumption breaks down in certain circumstances.
Below we provide a specific example to demonstrate its impact on the planetary system frequency.
The fractions of Sun-like stars with cold giant planets and with planets that \emph{Kepler} is sensitive to are $10\%$ and $30\%$, respectively. The fraction of such stars with at least one planet in the joint parameter space would be $1-(1-30\%)\times(1-10\%)=37\%$ under the Poisson process assumption. However, this frequency is determined to be $\sim30\%$ as a result of the strong correlation between the inner and the outer planets (Section~\ref{sec:correlation}).
The correlations (or sometimes anti-correlations) between the occurrences of planets around the same host also suggest that one may not be able to extrapolate a parameterized distribution of the planetary system frequency to a parameter space that is not covered by the data.

A number of studies have reported $F_{\rm p}$ by using the detectability of the first detected (or the most detectable)  planet in the system as that of the whole system \citep[e.g.,][]{Cumming:2008,Mayor:2011,Fressin:2013,Petigura:2013}. This approach does not require assumptions on planet multiplicity or architecture. However, as the detectability of any planet is no greater than the detectability of the system it resides in, this approach typically tends to overestimate $F_{\rm p}$ \citep{Zhu:2018}.

\subsection{On inferring the frequency of planets} \label{sec:idem}

In this section, we discuss the commonly used methods of inferring the frequency of planets $\bar{n}_{\rm p}$ from a statistical survey of $N_\star$ target stars.

A popular method is the so-called inverse detection efficiency method (IDEM), which has been used extensively in the literature, including many influential studies \citep[e.g.,][]{Mayor:2011,Howard:2012,Fressin:2013,Petigura:2013,Dressing:2013,Dressing:2015}. For our illustrative survey, the average number of planets per star according to IDEM is
\begin{equation} \label{eqn:idem}
\bar{n}_{\rm p}^{\rm IDEM} = \frac{1}{N_\star} \sum_{i=1}^{N_{\rm p}} \frac{1}{p_i} = \frac{N_{\rm p}}{N_\star} \left\langle \frac{1}{p} \right\rangle ~.
\end{equation}
Here $p_i$ is the survey detection efficiency of the $i$-th of $N_{\rm p}$ detected planets and $\langle \cdot \rangle$ is the average over all detected planets. IDEM is intuitive, simple to perform, and computationally efficient, as it does not require computing the detection efficiencies of null detections (which are usually the majority of the targets), so it is useful in getting a rough estimate of the underlying frequency. However, this method is not rigorously established in the probability theory and can potentially lead to biased results \citep{ForemanMackey:2014, Hsu:2018}. Specifically, with a low detection efficiency and a small number of detections, \citet{Hsu:2018} found that IDEM often leads to underestimated $\bar{n}_{\rm p}$ since the actual detections typically come from targets with larger-than-average sensitivities. IDEM can also suffer substantial fluctuations because of the inversion of the (typically small) detection efficiency. 

An approach with sound statistical basis is modeling planet occurrence as a Poisson process and performing maximum likelihood analysis \citep[e.g.,][]{Tabachnik:2002, Cumming:2008, Gould:2010, Youdin:2011, Dong:2013, Burke:2015}. In a given bin that has $N_{\rm p}$ planet detections, \citet[see their Section 3.1]{Youdin:2011} and  \citet[see their Appendix A]{ForemanMackey:2014} show that the maximum likelihood (ML) estimator for planet frequency is
\begin{equation} \label{eqn:ml}
\bar{n}_{\rm p}^{\rm ML} = \frac{N_{\rm p}}{\sum_{j=1}^{N_\star} p_j} = \frac{N_{\rm p}}{N_\star} \frac{1}{\langle p \rangle} = \frac{N_{\rm p}}{N_\star^{\rm eff}} ~;\quad
N_\star^{\rm eff} \equiv N_\star \langle p \rangle ~.
\end{equation}
Here $p_j$ is the planetary detection efficiency in the bin for the $j$-th star, regardless of whether the star yields any actual planet detection or not, and $N_\star^{\rm eff}$ is the effective sample size. Unlike in \textbf{Equation~(\ref{eqn:idem}), the average here is performed among all stars in the sample.} Compared to IDEM, this method is computationally more expensive, while being statistically superior. It is more robust against fluctuations in the efficiencies of individual detections (as well as null detections) because the averaging is performed on $p$ rather than $1/p$. 

Next we elaborate on incorporating the above approach into the Bayesian framework following the simplified Bayesian model of \citet[see their Appendix B]{Hsu:2018} but with some corrections. The posterior probability distribution of planet frequency $\bar{n}_{\rm p}$ for the statistical sample is given by
\begin{equation}
P(\bar{n}_{\rm p}|N_{\rm p},~N_\star^{\rm eff}) \propto P(N_{\rm p}|\bar{n}_{\rm p},~N_\star^{\rm eff}) P_{\rm pri}(\bar{n}_{\rm p}) .
\end{equation}
 The first term on the right-hand side quantifies the probability (or likelihood) of having the $N_{\rm p}$ detections for a given rate $\bar{n}_{\rm p}$, which under the Poisson process assumption is described by a Gamma distribution
\footnote{A Gamma distribution can be parameterized in terms of a shape parameter $\alpha$ ($>0$) and a rate parameter $\beta$ ($>0$). The probability density function of a variable $x$ is $f(x;~\alpha,~\beta)= (\beta^\alpha x^{\alpha-1} e^{-\beta x})/\Gamma(\alpha) \propto x^{\alpha-1} e^{-\beta x} $, where $\Gamma(\alpha)$ is the Gamma function evaluated at $\alpha$.}.
The second term, $P_{\rm pri}(\bar{n}_{\rm p})$, is the prior distribution of $\bar{n}_{\rm p}$. If a conjugate prior is assigned as a Gamma distribution with a shape parameter $\alpha_0$ and a rate parameter $\beta_0$, the resulting posterior distribution is then a Gamma distribution with the shape parameter $\alpha_0+N_{\rm p}$ and the rate parameter $\beta_0+N_\star^{\rm eff}$
\begin{equation} \label{eqn:gamma}
P(\bar{n}_{\rm p}|N_{\rm p},~N_\star^{\rm eff}) \propto \bar{n}_{\rm p}^{\alpha_0+N_{\rm p}-1} e^{-\bar{n}_{\rm p}(\beta_0+N_\star^{\rm eff})}.
\end{equation}
For a flat prior on $\bar{n}_{\rm p}$, the two parameters are $\alpha_0=1$ and $\beta_0=0$, respectively. For completeness, the mean and standard deviation of this Gamma distribution posterior are
\begin{equation} \label{eqn:mean_std}
\mu(\bar{n}_{\rm p}) = \frac{\alpha_0+N_{\rm p}}{\beta_0+N_\star^{\rm eff}} ,\quad
\sigma(\bar{n}_{\rm p}) = \frac{\sqrt{\alpha_0+N_{\rm p}}}{\beta_0+N_\star^{\rm eff}}.
\end{equation}
The first expression reduces to the ML estimator of \textbf{Equation~\ref{eqn:ml}} if a log-flat prior on the planet frequency $\bar{n}_{\rm p}$ is assumed (i.e., $\alpha_0=\beta_0=0$).
The expressions given by \textbf{Equation~\ref{eqn:mean_std}} provide easy-to-use estimates to report when the number of detections is relatively large.
However, when small or null detections are involved, the posterior probability distribution is fairly non-Gaussian. It is then more appropriate to report the median value, the 68\% credible interval, and/or the 95\% upper limit, all of which can be derived from the cumulative posterior probability distribution. It is also worth noting that, in the case of null detections, a meaningful upper limit on $\bar{n}_{\rm p}$ cannot be derived with the log-flat prior because the shape parameter becomes zero and the Gamma distribution is undefined. We show in Section~\ref{sec:kepler_planets} an application of the Bayesian approach to derive the frequency of planets in the \emph{Kepler} parameter space.

%% file: sec2.tex
\section{THE INNER PLANETARY SYSTEM} \label{sec:inner_planets}

We review in this section planet statistics in the inner region ($\lesssim1~$AU of Sun-like stars), which is well explored thanks to thousands of planets detected by the RV and transit techniques. We focus on the best statistical probe by far of the inner region---the large and uniform sample from the \emph{Kepler} mission, which is sensitive to transiting planets with radii $R_{\rm p}$ down to $\sim R_\oplus$ and orbital periods $P$ up to $\sim 1~$yr \citep{Borucki:2010}. 

We first derive a clean baseline sample based on the final \emph{Kepler} data release \citep[DR25,][]{Thompson:2018} and the improved stellar parameters from \citet{Berger:2020a}. The latter work combines the astrometric measurements from \emph{Gaia} DR2 \citep{Gaia_dr2} with the available photometric and spectroscopic information to yield stellar radii with a median uncertainty of $4\%$. Starting from the DR25 planet catalog, we have removed planet candidates with: a) transit signal-to-noise ratio (S/N) below the nominal threshold (S/N$=7.1$), b) NASA Exoplanet Archive \footnote{exoplanetarchive.ipac.caltech.edu} disposition flag being false positive, c) the derived planetary radius $R_{\rm p}>20\,R_\oplus$, d) the orbital period $P>400\,$days, and e) the best-fit transit impact parameter $b>1$. We restrict to Sun-like stars that are defined as main-sequence stars (as classified by \citealt{Berger:2018}) with effective temperatures between 4700\,K and 6500\,K. The bulk of this section is about planets around Sun-like hosts, and topics such as correlations with various stellar properties, such as stellar mass, metallicity and binarity, are discussed in Section~\ref{sec:stellar_dependence}.

\begin{figure}
\includegraphics[width=\textwidth]{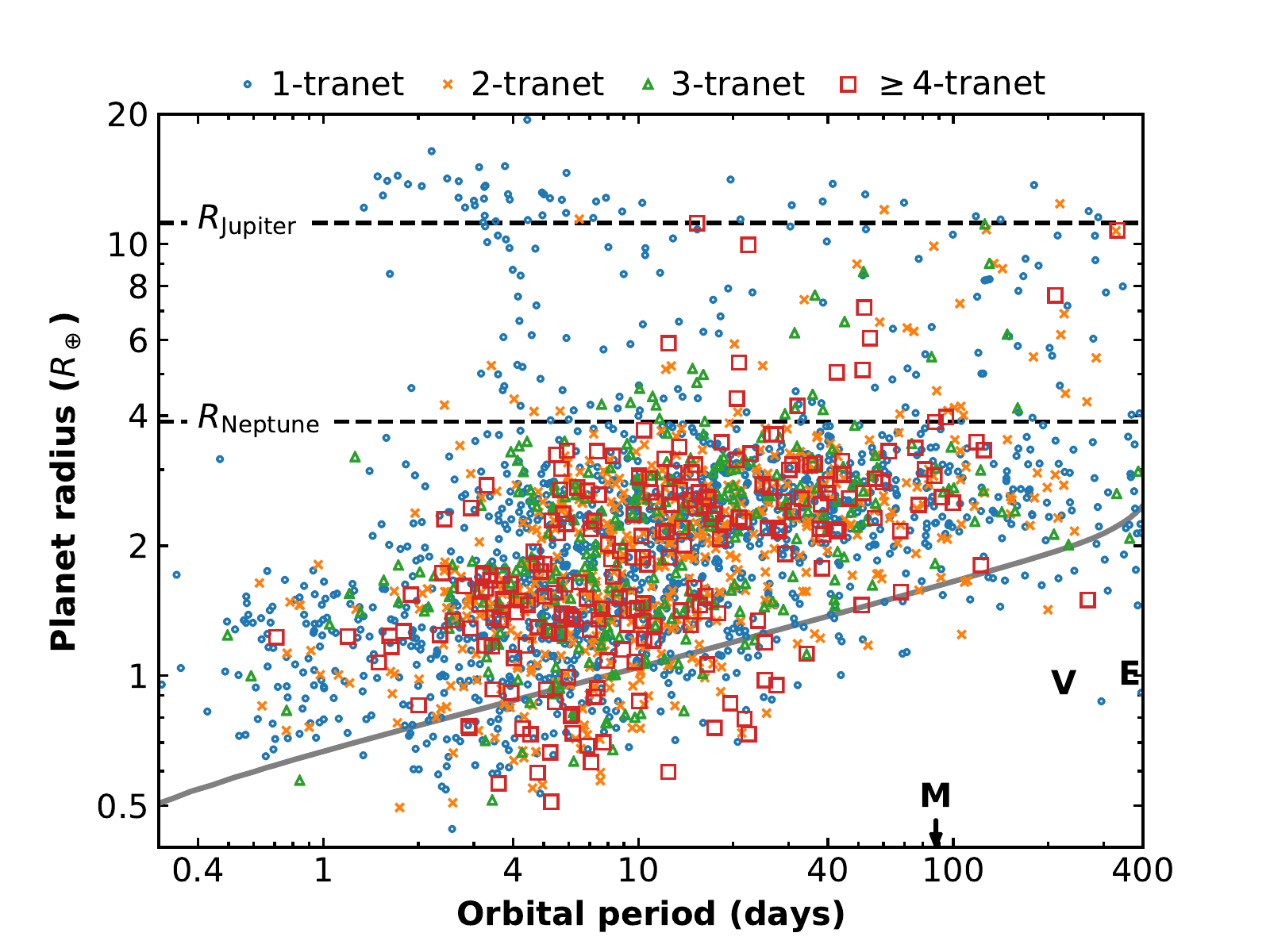}
\caption{The close-in ($P<400\,$days) \emph{Kepler} planets in the radius--period plane, with various symbols and colors indicating the observed multiplicity (note that we use the contraction ``tranet'' to refer to ``transiting planet'' in the  legend). The gray solid curve indicates the median detection efficiency of the planet search pipeline. The solar system planets in the inner region, namely Mercury, Venus, and Earth, are denoted with their first letters. The median precision on the planetary radius is $\sim 7\%$. The radii of the Jupiter and Neptune are shown with horizontal dashed lines, respectively. The radius valley at $\sim 2\,R_\oplus$ (see Section~\ref{sec:valley}) is visible.
\label{fig:kepler_planets}}
\end{figure}

The baseline sample contains 2,525 planet detections around 98,213 Sun-like stars. Of all the transiting planets, 1451 are found in systems with only one detected transiting planet \footnote{We sometimes use the contraction ``tranet'' to stand for ``transiting planet'' in the text and figure legends and captions.} and the remaining 1074 are from systems with multiple detected transiting planets. The average observed multiplicity rate, namely the average fraction of planets from known multi-planet systems, is $42.5\%$. This is a lower limit on the intrinsic multiplicity rate, as many of the single-planet systems seen in \emph{Kepler} are likely part of intrinsic multi-planet systems (see details in Section~\ref{sec:multiplicity}). The observed transit multiplicity distribution in the sample is
\begin{equation}
(N_1,~N_2,~N_3,~N_4,~N_5,~N_6,~N_7) = (1451,~278,~97,~37,~12,~2,~1) 
\end{equation}
and no system has more than seven transiting planets.\footnote{Note that the only system in our sample with seven transiting planets, Kepler-90, has been found to contain one additional planet candidate \citep{Shallue:2018}. However, this additional candidate was not found by the \emph{Kepler} DR25 pipeline and thus not included.}
\textbf{Figure~\ref{fig:kepler_planets}} illustrates the planets in our sample in the radius--period plane. Different multiplicities of transiting planets are shown with different symbols.

\subsection{Planet distribution in the radius--period plane} \label{sec:kepler_planets}

With the above statistical sample we derive the planet frequencies in the Bayesian framework of Section~\ref{sec:idem}. The parameter space in the radius--period plane is divided into logarithmically equally-spaced cells 
\footnote{Since the typical precisions of planetary period and radius are much smaller than the cell sizes, we ignore the uncertainties of planetary parameters. See \citet{ForemanMackey:2014} for how to incorporate the planet parameter errors in the analysis.}. In each cell, the number of planet detections, $N_{\rm p}$, is found and the average detection efficiency, $\langle p \rangle$, is computed via
\begin{equation}
\langle p \rangle = \frac{\int_{R_{\rm p,min}}^{R_{\rm p,max}} \int_{P_{\rm min}}^{P_{\rm max}} (R_\odot/a) S(P,~R_{\rm p}) d\ln{P} d\ln{R_{\rm p}}} {\int_{R_{\rm p,min}}^{R_{\rm p,max}} \int_{P_{\rm min}}^{P_{\rm max}} d\ln{P} d\ln{R_{\rm p}}} .
\end{equation}
Here $R_{\rm p,min}$, $R_{\rm p, max}$, $P_{\rm min}$, and $P_{\rm max}$ denote the boundaries of the cell, and $R_\odot/a$ is approximately the transit geometric probability at semi-major axis $a$ around a Sun-like host. The sensitivity due to survey detection thresholds at a given period and radius, $S(P,~R_{\rm p})$, is computed with the \texttt{KeplerPORTs} code,
\footnote{The code is publicly available at \url{https://github.com/nasa/KeplerPORTs}.}
which was first developed in \citet{Burke:2015} and further updated for \emph{Kepler} DR25 \citep{Burke:2017} by incorporating results of transit injection and recovery tests for the final \emph{Kepler} pipeline \citep{Burke:2017b,Christiansen:2020}. Updated stellar parameters were used to derive the mean sensitivity curve.

\begin{figure}
\includegraphics[width=\typewidth]{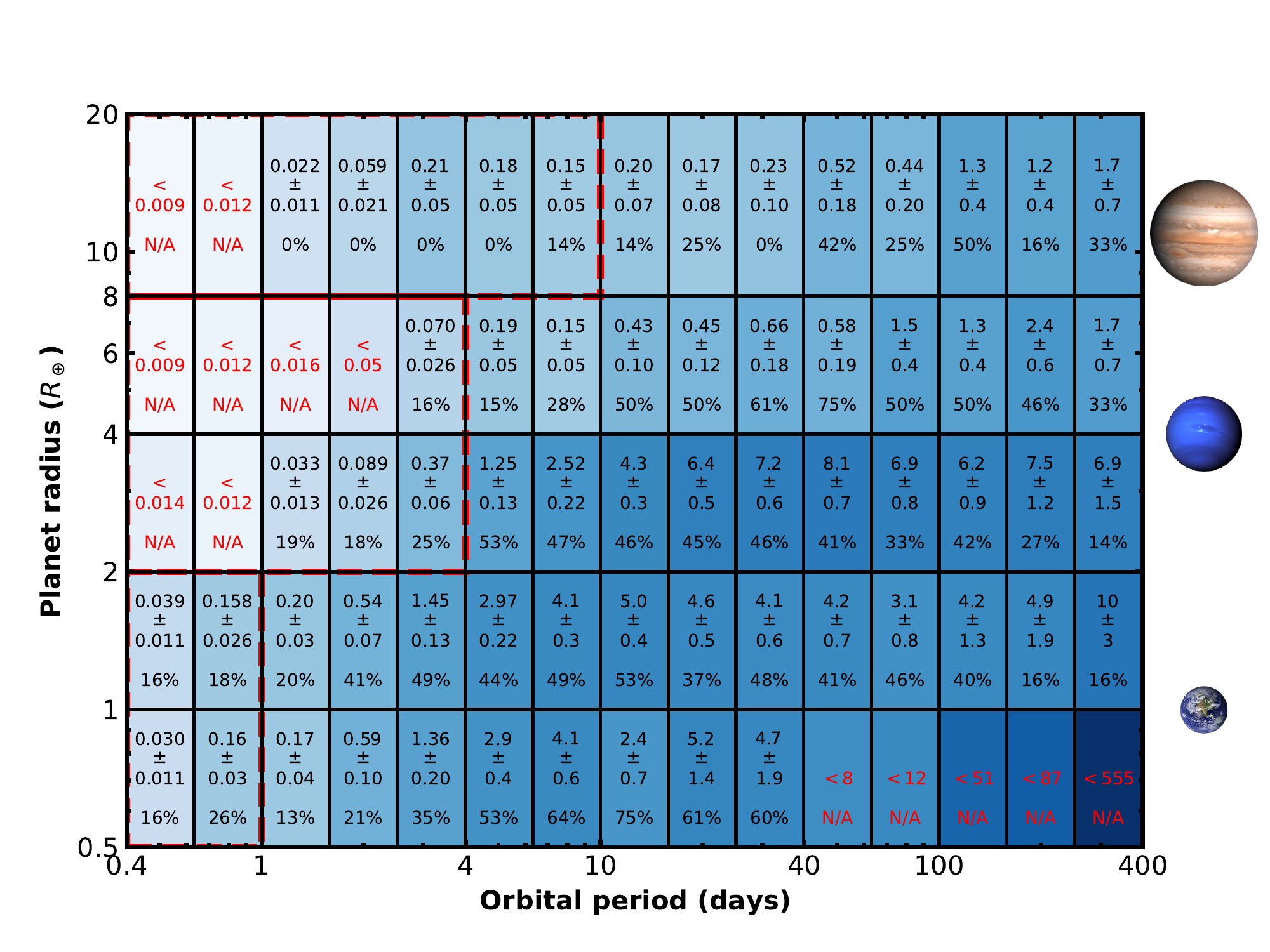}
\caption{This figure illustrates the planet frequencies ($\bar{n}_{\rm p}$) and the observed multiplicity fractions based on the planet sample in \textbf{Figure~\ref{fig:kepler_planets}}. The numbers and error bars are the average number of planets with periods and radii within the given cell per 100 Sun-like stars. If there are less than three detections found within the cell, then the $95\%$ upper limit is reported instead. These upper limits are highlighted in red. The fraction in each cell denotes the observed multiplicity fraction, namely the fraction of planets in that cell found to reside in multi-planet systems. We use ``N/A'' for cells with less than three detections. The red dashed lines mark the regions corresponding to hot Jupiters, hot Neptune ``desert,'' and USPs.
\label{fig:rate_map}}
\end{figure}

We adopt a flat prior on $\bar{n}_{\rm p}$, and its posterior distribution is then described by the Gamma distribution of \textbf{Equation~\ref{eqn:gamma}} with $\alpha_0=1$ and $\beta_0=0$. For cells with $\le2$ detections we report the 95\% upper limits, whereas for the rest the means and the standard deviation given by \textbf{Equation~\ref{eqn:mean_std}} are reported as the measurements and associated uncertainties, respectively. We have verified that the deviation between the mean and the median is substantially smaller than the uncertainty for all relevant cells.

The derived planet frequency map is shown in \textbf{Figure~\ref{fig:rate_map}}. For cells with more than two detections, we also indicate the observed multiplicity rate of planets in the cell. Again, these multiplicity rates represent the lower limits on the fraction of planets in those cells that reside in multi-planet systems. We summarize several key results below:
\begin{itemize}
\item The integrated planet frequency is $\bar{n}_{\rm p}=1.23\pm0.06$ for planets with radii in the range 1--$20\,R_\oplus$ and orbital periods up to $400\,$days. This is broadly consistent with results from previous studies \citep[e.g.,][]{Fressin:2013,Petigura:2018,Hsu:2019}. As stressed in Section~\ref{sec:occurrence_rate}, statistical analyses like this one do not yield the fraction of stars with planets $F_{\rm p}$, as the impact of the multiplicity and the mutual inclination has not been taken into account (see Section~\ref{sec:multiplicity}).
\item As it has been clear since the earliest {\it Kepler} statistical studies, there are generally many more small planets with radii $R_{\rm p} \lesssim 4\,R_\oplus$ than larger ones, for orbital periods $P<400\,$days. Planet frequencies tend to increase from the upper left (large $R_{\rm p}$ and small $P$) toward the lower right (small $R_{\rm p}$ and large $P$). In other words, the intrinsic radius distribution is dependent on the orbital period \citep[e.g.,][]{Dong:2013,ForemanMackey:2014,Hsu:2018}. There exist some local regions where the general trends break down, such as the radius valley (see Section~\ref{sec:valley}). 
\item Sub-Earths ($R_{\rm p}<1\,R_\oplus$) and Earth-sized planets in Earth-like orbits are not well probed by \emph{Kepler}, and thus estimates of their frequencies are most susceptible to the uncertainties of survey sensitivity estimates. As a result, there remain large discrepancies on their intrinsic frequencies in the literature (see Table~2 of \citealt{WinnFabrycky:2015} and Figure~17 of \citealt{Burke:2015}). 
\item  \emph{Kepler} planets commonly reside in multi-planet systems in most parts of the radius--period plane, with some notable exceptions such as the hot Jupiter region (\citealt{Steffen:2012}; see Section~\ref{sec:hot_Jupiters} for more discussion). The intrinsic multiplicity rates are likely higher than the observed multiplicity rates shown in \textbf{Figure~\ref{fig:rate_map}}. We defer to Section~\ref{sec:multiplicity} for further discussions.
\end{itemize}

The above method to derive the planet frequency $\bar{n}_{\rm p}$ is non-parametric. An alternative approach employs a parameterized planet distribution function and then constrains the associated parameters. The parametric approach has been widely used in statistical studies of various detection techniques, including transit \citep[e.g.,][]{Youdin:2011, Howard:2012, Dong:2013, Burke:2015}, RV \citep[e.g.,][]{Tabachnik:2002, Cumming:2008}, and microlensing studies \citep[e.g.,][]{Gould:2010, Suzuki:2016, Clanton:2016}. It is also commonly used in simulations of generating synthetic planetary systems \citep[e.g.,][]{Mulders:2018, He:2019}. The commonly adopted planet distribution function is separable between the orbital period (or semi-major axis) and the planetary radius (or mass)
\begin{equation}
\frac{d^2 N}{d\ln{P} d\ln{R_{\rm p}}} \propto \frac{dN}{d\ln{P}} \frac{dN}{d\ln{R_{\rm p}}} .
\end{equation}
The distributions of the orbital period and the planetary radius are usually parameterized as power laws or broken power laws. The use of such a separable function implicitly assumes that the period (radius) distribution is independent of the planetary radius (period). As discussed above, such an assumption is not valid for the inner planetary system. It is likely not valid for planet distributions in other regions of the parameter space, either. The implications of this failure on the derived occurrence rates from the parametric method and on the theoretical interpretations of the underlying population have not been fully explored. 

In what follows, we provide brief discussions about selected regions in the radius--period plane.

\subsubsection{Hot Jupiters} \label{sec:hot_Jupiters}
As the first type of exoplanets found around solar-type stars \citep{Mayor:1995}, hot Jupiters ($8\,R_\oplus<R_{\rm p}<20\,R_\oplus$ and $P<10\,$days) remain interesting and exciting targets for both observational and theoretical purposes. Here we only review the occurrence and multiplicity rates of hot Jupiters in the current context and refer interested readers to the recent review by \citet{Dawson:2018} for more in-depth discussions about the hot Jupiter population.

There is a long-standing discrepancy between the hot Jupiter frequency inferred from RV and transit surveys (e.g., \citealt{Gould:2006a,Wright:2012}; see Table~B9 of \citealt{Santerne:2016} for an incomplete list). For example, our statistical sample yields a rate of $0.62\pm0.09\%$, which is in good agreement with previous studies of the hot Jupiter frequency in the \emph{Kepler} field \citep[e.g.,][]{Howard:2012,Fressin:2013,Santerne:2016}, whereas the RV surveys of stars in the Solar Neighborhood report rates that are typically a factor of $\sim$2 higher ($0.9$--$1.2\%$; \citealt{Mayor:2011,Wright:2012}). It was suggested that the discrepancy could be caused by the different stellar properties, such as age, metallicity, and binary fraction, between the RV and transit samples. This has been tested by several follow-up studies of the \emph{Kepler} sample. The \emph{Kepler} stars are only slightly sub-solar on average ($\left\langle {\rm [Fe/H]} \right\rangle_{\rm Kepler} \approx-0.04$; \citealt{Dong:2014a}), and their metallicity differences with the RV targets ($\left\langle {\rm [Fe/H]} \right\rangle_{\rm RV} \approx0.0$) seem to be too small to fully account for the discrepancy even given the steep dependence of hot Jupiter frequency with metallicity \citep{Guo:2017}. The unresolved binaries are also unlikely to substantially change the hot Jupiter frequency in the \emph{Kepler} sample \citep{Bouma:2018}. However, because RV surveys preferentially exclude close ($\sim1$--$50\,$AU) stellar binaries from their sample, this discrepancy in hot Jupiter frequencies between transit and RV surveys can potentially be resolved if the formation of hot Jupiters are suppressed in such close binary systems \citep{Moe:2020}. Searching for stellar companions of transiting hot Jupiters \citep[e.g.,][]{Ngo:2016} and making comparisons with field stars is a promising way to further test this possibility. 

As shown in \textbf{Figure~\ref{fig:kepler_planets}}, one out of the 49 hot Jupiters in our statistical sample, Kepler-730b, has a nearby small planet companion \citep{Zhu_note,Canas:2019}. As of writing, only two other hot Jupiters, WASP-47b \citep{Becker:2015} and TOI-1130c \citep{Huang:2020} are known to share the same property. Our statistical sample suggests that $\sim2\%$ ($<9.7\%$, $95\%$ upper limit) of hot Jupiters have nearby ($\lesssim20$ days), small ($\sim$1--$4\,R_\oplus$) and nearly coplanar companions (see also \citealt{Steffen:2012} for the constraint on non-coplanar companions). This low multiplicity rate of hot Jupiters supports the general idea that most of them have undergone some large-scale migrations to arrive at current locations \citep[e.g.,][]{Lin:1996,Rasio:1996,Weidenschilling:1996}. We refer to \citet{Dawson:2018} for more in-depth discussions on this topic.

\subsubsection{Hot Neptune ``desert''}

The region $P\lesssim4\,$days and $2\,R_\oplus \lesssim R_{\rm p} \lesssim 8\,R_\oplus$ lands in the so-called hot Neptune (or sub-Jovian) ``desert'' (e.g., \citealt{Szabo:2011, Beauge:2013, Mazeh:2016} and references therein), which is considered underpopulated, especially when inspecting mixed planet samples found in surveys with different detection sensitivities (e.g., ground-based transits and \emph{Kepler}). This ``desert'' is however not that barren: the total planet frequency enclosed in the above region is $0.61\pm0.07\%$ from our statistical analysis (see \textbf{Figure~\ref{fig:rate_map}}), making this hot Neptune ``desert'' similarly populated as the hot Jupiter region \citep[see also][]{Dong:2018}. While the above frequency is derived for a rectangular region in the radius--period plane, it is worth noting that the boundaries of this ``desert'' region are better described as a triangle and extend out to $5$--$10$\,d in $m_{\rm p}$ vs.\ $a$ and $R_{\rm p}$ vs.\ $P$ planes (see Figure~1 and Figure~4 of \citealt{Mazeh:2016}). 
 \citet{Dong:2018} found that the frequency of planets inside this region depends on the host star metallicity in a way similar to the frequency of hot Jupiters, and they dubbed this population as ``Hoptunes'' (rather than ``hot Neptunes'') to reflect that not all of them were known Neptune-like physically. Out of our baseline sample of 61 planets in this region, 14 are observed to have planetary companions, and the periods for majority of these companions are within 10\,d. The observed multiplicity rate is thus $23\%$, which is lower than the \emph{Kepler} average while higher than that of hot Jupiters \citep[see also][]{Dong:2018}. We refer to \citet{Dawson:2018} for more discussions on the connection of this population with close-in Jupiters and related theoretical implications. 

A number of theories have been proposed to explain the formation of planets in this region  \citep[e.g.,][]{Kurokawa:2014,Matsakos:2016,Lundkvist:2016,Bailey:2018,OwenLai:2018}. The leading explanations of its triangular boundaries invoke photoevaporation (see more discussion in Section~\ref{sec:valley}) and tidal effects following the high-eccentricity migration. The upper boundary is best explained as the tidal disruption barrier for gas giants following their high-eccentricity migrations \citep{Matsakos:2016, OwenLai:2018}. More massive planets can be tidally circularized closer to the star without tidal disruption, resulting in the negative slope of the upper boundary. It has been proposed that the same mechanism also produces the lower boundary, with the positive slope resulting from a mass--radius relation of small planets that is different from the relation of giant planets \citep{Matsakos:2016}. However, this mechanism may not be able to explain the planets that are in or near the ``desert'' region and reside in multi-planet systems. An alternative theory, proposed by \citet{OwenLai:2018}, suggests that the lower boundary is better explained by the photoevaporation of highly irradiated planets, and that the positive slope results from the fact that the photoevaporation mechanism is more effective if the planet is closer to the host star. There has been a growing interest of planets in this region with the TESS mission \citep[e.g.,][]{Armstrong:2020,Burt:2020}, and the follow-up studies of such planets will soon allow for a better understanding of their physical properties and formation mechanisms.

\subsubsection{Ultra-short-period planets} \label{sec:usp}
Planets with radii between $0.5$--$2\,R_\oplus$ and periods $P\lesssim1\,$d, known as ultra-short-period planets (USPs), represent a rather extreme planet population. The period threshold for USPs at one day corresponds to an equilibrium temperature $\sim2000\,$K for a Sun-like host, which is hot enough to sublimate dust grains. Below we briefly summarize several key properties of USPs and refer interested readers to the recent comprehensive review by \citet{Winn:2018} for more discussions about this extreme planet population.

Our statistical analysis yields $\bar{n}_{\rm p}=(0.39\pm0.04)\%$ for USPs. This is in general agreement with the result of \citet{SanchisOjeda:2014}, whose specialized pipeline yields $\bar{n}_{\rm p}=(0.51\pm0.07)\%$ for planets with radii in the range $0.8$--$2\,R_\oplus$ and $P<1\,$d. Out of the 81 USPs in our sample, 16 are found with outer planetary companions, indicating an observed multiplicity rate of $20\%$. The true multiplicity rate is probably much higher, since USPs can be largely misaligned relative to the outer planetary companions \citep{Dai:2018,Petrovich:2019}. In 13 of the 16 multi-planet systems involving USPs, the closest outer companion has $P_{\rm c}\lesssim 10\,$d, and the USP is usually farther apart in terms of the period ratio from the rest of the planets in the same system (see also \citealt{Steffen:2013}).

The highly irradiative environment at sub-day orbit implies that USPs are unlikely to have formed \textit{in situ}. Partially because of the comparable rates between hot Jupiters and USPs, it had been suggested that USPs could be the surviving cores of tidally disrupted hot Jupiters \citep{Jackson:2013}, but this was not supported by several pieces of evidence including the lack of strong host metallicity dependence \citep{Winn:2017} and the relatively high multiplicity rate compared to hot Jupiters. A more plausible scenario is that the USPs have arrived at their current locations without losing much of their initial mass. One way of achieving this is the gradual decay of the orbit due to the tidal dissipation within the host star \citep{Lee:2017}. Alternatively, the proto-USP planet may have been sent to an eccentric (and misaligned) orbit following the dynamical interactions with other planets in the system, and then the orbit decays and circularizes due to the tidal dissipation within the planet \citep{Schlaufman:2010,Petrovich:2019,PuLai:2019}. This latter model sees its support in the relatively large mutual inclinations of USPs \citep{Dai:2018}. Additionally, in order for the tidal inspiral model to produce USPs, the tidal dissipation in USP hosts needs to be efficient, but the population analysis on stellar kinematic ages seems to suggest otherwise \citep{Hamer:2020}.

\subsubsection{Radius valley} \label{sec:valley}

\begin{figure}
\includegraphics[width=\typewidth]{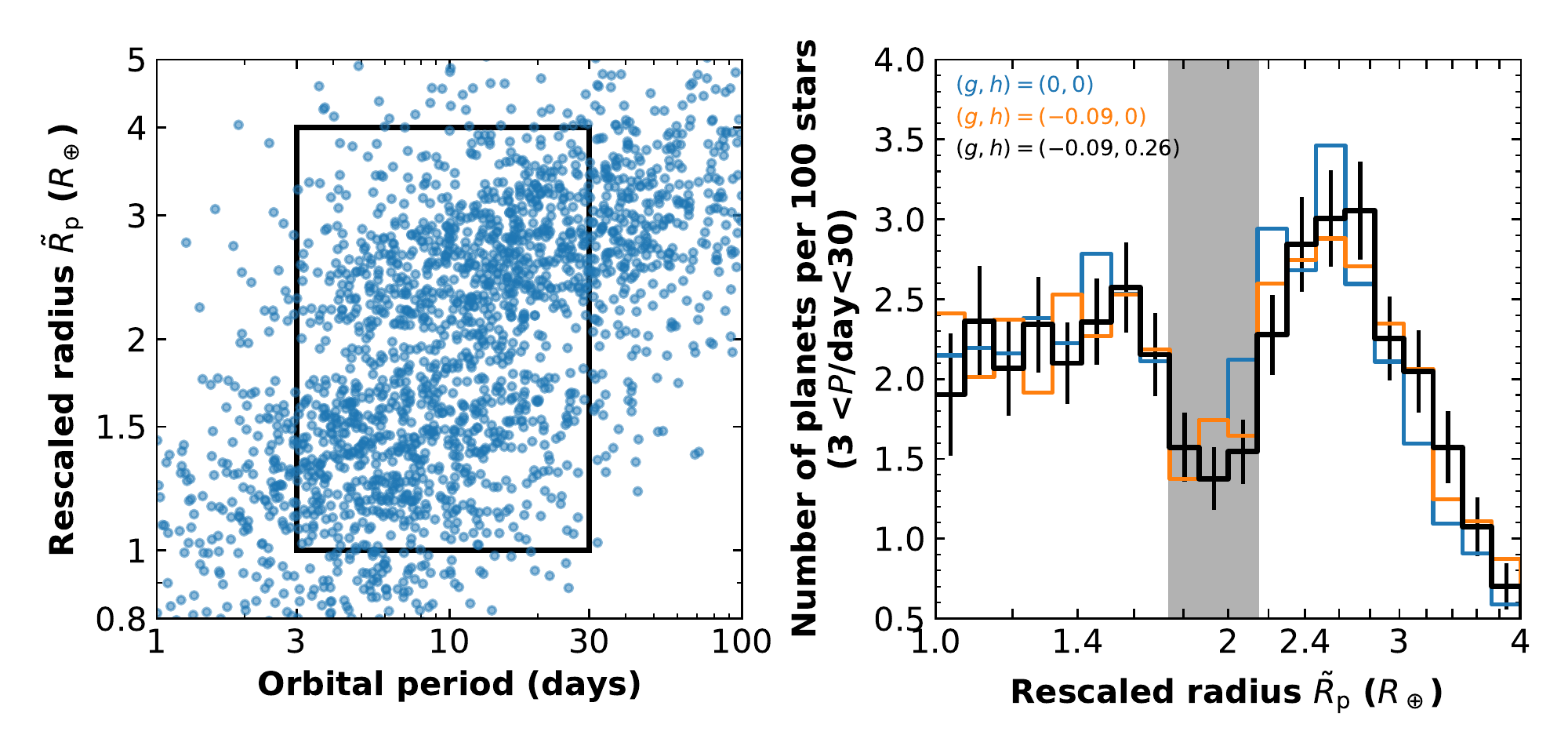}
\caption{(\textit{a}) The zoom-in view of the \emph{Kepler} planets in our sample centered at the radius valley. The $y$-axis shows the rescaled radius $\tilde{R}_{\rm p} \equiv R_{\rm p} (P/{10\,\rm days})^{-g} (M_\star/M_\odot)^{-h}$ (see \textbf{Equation~\ref{eqn:valley}}) and we adopt the best-fit $g=-0.09$ \citep{VanEylen:2018} and $h=0.26$ \citep{Berger:2020b}. The black box marks the boundary within which planets are used to derive the intrinsic radius distribution. (\textit{b}) The intrinsic distribution of the rescaled radius $\tilde{R}_{\rm p}$. The radius gap, highlighted in the gray band, is most prominent when both period and stellar mass dependences are taken into account.
\label{fig:valley}}
\end{figure}

An important discovery in the field of exoplanet in recent years is the radius valley, which refers to a region in the radius--period plane at radii $R_{\rm p}\sim2\,R_\oplus$ and periods between $\sim$3--$30\,$days \citep{Fulton:2017,VanEylen:2018,Fulton:2018}. This radius valley is visible in our statistical sample (see \textbf{Figure~\ref{fig:kepler_planets}}). The position of the valley in radius is reported to decrease with the orbital period \citep{VanEylen:2018} and increase with the stellar mass \citep{Wu:2019,Berger:2020b}. The two dependences can be parameterized as
\begin{equation} \label{eqn:valley}
\frac{R_{\rm p}}{R_{\rm p}^{\rm valley}} = \left( \frac{P}{10\,\rm days} \right)^g \left(\frac{M_\star}{M_\odot}\right)^h .
\end{equation}
The valley position at orbital period $P=10\,$days and host mass $M_\star=M_\odot$ is found to be $R_{\rm p}^{\rm valley} = 1.9\pm0.2\,R_\oplus$ and the slope quantifying the period dependence is $g=-0.09^{+0.02}_{-0.04}$ \citep{VanEylen:2018}. The slope quantifying the stellar mass dependence is $h=0.26^{+0.21}_{-0.16}$ \citep{Berger:2020b}.
With the above relation one can then highlight the radius valley by rescaling the radius to $\tilde{R}_{\rm p} \equiv R_{\rm p} (P/{10\,\rm days})^{-g} (M_\star/M_\odot)^{-h}$.  \textbf{Figure~\ref{fig:valley}a} illustrates our sample in this rescaled radius (with $g=-0.09$ and $h=0.26$) vs.\ orbital period plane. We also show the intrinsic distribution of the rescaled radius in \textbf{Figure~\ref{fig:valley}b} for planets with $\tilde{R}_{\rm p}$ in the range $1$--$4\,R_\oplus$ and $P$ in the range $3$--$30\,$days. Our choice of the period upper boundary is motivated by \textbf{Figure~\ref{fig:kepler_planets}}: beyond $\sim30\,$days the number of detections in the relevant region and thus the statistical power drops significantly. The peak-to-dip contrast in our ``radius'' distribution is not as significant as that shown in \citet{Fulton:2017} and \citet{Fulton:2018}. In particular, our rescaled radius distribution does not show an obvious single peak at $\tilde{R}_{\rm p}<R_{\rm p}^{\rm valley}$. We have tried with the same period range as used in those studies and confirm that our specific choice of the period range is not the cause of this difference. One possible reason is the different statistical methods used to infer the occurrence rate: As discussed in Section~\ref{sec:idem}, the IDEM approach used in \citet{Fulton:2017} and \citet{Fulton:2018} tends to underestimate the occurrence rates at low sensitivity regions ($R_{\rm p} \sim R_\oplus$). 
The fact that the radius distribution does not seem to decrease at sub-Earth sizes suggests the presence of many undiscovered sub-Earths. The broader radius distribution may also imply that the planetary mass distribution is not as narrowly peaked as some previous studies inferred \citep[e.g.,][]{Wu:2019}.

The leading theory for the radius valley is the atmospheric evaporation driven by high-energy photons from the host star \citep[photoevaporation;][]{OwenWu:2013,Lopez:2013,OwenWu:2017}. In fact, the existence of the radius valley at approximately the discovered position had been predicted years before its discovery (\citealt{OwenWu:2013,Lopez:2013}; see a historic overview in \citealt{Owen:2019}), which is exceptional in exoplanetary science. 
The photoevaporation of the atmosphere is thought to mostly take place during the early ages of the system when the star emits a higher fraction of its total luminosity at high energy ($\lesssim100\,$Myr; e.g., \citealt{Jackson:2012,Tu:2015}, but also see \citealt{King:2020}). For close-in ($\sim3$--$30\,$days) planets with core masses of a few $M_\oplus$, the high-energy radiation is sufficient to unbind the entire hydrogen/helium atmosphere if its initial mass fraction is below some critical value (a few percent; \citealt{OwenWu:2017}). The radius valley thus emerges, separating planets with and without extended atmospheres \citep{OwenWu:2013,Lopez:2013,OwenWu:2017}. The observed period and stellar mass dependences can also be well explained by photoevaporation. As the orbital period increases and/or the host mass decreases, the amount of high-energy radiation the planet receives decreases and thus the valley moves to smaller radii \citep{OwenWu:2017,Wu:2019}.\footnote{Although later-type stars have higher fractions of the total luminosity emitted in higher energy ($\propto M_\star^{-3}$; \citealt{Lopez:2018}) and remain active for a longer period of time, these lower-mass stars have much lower total luminosities ($\propto M_\star^4$ for Solar and later-type stars). The lifetime-integrated high-energy radiation at a certain orbital separation is shown to decrease with decreasing stellar mass \citep[see Figure~4 of][]{McDonald:2019}.}
We refer interested readers to \citet{Owen:2019} for a comprehensive review on the photoevaporation mechanism.

According to the photoevaporation theory, the properties (e.g., location and shape) of the radius valley depend on the underlying planetary properties, especially distributions of the core mass, core composition, and atmospheric mass fraction \citep{OwenWu:2013,Lopez:2013}. Therefore, the observed radius valley opens up a venue to statistically infer the properties of close-in low-mass planets at birth \citep{OwenWu:2017,Jin:2018,Wu:2019,RogersOwen:2020}. Assuming that photoevaporation is the underlying mechanism, these studies collectively point to a typical core mass of a few $M_\oplus$, a core composition similar to that of the Earth (i.e., rich in silicate/iron and poor in water/ice), and a typical atmosphere mass fraction at birth of a few percent. These inferred properties have important implications to the formation and migration history of these close-in planets (see Section~\ref{sec:theory}).

While photoevaporation has seen its success in predicting and explaining the radius valley, alternative theories exist that can also explain the observed valley \citep[e.g.,][]{Ginzburg:2018,Lee:2020}, of which the core-powered mass-loss mechanism is considered the main competing theory. Unlike photoevaporation, the energy source for atmosphere stripping in core-powered mass-loss mechanism is the internal luminosity of the cooling core, and this process is expected to operate on much longer timescales ($\sim$ Gyr) \citep{Ginzburg:2018}. The observed period and stellar mass dependences of the radius valley (\textbf{Equation~\ref{eqn:valley}}) are also consistent with this mechanism \citep{Gupta:2019,Gupta:2020}. Similar to photoevaporation, core-powered mass-loss mechanism also supports that the close-in low-mass planets have predominantly rocky cores with low water-ice fractions \citep{Gupta:2019}.

Attempts have been made to identify which of the two mechanisms discussed above is more responsible for the observed features. These studies made use of either the different stellar mass or age dependences of the two mechanisms \citep[e.g.,][]{Hirano:2018,Berger:2020b}. However, the currently available data provide no conclusive result to distinguish between the two. Larger samples and/or more precise measurements of stellar properties will be needed.

\subsection{Mutual inclinations and the intrinsic multiplicity} \label{sec:multiplicity}

The mutual inclination distribution of planets in multi-planet systems conveys important information on the formation and dynamical evolution of planetary systems. However, currently employed detection techniques are usually incapable of directly measuring mutual inclinations. This is particularly true for RV and microlensing. The transit technique is strongly biased toward (nearly) coplanar systems. Nevertheless, advancements have made it possible to statistically infer the mutual inclination distribution from the \emph{Kepler} data.

The key issue in constraining the mutual inclination distribution with transit is the strong degeneracy with the intrinsic multiplicity \citep[e.g.,][]{Lissauer:2011,Tremaine:2012}. Specifically, with the observable multiplicity function of transit alone one cannot distinguish between high-multiplicity systems with large mutual inclinations and low-multiplicity systems with small mutual inclinations. We therefore combine mutual inclinations and intrinsic multiplicity in the same discussion.

Before discussing the statistically inferred mutual inclinations, we briefly overview a handful of systems with measured large mutual inclinations. By combining \emph{HST} astrometry and ground-based RV measurements, \citet{McArthur:2010} measured the mutual inclination between two of the three planetary companions in the Upsilon Andromeda system to be about $30^\circ$. \citet{Mills:2017} performed photo-dynamical modeling of the transit timing variation (TTV) and transit duration variation (TDV) signals of the Kepler-108 system and found the mutual inclination to be $\Delta I=24_{-8}^{+11}\,^\circ$ between the two transiting planets. The pi Mensae system, which hosts a long-period giant planet and a TESS transiting super Earth \citep{Huang:2018,Gandolfi:2018}, is reported to have significant mutual inclinations ($\sim30$--$150^\circ$) from joint analyses of the \emph{Hipparcos} and \emph{Gaia} DR2 astrometry \citep{Xuan:2020,Damasso:2020,DeRosa:2020}. Additionally, some USP systems have also been determined to have large mutual inclinations \citep[e.g.,][]{Dai:2018}. More planetary systems with large mutual inclinations are expected to be found in the following years, especially with \emph{Gaia}'s capability to determine the 3D orbital configurations \citep{Perryman:2014}.

\subsubsection{The weighted transit duration method}

A popular method to statistically infer the mutual inclination of \emph{Kepler} muti-planet systems makes use of the ratio of transit chord lengths \citep{Steffen:2010}
\begin{equation} \label{eqn:xi}
\xi \equiv 
\frac{T_{\rm in} P_{\rm in}^{-1/3}}{T_{\rm out} P_{\rm out}^{-1/3}} = \sqrt{\frac{(1+r_{\rm in})^2-b_{\rm in}^2}{(1+r_{\rm out})^2-b_{\rm out}^2}} ~.
\end{equation}
The subscripts ``in'' and ``out'' denote values of the inner and the outer transiting planets, respectively. Here $T$ measures the time from the first to the last contact points of transit, $r$ is the planet-to-star radius ratio, and $b$ is the transit impact parameter. As both $T$ and period $P$ are precisely measured from transit data \citep{Seager:2003}, the parameter $\xi$ is well determined from observations. The last expression in \textbf{Equation~\ref{eqn:xi}} is used to construct the $\xi$ distribution from models with assumed mutual inclination distributions. When two transiting planets are exactly coplanar, the ratio $b_{\rm in}/b_{\rm out} = a_{\rm in}/a_{\rm out} = (P_{\rm in}/P_{\rm out})^{2/3}$ is precisely measured and thus the parameter $\xi$ only concerns one poorly constrained fiducial parameter (either $b_{\rm in}$ or $b_{\rm out}$, since both $r_{\rm in}$ and $r_{\rm out}$ are reasonably well measured). The distribution of $\xi$ for coplanar systems is thus expected to narrowly peak at unity. In practice, the observed distribution is not so narrow, because of the introduction of the mutual inclination (see \textbf{Figure~\ref{fig:durations}a}).\footnote{The orbital eccentricity $e$ in principle also affects the $\xi$ distribution, but its contribution is relatively minor and thus $e$ cannot be well constrained with this method \citep{Fabrycky:2014}.}
Applying this weighted transit duration method to large samples of \emph{Kepler} planet pairs, \citet{FangMargot:2012} and \citet{Fabrycky:2014} found that the mutual inclinations between transiting planets in the \emph{Kepler} multi-planet systems could be well described by a Rayleigh distribution with a dispersion of a few degrees ($\lesssim3$--$5^\circ$). This has been frequently interpreted as multi-planet systems being nearly coplanar.
However, with the use of only transiting planet pairs, which preferentially have small mutual inclinations, the weighted transit duration method cannot well determine the higher end of the mutual inclination distribution. As an extreme case, even the isotropic distribution of orbital inclinations cannot be reliably ruled out with the use of transit data alone \citep{Tremaine:2012}.

\subsubsection{``\emph{Kepler} dichotomy''}

To recover the true mutual inclination distribution, one needs to break its strong degeneracy with intrinsic multiplicity. The first attempt was carried out by \citet{Lissauer:2011}. The authors tried different functional forms for the intrinsic multiplicity distribution (uniform, Poisson, and exponential) as well as for the mutual inclination distribution (uniform and Rayleigh; see also \citealt{Sandford:2019}). By modeling the intrinsic multiplicity as a uniform (or Poisson) distribution and the mutual inclination as a Rayleigh distribution, \citet{Lissauer:2011} were able to find matches to all observed transit multiplicities except the transit singles. Specifically, their models would under-predict the number of systems with only one transiting planets by nearly $50\%$. This signals the failure of their simplified model. Nevertheless, this feature was picked up by many others and phrased as the evidence for two distinct populations of planetary systems (the so-called ``\emph{Kepler} dichotomy''): In one population planetary systems have small mutual inclinations and relatively compact configurations, whereas in the other population planetary systems have either only one planet or at least two largely mutually inclined planets \citep[e.g.,][]{Johansen:2012,Ballard:2016,Mulders:2018,He:2019}. Taking the \emph{Kepler} sample as a whole, in terms of distributions of many properties of stars (e.g., stellar mass, metallicity) and planets (e.g., period), transit singles and transit multis are statistically consistent with being drawn from the same parent population \citep[e.g.,][]{Xie:2016,MunozRomero:2018,Zhu:2018,Weiss:2018b}, suggesting that they probably have the same origin.

While modeling the mutual inclination as a Rayleigh distribution (or more generally, Fisher distribution; \citealt{Tremaine:2012,Zhu:2018}) seems a reasonable choice \citep[see also][]{Tremaine:2015}, the proper functional form for the intrinsic multiplicity distribution remains an open question. Nevertheless, it is certainly oversimplified to assume that all planetary systems have the same number of planets \citep[e.g.,][]{Ballard:2016,Mulders:2018}. Having a Poisson distribution for the intrinsic multiplicity \citep[e.g.,][]{Lissauer:2011,Zink:2019,Sandford:2019,He:2019} is likely not justified, either. The underlying assumption behind the Poisson distribution is that occurrences and properties of individual planets around the same host are independent from each other. While it has not been proved invalid for \emph{Kepler} planets, there is emerging evidence that the presence and properties of planets inside the same system may be correlated due to the shared formation environment and/or host properties (see Sections~\ref{sec:occurrence_rate}, \ref{sec:intra-system} and \ref{sec:stellar_dependence}). Furthermore, the exponential or power-law (i.e., Zipfian distribution; \citealt{Sandford:2019}) forms can be securely ruled out. These distributions predict overly abundant intrinsic single-planet systems, which is not supported by TTV observations (see below).

Given the strong degeneracies, disentangling the intrinsic multiplicity function and the mutual inclination distribution therefore requires external information. To this end, \citet{Tremaine:2012} developed a general statistical framework to account for observational biases of different techniques. Applying their method to planetary systems found by \emph{Kepler} and RV, \citet{Tremaine:2012} found that the mean mutual inclination dispersion, which was assumed to be the same for all multiplicities, should be $\lesssim 5^\circ$ and that the intrinsic multiplicity function could not be constrained. See \citealt{Figueira:2012} for a different attempt in combining \emph{Kepler} and RV data.

\citet{Tremaine:2012} also pointed out an observational feature that was difficult for their models to explain. As originally noticed by \citet{Ford:2011}, the fraction of systems showing TTV signals does not seem to vary significantly with the transit multiplicity, except perhaps for very high ($\ge 4$) multiplicities (see also \citealt{Xie:2014}). A similar feature also shows up in later large and uniform TTV searches, which consistently found that nearly half of the TTV detections were from systems with only one transiting planets \citep{Holczer:2016,Ofir:2018}. This indicates that planets in transit singles have almost the same probability to show TTV signals as planets in transit multis.

\subsubsection{Multiplicity-dependent mutual inclinations}

\begin{figure}
\includegraphics[width=\typewidth]{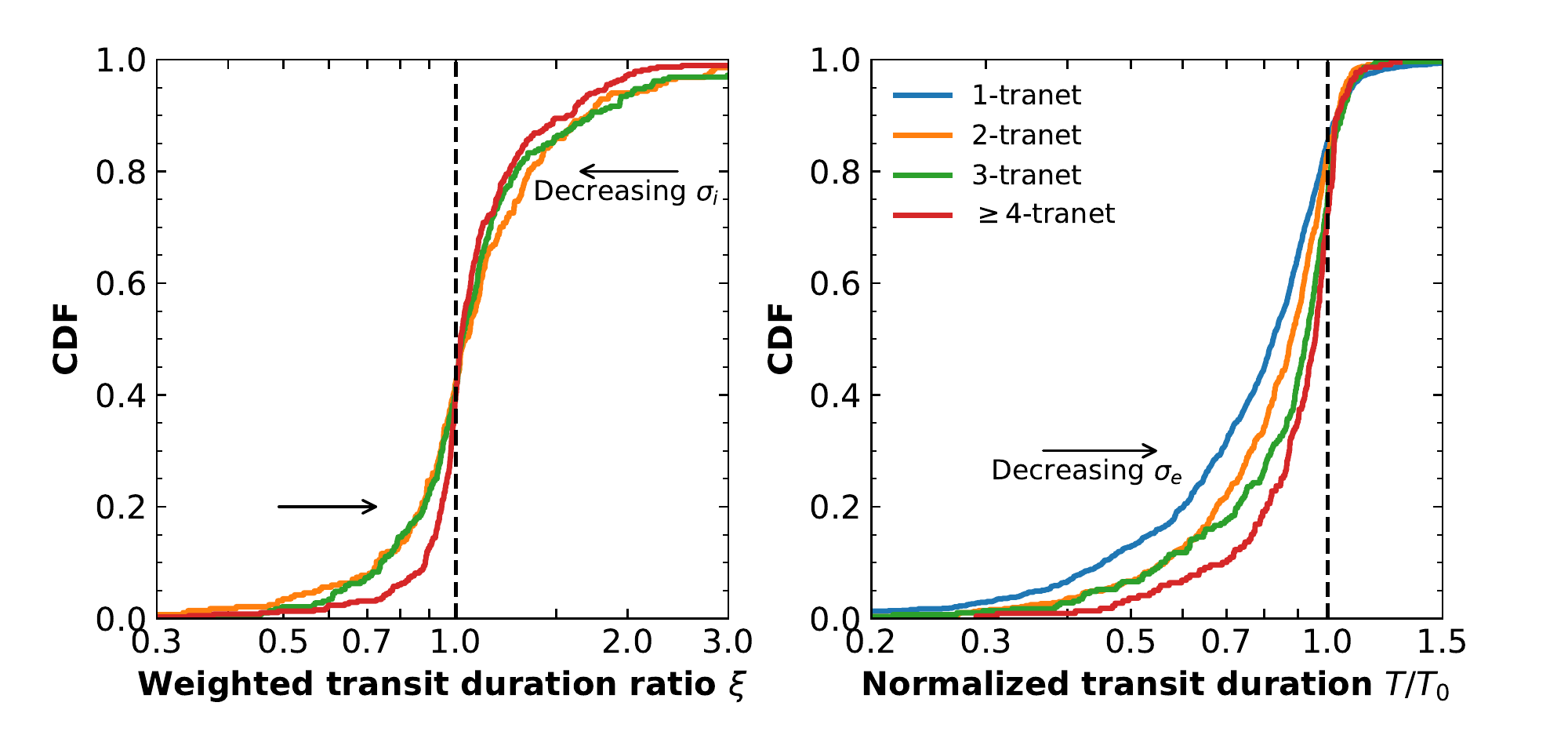}
\caption{(\textit{a}) The cumulative distribution functions (CDFs) of the weighted transit duration ratio, $\xi$ (\textbf{Equation~\ref{eqn:xi}}), for different transit multiplicities. Larger transit multiplicities tend to have narrower $\xi$ distributions, suggesting smaller mutual inclination dispersions $\sigma_i$. (\textit{b}) The CDFs of the normalized transit duration, $T/T_0$ (\textbf{Equation~\ref{eqn:ecc}}), for different transit multiplicities. Here $T_0$ is the transit duration for a circular and coplanar orbit. Larger transit multiplicities have narrower $T/T_0$ distributions, indicating smaller eccentricity dispersions $\sigma_e$.
\label{fig:durations}}
\end{figure}

\begin{figure}
\includegraphics[width=\typewidth]{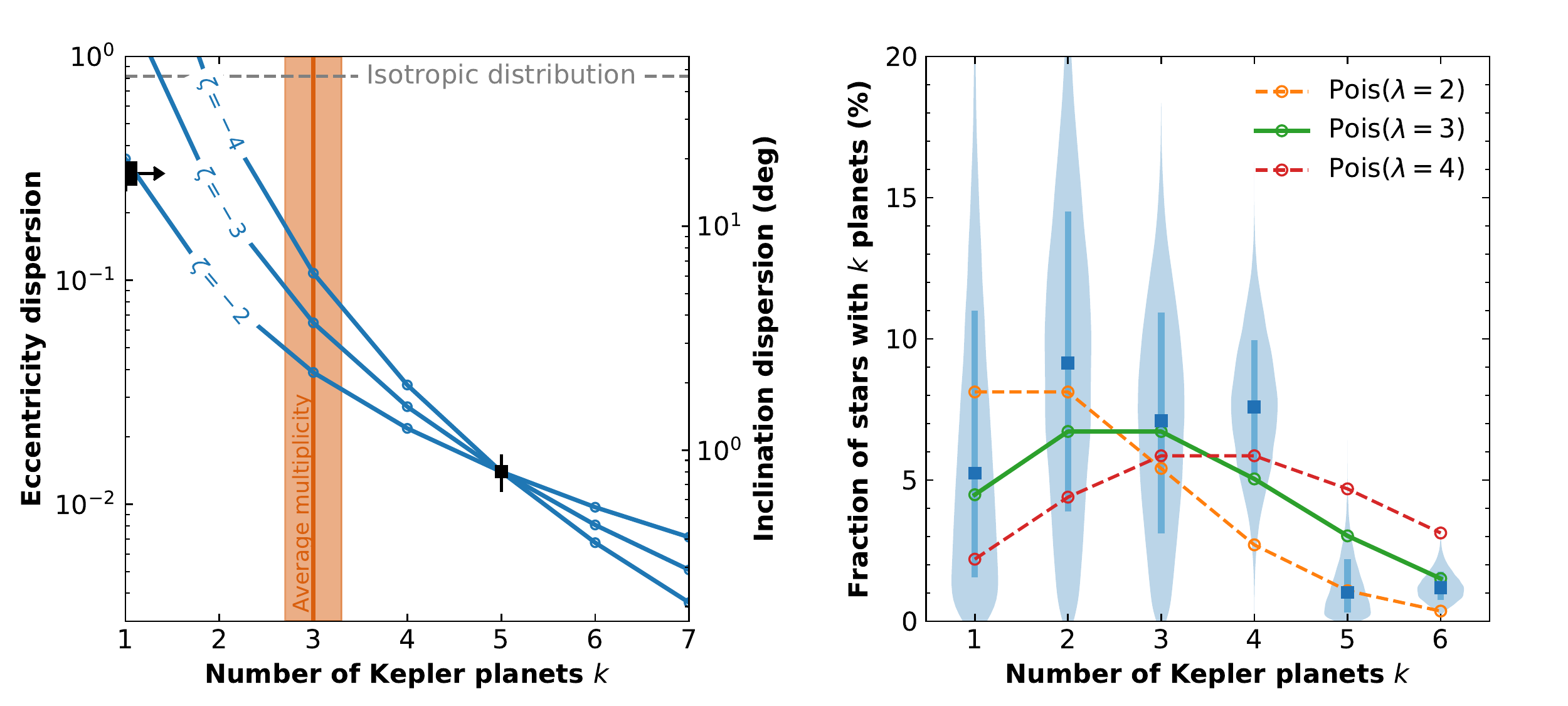}
\caption{(\textit{a}) Distributions of eccentricity and mutual inclination dispersions as functions of the intrinsic multiplicity. Here \emph{Kepler} window is roughly the region above the gray curve in \textbf{Figure~\ref{fig:kepler_planets}}. The relation $\sigma_e=\sigma_i$ is assumed (see Section~\ref{sec:ecc_sub}). The $\zeta$ parameter quantifies the strength of the correlation with the intrinsic multiplicity (\textbf{Equation~\ref{eqn:multiplicity_dependence}}). The orange band denotes the inferred average multiplicity $\bar{m}_{\rm p}$ for \emph{Kepler} systems. The eccentricity dispersion inferred from transit singles (i.e., $k\ge1$) and the mutual inclination dispersion constraint from systems with five \emph{Kepler} transiting planets are shown as black squares. 
(\textit{b}) The inferred intrinsic multiplicity vector and the associated probability distribution functions. The fraction of Sun-like stars with more than seven planets in the \emph{Kepler} window is limited to $<2.2\%$ ($95\%$ upper limit). The medians and the $16$--$84\%$ ranges of individual components are denoted with squares and error bars, respectively. Poisson distributions with different values of mean parameter $\lambda$ are shown for references.
Both plots are adapted from \citet{Zhu:2018}.
\label{fig:Zhu_2018}}
\end{figure}

The assumption that the mutual inclination distribution is independent of the intrinsic multiplicity may not be valid. With all else being equal, the critical mutual inclination for long-term instability is probably dependent on the number of planets in the system (e.g., \citealt{PuWu:2015}; see also Section~\ref{sec:spacings}). Observationally, one also finds that the distribution of the $\xi$ parameter appears statistically different for different transit multiplicities. As shown in \textbf{Figure~\ref{fig:durations}a}, lower transit multiplicities have broader $\xi$ distributions that are suggesting larger mutual inclinations (see also \citealt{He:2020}).

\citet{Zhu:2018} introduced the following relation between the mutual inclination dispersion, $\sigma_i$, and the intrinsic multiplicity (within \emph{Kepler} window), $k$,
\begin{equation} \label{eqn:multiplicity_dependence}
\sigma_i(k) = 0.8^\circ \left(\frac{k}{5}\right)^\zeta ~.
\end{equation}
They applied the statistical framework of \citet{Tremaine:2012} and combined the transit and TTV statistics to infer the intrinsic multiplicity and mutual inclination distributions. TTV, as a detection technique \citep{Agol:2005,Holman:2005}, applies to the same population of planetary systems as transit, and thus the combination of TTV and transit is free from many assumptions and selection biases (compared to the use of RV; e.g., \citealt{Tremaine:2012}).
\citet{Zhu:2018} found that the intrinsic multiplicity and the mutual inclination dispersion should be strongly correlated, with $-4<\zeta<-2$ at the $2\,\sigma$ confidence level (see \textbf{Figure~\ref{fig:Zhu_2018}a} for an illustration). In other words, systems with fewer planets are dynamically hotter. This result also points to large mutual inclinations ($\gtrsim10^\circ$) for 2-planet and 3-planet systems. A recent work by \citet{He:2020} found a qualitatively similar (although statistically different) result with a best-fit $\zeta=-1.7$ from modeling a collection of \emph{Kepler} statistics (including transit multiplicities, the period distribution, period ratio distribution, etc) and imposing the angular momentum deficit (AMD) stability criterion \citep{Laskar:1997,Laskar:2017} in simulated planetary systems. It is also worth noting that such a relation is steeper than the similar relation inferred from RV eccentricities \citep{Limbach:2015}, ergodic models \citep{Tremaine:2015}, or the extrapolations of the empirical stability boundary \citep[e.g.,][]{PuWu:2015}.

\citet{Zhu:2018} also reported constraints on the intrinsic multiplicity vector, which is reproduced in \textbf{Figure~\ref{fig:Zhu_2018}b}. Although the individual components of the multiplicity vector are not well constrained, the summed fraction is well measured to be $30\pm3\%$ and does not rely on many assumptions like the other measurements do (see Section~5.1 of \citealt{Zhu:2018}; see also Section~\ref{sec:occurrence_rate}). The resulting average multiplicity in the \emph{Kepler} parameter space is $\bar{m}_{\rm p}=3.0\pm0.3$. This serves a lower bound on the average multiplicity in the inner ($\lesssim1\,$AU) region, as smaller planets below the detection threshold of \emph{Kepler} are unconstrained.

\subsection{Eccentricity distribution} \label{sec:eccentricity}

Similar to mutual inclinations, orbital eccentricities also provide important information on the formation and dynamical evolution of planetary systems. Here we focus on the eccentricity results from the \emph{Kepler} sample. Readers can find discussions about eccentricities from RV in the Section 3.1 of \citet{WinnFabrycky:2015}.

The majority of the eccentricity measurements of individual \emph{Kepler} planets were made through modeling the TTV and TDV signals \citep[e.g.,][]{Lithwick:2012,Wu:2013,Hadden:2014,Hadden:2017}. These studies have found that the eccentricities of \emph{Kepler} planets in near-resonance pairs are typically small, with a Rayleigh dispersion of up to a few percent. However, the planets selected for such dynamical modelings are probably a biased sample, and thus the derived eccentricity distribution may not be representative of the more general population.

\subsubsection{The transit duration method}

The transit duration (between the first and the fourth contact points) is given by
\footnote{Note that our definition of the transit duration follows that of \citet{Seager:2003} and is different from that of \citet{WinnFabrycky:2015}. The latter measures the duration between two points where the planetary center sits on the edge of the projected stellar surface (see Figure~2 of \citealt{Winn:2010}).}
\begin{equation} \label{eqn:ecc}
\frac{T}{T_0} = \sqrt{(1+r)^2-b^2} \frac{\sqrt{1-e^2}}{1+e\sin{\omega}} ~.
\end{equation}
Parameters $r$, $b$, and $T$ are the same as those in \textbf{Equation~\ref{eqn:xi}}, and $\omega$ is the argument of periapsis. The quantity $T_0$ measures the transit duration between the first (second) and the third (fourth) contact points of a planet with the same period but circular ($e=0$) and edge-on ($b=0$) orbit and is related to the mean density of the host star, $\rho_\star$, via
\begin{equation}
T_0 \equiv \frac{R_\star P}{\pi a} = 13~{\rm hr} \left(\frac{P}{\rm yr}\right)^{1/3} \left(\frac{\rho_\star}{\rho_\odot}\right)^{-1/3} ~.
\end{equation}
With known parameters from transit modeling ($b$, $r$, $P$, and $T$) and the nuisance parameter $\omega$ assumed to follow a uniform distribution, the quantity $T/T_0$ can be used to constrain the statistical distribution of $e$, provided that the stellar mean density is precisely measured \citep{Ford:2008}. With other parameters being the same, larger eccentricities lead to broader distributions of the $T/T_0$ ratio (see \textbf{Figure~\ref{fig:durations}b}).
The successful application of this method heavily depends on the accurate characterizations of the host stars. As a result, early attempts to study the \emph{Kepler} sample were all limited by the systematic uncertainties in the stellar properties \citep[e.g.,][]{Moorhead:2011,Kane:2012,Plavchan:2014}.

\subsubsection{Multiplicity-dependent eccentricity distribution} \label{sec:ecc_sub}

\citet{VanEylen:2015} applied a variant of the transit duration method to a carefully selected sample of \emph{Kepler} multi-planet systems whose host stars were precisely characterized via asteroseismology. These authors found that the eccentricities of planets in their sample could be well described by a Rayleigh distribution with $\sigma_e\approx0.05$. Using  accurate spectroscopic stellar parameters from LAMOST, \citet{Xie:2016} found similar nearly-circular orbits for planets in the \emph{Kepler} multis, and they reported a much larger eccentricity dispersion ($\sigma_e\approx0.3$) for \emph{Kepler} planets in systems with single transiting planets. Both results have been confirmed by later works \citep{VanEylen:2019,Mills:2019}.

The multiplicity-dependent eccentricity distribution goes beyond the single vs.\ multiple bifurcation. This is demonstrated in \textbf{Figure~\ref{fig:durations}b}, where we show the cumulative distributions of the $T/T_0$ ratios derived from our planet sample for different transit multiplicities. Here we have used the stellar mean densities from isochrone fits by \citet{Berger:2020a} and the values of $T$ from the \emph{Kepler} DR25 MCMC chains \citep{Hoffman:2017}.
As \textbf{Figure~\ref{fig:durations}b} indicates, the distribution of $T/T_0$ ratio becomes narrower with increasing transit multiplicities. As the transit multiplicity can be viewed as a rough proxy of the intrinsic planet multiplicity, it is suggestive that planetary systems with more planets have smaller eccentricity dispersions. This is also qualitatively consistent with studies of the RV planets \citep{Limbach:2015,Zinzi:2017}. Based on observations of solar system and the general expectation that the dispersions of orbital eccentricity and mutual inclination are proportional to each other \citep{Ida:1993,Tremaine:2012,Xie:2016}, we may use the same relation between intrinsic multiplicity and mutual inclination dispersion (\textbf{Equation~\ref{eqn:multiplicity_dependence}}) for the relation between intrinsic multiplicity and orbital eccentricity dispersion. The multiplicity-dependent eccentricity dispersion is also shown in \textbf{Figure~\ref{fig:Zhu_2018}a} (see also \citealt{He:2020}). 

The large eccentricities and mutual inclinations of \emph{Kepler} low-multiples have important theoretical implications. The largest eccentricity that can be achieved via scatterings among small \emph{Kepler} planets themselves can be roughly estimated as:
\begin{equation}
e_{\rm max} \sim \frac{v_{\rm esc}}{v_{\rm orb}} \approx \sqrt{\frac{2 m_{\rm p} a}{M_\star R_{\rm p}}} = 0.15 \left(\frac{m_{\rm p}/M_\star}{10^{-5}}\right)^{1/2} \left(\frac{a}{0.1~\rm AU}\right)^{1/2} \left(\frac{R_{\rm p}}{2~R_\oplus}\right)^{-1/2} ~.
\end{equation}
Here $v_{\rm esc}$ and $v_{\rm orb}$ are the surface escape velocity and orbital velocity of the planet, respectively. The evaluation takes the typical values of a \emph{Kepler} planet. While the above scaling relation bears some significant uncertainties, the large eccentricities ($\sigma_e\approx0.3$) and mutual inclinations ($\sigma_i\gtrsim10^\circ$) observed in the low-multiplicity planetary systems are probably on the high end of the distribution. It suggests that these planetary systems may have undergone significant dynamical interactions among the inner planets themselves. Alternatively, other mechanisms may have been invoked to excite eccentricities and mutual inclinations to values larger than what the self-scatterings can achieve. One promising mechanism is the interaction between the inner system and the outer massive planets \citep[e.g.,][and references therein]{Johansen:2012,Huang:2017,PuLai:2020}. We return to this point in Section~\ref{sec:correlation}.

\subsection{Intra-system variation} \label{sec:intra-system}

The intra-system variation, which is about the relative properties of planets around the same host, is useful in constraining the formation and evolution processes of planetary systems.
It also concerns the statistical inference of exoplanets in general: In some statistical studies, planet detections from the same star are treated as independent events (see Sections~\ref{sec:idem} and \ref{sec:multiplicity}); in some others, specific assumptions about the relative properties of planets in multi-planet systems must be made when synthetic systems are generated \citep[e.g.,][]{Mulders:2018,He:2019}. The derived statistics to some extent are subject to the validity of such assumptions.

\subsubsection{``Peas in a pod?''} \label{sec:peas}

Transiting planets in the same \emph{Kepler} multi-planet systems preferentially have similar sizes. This feature has been noticed since the early days of the \emph{Kepler} mission \citep{Lissauer:2011,Ciardi:2013}. Follow-up observations that provided improved characterizations of the host stars enabled further studies that tried to understand the nature of this feature \citep{Weiss:2018a,He:2019,Zhu:2020,Weiss:2020,Murchikova:2020}. In particular, \citet{Weiss:2018a} quantified the correlation between sizes of neighboring planets around the same host in their sample. To check the statistical significance of this correlation, they generated synthetic systems by randomly drawing planetary radii from the observed size distribution and then performed the same correlation test. The size correlations in their synthetic systems were much weaker than what they saw in real systems, and thus they concluded the pattern was astrophysical. Together with a similar result on the spacings between planets, \citet{Weiss:2018a} concluded that planets in \emph{Kepler} multi-planet systems have similar sizes and regular spacings, a pattern they termed ``peas in a pod'' (see also \citealt{Millholland:2017} for a similar claim about \emph{Kepler} planet masses). A later study by \citet{He:2019} reached a similar conclusion. According to these authors, planetary systems that contain clusters of planets whose sizes and orbital periods are correlated produce a better match to the observed \emph{Kepler} systems in terms of the joint statistics of transit depth distribution, period distribution, period ratio distribution, etc.\footnote{The clustered model of \citet{He:2019} has more free parameters than their non-clustered model. However, the authors did not perform model comparisons to justify the introduction of more flexibilities. See \citet{Zhu:2020} for more discussions.}

Different opinions exist about the nature of the observed correlations. \citet{Zhu:2020} pointed out a detection bias that was underestimated in the statistical method of \citet{Weiss:2018a}. Because small planets can be detected around bright and quiet stars whereas large planets are only detectable around faint or noisy stars, the same transit detection threshold (i.e., a fixed S/N) naturally leads to varying planetary size thresholds in different systems. This, combined with the fact that smaller planets are more abundant, naturally leads to a size correlation in the observed transit pairs (see also \citealt{Murchikova:2020}). However, it appears that the apparent correlation in planetary sizes is too strong to be explained entirely by this detection bias alone \citep{Zhu:2020}.

Another factor that has not been fully explored is the contribution of the planets that are missing, due to large impact parameters or sub-threshold values of transit S/N, in known \emph{Kepler} multi-planet systems. Our solar system is an excellent example to demonstrate this point. The four outer giant planets would be unlikely to be detected by a transit mission similar to \emph{Kepler} because of their long orbital periods. Of the four terrestrial planets, Mercury and Mars are almost impossible to detect in transit due to their small sizes. Therefore, a \emph{Kepler}-like mission would, if possible at all, most likely detect the Venus--Earth planet pair, which shows very similar sizes ($0.95\,R_\oplus$ vs.\ $1\,R_\oplus$) and masses ($0.82\,M_\oplus$ vs.\ $1\,M_\oplus$). However, this level of similarity is not representative among the solar system planet pairs. 

The physical interpretation of the size correlation (if any) is also unclear. One interpretation is that planets ``know'' about their siblings, namely the formations of two neighboring planets are directly correlated \citep[e.g.,][]{Kipping:2018,Mulders:2018,Sandford:2019,He:2019,Gilbert:2020}. Another interpretation is that planets ``know'' about the system and the environment they formed in, namely the formations of planets in the same system are all related to some global properties \citep{Murchikova:2020}. In this latter case, the apparent correlation between planetary sizes is only a projection of the correlation between the individual planets and the host star (or the birth disk).
This latter interpretation has some observational evidence. For example, the planet distribution is shown to depend on the orbital period (see Section~\ref{sec:kepler_planets}) and stellar properties (see Section~\ref{sec:stellar_dependence}). \citet{Murchikova:2020} demonstrated with a toy model that the observed size correlation could be well reproduced if the planets ``know'' about the host star but do not ``know'' about their neighbor planets.

\subsubsection{Orbital spacings} \label{sec:spacings}

\begin{figure}
\includegraphics[width=\textwidth]{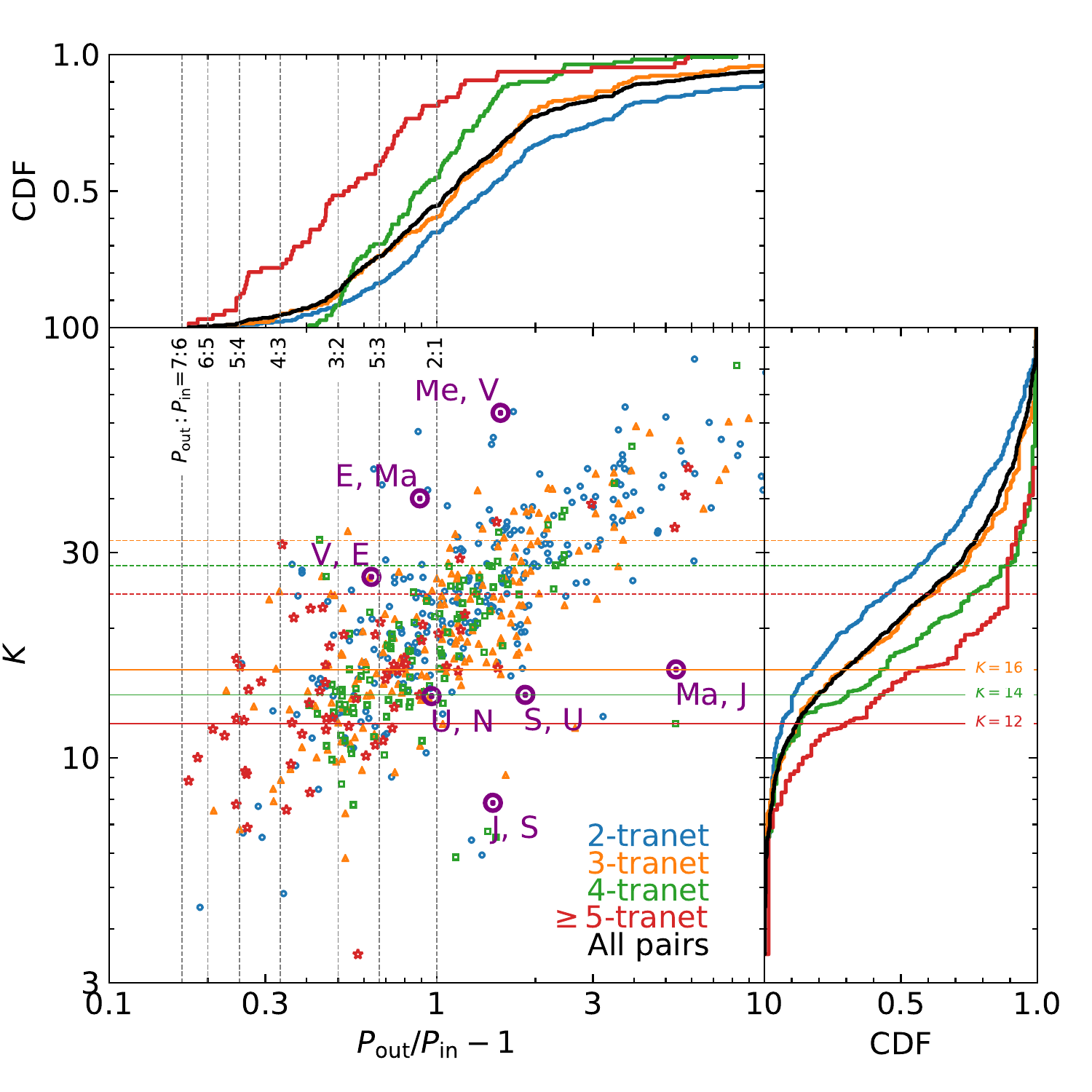}
\caption{Spacings between the apparently adjacent planets in \emph{Kepler} multi-planet systems, with different colors indicating planet pairs from different transit multiplicities. The $x$-axis of the main panel (lower left) shows the spacing in terms of the orbital period ratio, and the upper panel shows the corresponding cumulative distributions. A few example period commensurabilities are indicated in both panels.
The $y$-axis of the main panel shows the spacing in terms of the mutual Hill radii (\textbf{Equation~\ref{eqn:hill_spacing}}), and the right panel shows the corresponding cumulative distributions. The stability thresholds for 3-tranet, 4-tranet, and 5-tranet systems, derived according to \textbf{Equation~\ref{eqn:kepler_spacing}}, are indicated with solid horizontal lines. Values corresponding to twice of the thresholds are also shown as dashed horizontal lines.
The code \texttt{Forecaster} from \citet{ChenKipping:2017} is used to predict the planet mass based on the planetary radius, and we have revised the upper mass limit to $10^3\,M_\oplus$ ($\sim3~M_{\rm J}$) to avoid masses beyond the planetary regime. 
Solar system planet pairs are also indicated for references.
\label{fig:spacings}}
\end{figure}

The relative positions of planets in \emph{Kepler} multi-planet systems have also drawn lots of interest. The majority of the early studies focused on the period ratio distribution.
As shown in \textbf{Figure~\ref{fig:spacings}}, \emph{Kepler} systems contain very few planet pairs near/in low-order mean-motion resonances (see also \citealt{Lissauer:2011,Fabrycky:2014}). This is in contrast with earlier RV results that a substantial fraction of well-characterized multi-planet systems contain pairs of giant planets close to mean-motion resonances \citep[e.g.,][]{Wright:2011}. We refer to Section~\ref{sec:theory_overview} for the theoretical implications of this feature.
Additionally, the asymmetry around exact period commensurabilities has also attracted lots of attention \citep{Fabrycky:2014}, and we refer interested readers to the fairly comprehensive overview by \citet{Terquem:2019} for this particular issue (see also the recent development by \citealt{Millholland:2019}). This review focuses on the dynamical compactness of the \emph{Kepler} multi-planet systems, which concerns the long-term stability and thus the dynamical evolution.

When the stability of the planetary system is concerned, the orbital spacing between planets is usually expressed in the dimensionless parameter $K$
\begin{equation} \label{eqn:hill_spacing}
K \equiv \frac{a_{\rm out}-a_{\rm in}}{R_{\rm H}}; \quad
R_{\rm H} \equiv \frac{a_{\rm in}+a_{\rm out}}{2} \left(\frac{m_{\rm in}+m_{\rm out}}{3M_\star}\right)^{1/3}\ .
\end{equation}
Here $R_{\rm H}$ is called the mutual Hill radius, $M_\star$ is the mass of the host star, and $a_{\rm in}$ ($a_{\rm out}$) and $m_{\rm in}$ ($m_{\rm out}$) are the semi-major axis and the mass of the inner (outer) planet, respectively. For two-planet systems, the condition for the long-term stability (and thus instability) has been well understood theoretically, and the instability arises when there are mean-motion resonance overlaps \citep{Wisdom:1980,Deck:2013,Hadden:2018}. For systems with more than two planets, we lack a good theoretical understanding on the origin of the dynamical instability (see attempts by \citealt{Chambers:1996,Zhou:2007,Quillen:2011,Yalinewich:2020}). Nevertheless, numerical studies have shown that the timescale before which close encounter occurs between planets, $t$, scales exponentially with the initial spacing $K$ \citep{Chambers:1996}. Details of this scaling relation depend on factors such as the number of planets, planet masses, orbital eccentricities and inclinations, as well as the inhomogeneity among planets (e.g., \citealt{Chambers:1996,Zhou:2007,Funk:2010}; see \citealt{PuWu:2015} for a recent summary).

In the context of \emph{Kepler} planetary systems, \citet{PuWu:2015} found through numerical simulations that the median spacing for stability could be approximated as
\begin{equation} \label{eqn:PuWu15}
\langle K \rangle = 2.87 + 0.7 \log_{10}{\tau} + 2.4 \left[ \left(\frac{\sigma_e}{e_{\rm H}}\right) + \left(\frac{\sigma_i}{4e_{\rm H}}\right) \right] ,
\end{equation}
where $\tau$ is the physical timescale $t$ scaled by the orbital period of the innermost planet, $e_{\rm H}$ is the mutual Hill radius scaled by the semi-major axis of the innermost planet, and $\sigma_e$ and $\sigma_i$ are the dispersions of orbital eccentricities and mutual inclinations among the planets, respectively. With the multiplicity-dependent $\sigma_e$ and $\sigma_i$ (\textbf{Equation~\ref{eqn:multiplicity_dependence}}) and the typical values for \emph{Kepler} systems ($t\gtrsim$ Gyr old and the innermost planet of planet-to-star mass ratio $q\approx10^{-5}$ at $0.1~$AU), \textbf{Equation~\ref{eqn:PuWu15}} yields
\begin{equation} \label{eqn:kepler_spacing}
\langle K \rangle \approx 10.2 + 2.2 \left( \frac{k}{5} \right)^\zeta .
\end{equation}
With $\zeta=-2$ \citep{Zhu:2018,He:2020}, planetary systems with (3, 4, 5) \emph{Kepler} planets should have critical spacings $\langle K \rangle=(16,~14,~12)$, respectively. 

We apply the above stability thresholds to the multi-planet systems from Section~\ref{sec:kepler_planets} and discuss the limitations.
After the use of Kepler's third law, the only unknown to determine the spacing parameter $K$ is the planet-to-star mass ratio. We estimate the planetary masses from the measured radii with the \texttt{Forecaster} code from \citet{ChenKipping:2017} and adopt the \emph{Gaia} stellar mass from \citet{Berger:2020a}. Systems without reported stellar mass measurements are excluded.
\textbf{Figure~\ref{fig:spacings}} illustrates the spacings between neighboring \emph{Kepler} planets of all systems and systems divided into different transit multiplicities. For transit multiplicities of $3$, $4$, and $5+$, the majority ($\sim70\%$) of planet pairs have spacings above the corresponding stability thresholds, confirming that they are indeed (most likely) long-term stable. The remaining $\sim30\%$ planet pairs, considered long-term unstable by the above empirical thresholds, are probably stable as well. While part of this misclassification is due to the choice of fixed \emph{Kepler} system parameters and the empirical (but sometimes unphysical) mass--radius relation \citep{ChenKipping:2017}, it nevertheless is a sign for the failure of the empirically determined stability criteria. In particular, these stability criteria do not take into account the impact of mean-motion resonances, which can be either protective or destructive to the involved planets.

Nevertheless, by applying the empirical stability thresholds to the data one finds that the majority of \emph{Kepler} planet pairs are not far from the empirical stability limits: The median spacing of all planet pairs is $K\approx20$, and about $80$--$90\%$ of planet pairs from systems with at least three transiting planets have spacings within twice of the empirical stability thresholds (horizontal dashed lines in \textbf{Figure~\ref{fig:spacings}}). These results are consistent with previous findings \citep[e.g.,][]{FangMargot:2013,PuWu:2015,Weiss:2018a} and also suggest that for the majority of \emph{Kepler} planet pairs there is no room for inserting another (undetected) planet in between \citep{FangMargot:2013}. In other words, the observed \emph{Kepler} planets are dynamically packed.
However, it does not necessarily mean that \emph{Kepler} systems do not contain additional planets. The space to the innermost and particularly the outermost \emph{Kepler} planet allows the existence of additional planets without risking instability. For example, seven planets with $q=10^{-5}$ are allowed per factor of 10\ in semi-major axis if mutually separated by $K=20$.
The observed dynamically packed structure is also probably due to the selection bias that it is increasingly difficult for both planets in a wider-spacing pair to transit the host star.

As part of the ``peas in a pod'' claim (see Section~\ref{sec:peas}), the spacings between \emph{Kepler} planets in the same multi-planet system are found to be statistically similar \citep{Weiss:2018a}. However, the observed correlation in spacings is driven by a small fraction ($\lesssim5\%$) of systems containing the highest multiplicities, and the majority of systems do not show such a regular spacing pattern \citep{Zhu:2020,Jiang:2020}. 

\subsection{Dependence of planet statistics on stellar properties} \label{sec:stellar_dependence}

\subsubsection{Impact of stellar companions} \label{sec:binary}

Stellar companions to the planet hosts affect the \emph{Kepler} planet statistics in several ways. In transit surveys like \emph{Kepler}, many of them appear unresolved and dilute the transit signals, potentially leading to misclassifications and erroneous planetary parameters \citep{Ciardi:2015,Bouma:2018}.\footnote{Here an ambient star that is not physically associated with the target is also considered a companion to the target star.}
Thankfully, follow-up high-resolution imaging observations have been performed for nearly all \emph{Kepler} planet candidates \citep[e.g.,][and references therein]{Furlan:2017,Ziegler:2018}. For bright targets that contain Jupiter-like transits, \citet{Santerne:2016} also performed systematic radial-velocity follow-up observations and identified a significant false positive rate ($55\%$) for Jovian planet candidates. These efforts have led to a much better understanding of the impact of transit dilution on the \emph{Kepler} planet statistics. In particular, \citet{Furlan:2017} reported that about $10\%$ ($30\%$) of the candidate host stars have observed companions within $1''$ ($4''$), the majority of which are fairly faint compared to the target stars. In the most likely scenario that the transit signals come from the primary stars \citep[see, e.g.][]{Bouma:2018}, the dilution effect overall only affects the planetary radii up to a few percent on average \citep{Furlan:2017}. This is within the uncertainty of \emph{Gaia}-derived radii, and thus one does not expect it to have a significant impact on the general planet statistics. However, Earth-sized planets $R_p\lesssim 2\,R_\oplus$ are much more susceptible to the dilution effect, and thus the relevant statistics may suffer a more dramatic impact \citep{Furlan:2017, Bouma:2018}.

Besides the transit dilution effect, stellar companions can also affect the presence of planets through many dynamical processes \citep[e.g.,][]{Artymowicz:1994,Holman:1999}. Very close ($a\lesssim0.5~$AU) stellar binaries can host circumbinary (i.e., planetary-type or P-type) planets, and over a dozen such systems have been found (see Section~6.2 of \citealt{WinnFabrycky:2015}). We limit our discussions to the observational aspects of circumstellar (i.e., satellite-type or S-type) planets and the implications on planet statistics. 

Studies based on RV and high-resolution imaging observations suggest that the existence of a close stellar-mass companion is usually associated with a lower frequency of circumstellar planets \citep[e.g.,][]{Wang:2014,Wang:2015,Ngo:2016,Kraus:2016,Moe:2020}. The effect of binary on the presence of close-in planets is quantified by a suppression factor $S_{\rm bin}$, which is the ratio between the fraction of planet hosts with stellar companions and the fraction of field stars with the same type of companions \citep{Kraus:2016,Moe:2020}. Circumstellar planets are almost completely suppressed ($S_{\rm bin}\lesssim15\%$) when the stellar companions are close (with separation $a_{\rm bin}\lesssim10\,$AU), regardless of the planetary size or observed multiplicities. Planets are nearly unaffected ($S_{\rm bin}\gtrsim85\%$) if the stellar companions are distant ($a_{\rm bin}\gtrsim100\,$AU). At intermediate separations ($\sim10$--$100\,$AU), the suppression effect gradually decreases with the increasing separation. See the Figure 3 of \citet{Moe:2020} for a compilation of observational studies and an illustration of the suppression factor $S_{\rm bin}$ as a function of the binary separation.

With the above suppression effect and the known binary separation distribution, one can then infer the planet formation efficiency from the measured planetary system frequency $F_{\rm p}$. \citet{Moe:2020} estimated that $F_{\rm bin}\approx43\%$ of Sun-like primaries in a magnitude-limited survey like \emph{Kepler} could not host close-in ($\lesssim1~$AU) planets simply because of the influence of binary companions \citep[see also][]{Kraus:2016}.
If these targets are excluded from the \emph{Kepler} statistics, one finds that the formation efficiency of close-in planets around single stars, a parameter directly related to formation theories, should be $1/(1-F_{\rm bin})=1.8$ times higher than the fraction of stars with planets $F_{\rm p}$. This additional factor also provides a plausible explanation to the discrepancy in hot Jupiter frequencies measured from RV and \emph{Kepler} (see Section~\ref{sec:hot_Jupiters}).

\subsubsection{Metallicity effect} \label{sec:metallicity}

\begin{figure}
\includegraphics[width=\typewidth]{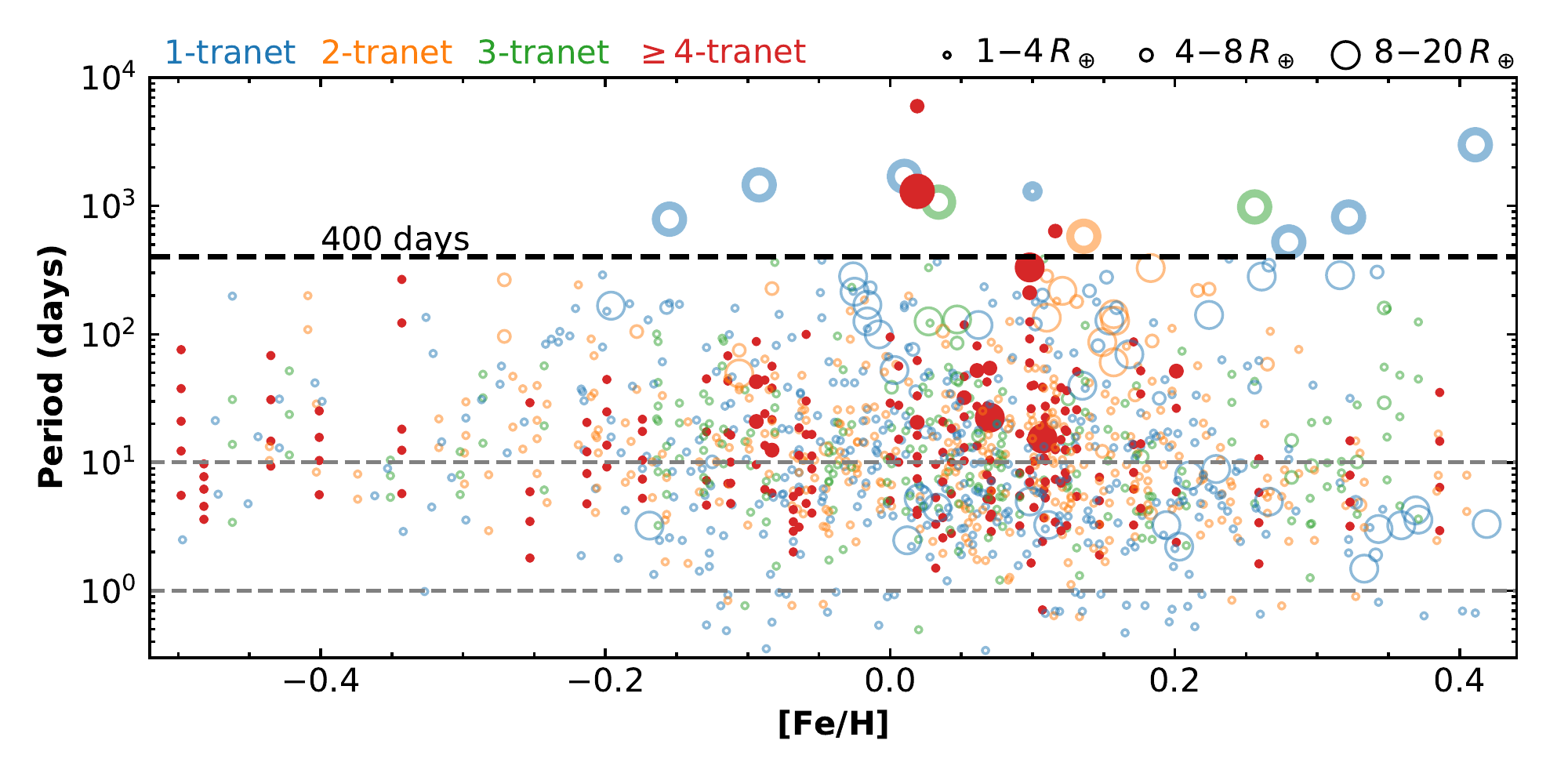}
\caption{An illustration of the \emph{Kepler} planetary systems in our baseline sample that have spectroscopic metallicity measurements. We use different colors to separate different observed multiplicities (out to 400$\,$days) and different label sizes to separate planets of different sizes (small as 1--4$\,R_\oplus$, intermediate as 4--8$\,R_\oplus$, and giant as 8--20$\,R_\oplus$). Cold ($P>400\,$days) planets found by RV (as tabulated in \citealt{ZhuWu:2018}) and long-period transit searches \citep{Kawahara:2019} are also indicated with thick circles. 
As the host metallicity increases, the system becomes more likely to contain giant planets at all periods and small planets at relatively close-in ($P\lesssim10\,$days) orbits \citep[e.g.,][]{Mulders:2016,Dong:2018,Petigura:2018}. At very high metallicities ([Fe/H]$\gtrsim0.2$), there seems to be a deficit of compact systems (with $\ge4$ transiting planets) and planets at intermediate orbits ($\sim$10--400$\,$days). These may be related to the emerging cold giants \citep{ZhuWu:2018}.
The median metallicity of \emph{Kepler} field stars is [Fe/H]$\approx0.0$ \citep{Dong:2014a}.
\label{fig:metallicity}}
\end{figure}

Under the general assumption that the bulk metallicity of the host star is correlated to the total mass of building blocks available for planet formation, it is reasonable to believe that the planetary occurrence rate and properties may be correlated with the host star metallicity. 
For giant planets ($R_{\rm p}\gtrsim8\,R_\oplus$ or $m_{\rm p}\gtrsim0.3\,M_{\rm J}$) found by RV, it has been well established that their presence correlates strongly with the host metallicity \citep[e.g.,][]{Santos:2001,Fischer:2005}. This giant planet--metallicity correlation lends support to the core accretion model as the leading theory for the formation of giant planets \citep[e.g.,][]{Pollack:1996,IdaLin:2004}. Some recent studies have also claimed that hosts of eccentric giant planets are more metal-rich than hosts of nearly circular giant planets \citep{Dawson:2013,Buchhave:2018}, but stronger statistical evidence is needed to fully establish this result.

Small planets, in particular those with radii $R_{\rm p}\lesssim4\,R_\oplus$, show weaker dependences on host metallicity \citep[e.g.,][]{Sousa:2008,Buchhave:2012}. While many studies have focused on the dependence of the planet frequency $\bar{n}_{\rm p}$ on host metallicity \citep[e.g.,][]{WangFischer:2015,Petigura:2018} and theoretical implications \citep[e.g.,][]{OwenMurrayClay:2018, Lee:2019}, one may argue that the planetary system frequency $F_{\rm p}$ is probably a more suitable parameter to characterize the efficiency of planet formation under such system-wide parameters like metallicity \citep{Zhu:2016,Zhu:2019}. If the general planet--metallicity relation
\begin{equation} \label{eqn:metallicity}
F_{\rm p} \propto 10^{\gamma \rm [Fe/H]}
\end{equation}
is applied, the result of \citet{Zhu:2019} suggests $\gamma\approx0.5$ for all \emph{Kepler}-type planets, which is much weaker than the giant planet--metallicity correlation ($\gamma\approx2$; \citealt{Fischer:2005}). The dependence is further reduced if the close binaries that show anti-correlation with stellar metallicity are excluded from the statistics \citep{Moe:2019,Kutra:2020}. Unlike the planetary system frequency $F_{\rm p}$, the planet frequency $\bar{n}_{\rm p}$ does not appear to have a monotonic relation with the host metallicity. In particular, it may start declining when the metallicity is high enough \citep{Zhu:2019}. It has been suggested that this behavior may be related to the formation of giant planets inside the same system: as the metallicity is high enough, the system has a significant probability to form giant planets, and these giants may reduce the multiplicity of the inner system either because they prohibit the formation of more small planets or because they dynamically remove some of the small planets out of the inner system. This scenario may also explain the increased diversity of planets around metal-rich \emph{Kepler} hosts \citep{Petigura:2018} and the over-abundant compact planetary systems around metal-poor stars \citep{ZhuWu:2018,Brewer:2018}. \textbf{Figure~\ref{fig:metallicity}} displays along the host metallicity [Fe/H] the \emph{Kepler} systems with metallicity measurements in our baseline sample.

While the stellar bulk metallicity measured in iron abundance [Fe/H] (or a mix of metals [m/H]) is usually used in studies of the planet metallicity dependence, other elemental abundances, in particular $\alpha$ elements and refractory elements, have also been looked for possible correlations with planet properties \citep[e.g.,][]{Adibekyan:2012,Liu:2016,Teske:2019}. No clear trends have been found so far, probably due to the limited sample size, the measurement precision, and/or the impact of Galactic chemical evolution. 

\subsubsection{Dependence on stellar mass} \label{sec:stellar_mass}

A number of studies have also investigated the dependence of planet frequency on host mass. A theoretical possibility is that, the stellar mass correlates with the total mass in the protoplanetary disk and thus the amount of solid materials available for planet formation. It is largely consistent with direct observations of protoplanetary disks in (sub-)millimeter wavelengths \citep{Andrews:2013,Ansdell:2016}, although at a fixed stellar mass the scatters of inferred disk masses remain substantial (up to an order of magnitude; \citealt{Ansdell:2016}).

We would like to start by pointing out several potential issues. Similar to the metallicity dependence (see Section~\ref{sec:metallicity}), the two frequencies, $F_{\rm p}$ and $\bar{n}_{\rm p}$, can behave differently, especially for the small planets with high multiplicity rates. Second, as more massive stars also tend to be more metal-rich, one may need to carefully disentangle possible correlation between stellar mass and metallicity in the sample \citep[e.g.,][]{Johnson:2010,Kutra:2020}. Furthermore, the choice of the parameter to study the correlation may matter. While planetary radius (or mass) and orbital period (or semi-major axis) are commonly used in statistical studies, Nature may prefer other physical units such as the planet-to-star mass ratio or the position of the water snow line (\citealt{Hayashi:1981,Kennedy:2008}; see \textbf{Figure~\ref{fig:q_temperature}} for an illustration). Last but not least, as far as the planet formation efficiency is concerned, one must correct for the suppression effect due to close stellar binaries (see Section~\ref{sec:binary}). It is established that the close binary fraction correlates with the primary mass \citep[e.g.,][]{Duchene:2013}, so the suppression effect is expected to affect the statistics of planets around different stellar masses differently \citep{Moe:2020}.

The dependence of giant planets on stellar mass has been investigated in many studies with different detection methods \citep[e.g.,][]{Johnson:2007,Johnson:2010,Howard:2012,Fressin:2013,Nielsen:2019}. To avoid many of the issues listed above, here we focus on the results from long-term RV surveys, as they cover a broad range of parameter space and are nearly free of close stellar binaries. In particular, \citet{Johnson:2010} analyzed a sample of 1266 stars with at least 3-year RV observations and masses spanning from $0.2\,M_\odot$ up to their estimated $1.9\,M_\odot$, and reported a linear relation between the occurrence rate
\footnote{Their derived occurrence rate is technically $\bar{n}_{\rm p}$, but because of the low multiplicity rate of giant planets it closely approximates the rate $F_{\rm p}$.}
and stellar mass. This result has been widely considered as a benchmark in both theoretical and observational studies of giant planets. The higher-mass part of the sample come from the so-called ``retired A-stars,'' and their spectroscopic mass estimates are controversial  (\citealt{Lloyd:2011,Schlaufman:2013,Malla:2020} and references therein). Recently, the asteroseismic study by \citet{Malla:2020} shows that the``retired A-stars'' with spectroscopic masses $>1.6\,M_\odot$ are overestimated, confirming the earlier reports by \citet{Lloyd:2011} and \citet{Schlaufman:2013}. Since such stars consist of the heavier half of the \citet{Johnson:2010} sample and contribute most of the statistical evidence to the reported stellar mass dependence (see the Figure 4 of \citealt{Johnson:2010}), a revisit of the mass correlation will be needed. Additionally, the result of \citet{Johnson:2010} is limited to the region with separation $a<2.5\,$AU. As shown by \citet{Clanton:2014} and \citet{Clanton:2016}, after those at (slightly) larger separations are taken into account, giant planets are almost as common around M-dwarfs as they are around Sun-like stars (see Section~\ref{sec:microlensing}).

For small planets, \emph{Kepler} survey provides the best sample to study their stellar mass dependence. Studies have shown that the planet frequency $\bar{n}_{\rm p}$ in the \emph{Kepler} parameter space is anti-correlated with stellar mass \citep[e.g.,][]{Howard:2012,Mulders:2015a,Mulders:2015b}. Using the \citet{Berger:2018,Berger:2020a} sample with stellar effective temperature between $4000$--$5000\,$K, we find a frequency of $\bar{n}_{\rm p}=3.3\pm0.4$ for planets in the radius range $1$--$20\,R_\oplus$ and period $<400\,$days, which is a factor of $\sim2.7$ higher than the rate for our baseline Sun-like sample (Section~\ref{sec:kepler_planets}). Later M-type stars have even more planets \citep{Dressing:2013,Dressing:2015}. The planetary system frequency $F_{\rm p}$ is also anti-correlated with stellar mass but likely at a weaker level, due to the increased average planet multiplicity around later-type stars \citep[e.g.,][]{Yang:2020}. There is some sign of increased observed multiplicity rate in our $4000$--$5000\,$K sample ($48.2\%$) compared to that ($42.5\%$) of our baseline Sun-like star sample (Section~\ref{sec:kepler_planets}). After the correction for the suppression effect due to close stellar companions, the difference in formation efficiencies of small planets between single Sun-like and later-type hosts is likely further reduced, although an anti-correlation probably remains \citep{Moe:2020}.

%% file: sec3.tex
\section{THE OUTER PLANET POPULATION} \label{sec:outer_planets}

In the earliest stage of planet formation, the region beyond $\sim1\,$AU is expected to contain most of the mass and the angular momentum of the protoplanetary disk. Therefore, the frequency and properties of planets in this outer region ($\sim1$--10\,AU) have important implications to the formation and evolution of the whole system, including the planets in the inner $\sim1\,$AU region. In this section, we review our current understanding of this outer planet population and discuss its connection with the inner planetary system.

We set the inner and outer boundary at $\sim1\,$AU partly because this is approximately the detection limit of the \emph{Kepler} mission, but also because it coincides with the position where giant planets show a rapid rise in frequency. RV surveys have found that giant planets ($m_{\rm p}\gtrsim M_{\rm Sat} \approx 0.3\,M_{\rm J}$) appear $\sim5$ times more often between $\sim1$--$3\,$AU than they do within $\sim1\,$AU \citep{Cumming:2008}. 

\subsection{Planet Frequency} \label{sec:cold_giants}

RV surveys have found that cold giant planets (0.3--$13\,M_{\rm J}$) at $\sim$1--$5\,$AU) appear around on the order of $\sim10\%$ of Sun-like stars. If the giant planet distribution is modeled as a parametric function that joins single power-law distributions of mass and orbital period \citep{Tabachnik:2002}, the integrated rate out to $P\approx5.5\,$years is found to be $\bar{n}_{\rm p}=0.105$ \citep{Cumming:2008}.
Such a single power-law period distribution tends to over-predict the number of giant planets at wider ($\gtrsim10\,$AU) separations. To better match the observed distribution, \citet{Fernandes:2019} replaced it with a broken power law and found a potential peak at $\sim$2--3$\,$AU (see also \citealt{Bryan:2016}). Extending their distribution function out to 100$\,$AU, \citet{Fernandes:2019} found $\bar{n}_{\rm p}=0.27_{-0.05}^{+0.08}$ and $0.062_{-0.012}^{+0.015}$ for planets in the mass range 0.1--20$\,M_{\rm J}$ and 1--20$\,M_{\rm J}$, respectively. If the frequency of the so-called Jupiter analogs, namely Jupiter-mass ($\sim$0.3--3$\,M_{\rm J}$) planets in Jupiter-like (a few AU) orbits around Sun-like hosts, are concerned, several independent studies have collectively pointed to a rate about a few percent (e.g., \citealt{Wittenmyer:2016} and references therein), suggesting that planetary systems similar to our own may be relatively uncommon (see also Section~\ref{sec:correlation}). Unlike our Jupiter, a significant fraction of cold giant exoplanets are on substantially eccentric orbits with typical eccentricities $e\sim0.3$ \citep[e.g.,][]{Wright:2009}. We refer to the review by \citet{WinnFabrycky:2015} for more discussions on these topics.

Although the region beyond $\sim1\,$AU is nominally out of the reach of \emph{Kepler}, studies have nevertheless systematically searched for and statistically studied the long-period transiting planets in the \emph{Kepler} data \citep[e.g.,][]{ForemanMackey:2016,Herman:2019,Kawahara:2019}. In particular, \citet{Herman:2019} reported an occurrence rate of $\bar{n}_{\rm p}=0.7_{-0.2}^{+0.4}$ for planets with sizes between 0.3--1$\,R_{\rm J}$ and orbital periods between 2--10$\,$years. The inferred radius distribution also suggests that cold Neptune-sized (3--5\,$M_\oplus$) planets are $\sim$4 times more common than cold Jupiter-sized (7.5--11\,$M_\oplus$) ones. This is broadly consistent with the result from microlensing surveys (see Section~\ref{sec:microlensing}), pointing to the potential existence of a large and unexplored low-mass planet population in the outer region.

\subsection{The inner--outer correlation} \label{sec:correlation}

The planetary systems inside and outside of $\sim1\,$AU appear strongly correlated. Such a strong inner--outer correlation has important implications to the formation and evolution of the system as a whole. Below we review the observational evidence and discuss briefly the implications of this correlation. More on the latter will be presented in Section~\ref{sec:constraints}. We highlight two classes of inner planets, hot Jupiters and super Earths, and discuss them separately below.

\subsubsection{``Friends'' with close-in Jupiters}

Hot Jupiters, while usually having no detectable planetary companions in the inner region, are frequently found to have distant massive companions (\citealt{Knutson:2014,Bryan:2016}; but see also \citealt{Schlaufman:2016}). Both of these features are important clues to the formation and evolution of hot Jupiters, and we refer interested readers to the review by \citet{Dawson:2018} for in-depth discussions.

For the completeness of the discussion about the inner--outer correlation, we briefly summarize here the key result of the ``friends of hot Jupiters'' search. \citet{Knutson:2014} conducted a systematic RV study of the distant companions to a sample of 51 hot Jupiters and reported that each hot Jupiter should have on average $0.51\pm0.10$ companions with masses between 1--13$\,M_{\rm J}$ and semi-major axes between 1--20$\,$AU. This sample was re-analyzed in \citet{Bryan:2016} with improved sensitivity calculations, and the companion rate was revised to $0.70\pm0.08$. Given the small fraction of systems with more than one cold companion, we take this average number to be approximately the fraction of hot Jupiter hosts with cold Jupiter companions. This fraction barely changes after we adjust to the parameter range used in this work (0.3--13$\,M_{\rm J}$ and 1--10$\,$AU) according to the planet distribution function of \citet{Bryan:2016}. We denote this fraction as $P({\rm CJ}|{\rm HJ})$. Additionally, given the known fractions of Sun-like stars with hot Jupiters and cold Jupiters, $P({\rm HJ})\approx1\%$ and $P({\rm CJ})\approx10\%$, respectively, the inversed conditional probability is $P({\rm HJ}|{\rm CJ})\approx7\%$. This is the fraction of cold Jupiter hosts with hot Jupiters. All the four fractions are shown in \textbf{Figure~\ref{fig:correlation}}.

Jupiter-sized planets in the inner region with known outer giant companions tend to have higher eccentricities, suggesting possible dynamical interactions in sculpting the architectures of these systems \citep[e.g.,][]{Bryan:2016}. In particular, warm ($\sim10$--$100\,$days) Jupiters on significantly eccentric orbits ($e\gtrsim0.4$) have much higher chances to possess relatively close ($\lesssim 3\,$AU) Jovian companions compared to those on nearly circular ($e\lesssim0.1$) orbits \citep{Dong:2014b}, and the existence of such companions are consistent with the high-eccentricity migration scenario to form eccentric warm Jupiters \citep{Dong:2014b, Dawson:2014, Petrovich:2016}. 
By contrast, warm Jupiters on nearly circular orbits show a weaker correlation with outer giant companions, and many of these warm Jupiters are found to have nearby small planetary companions \citep{Huang:2016}. These features cannot be easily reconciled in the high-eccentricity migration scenario, suggesting that the nearly circular warm Jupiters may have been formed \textit{in situ} or have undergone the disk-driven migration (e.g., \citealt{Raymond:2008,Hallatt:2020}).
We refer to \citet{Dawson:2018} for more comprehensive discussion on the observations and theories related to warm Jupiters.

\subsubsection{Super Earth--cold Jupiter relation}

\begin{figure}
\includegraphics[width=\textwidth]{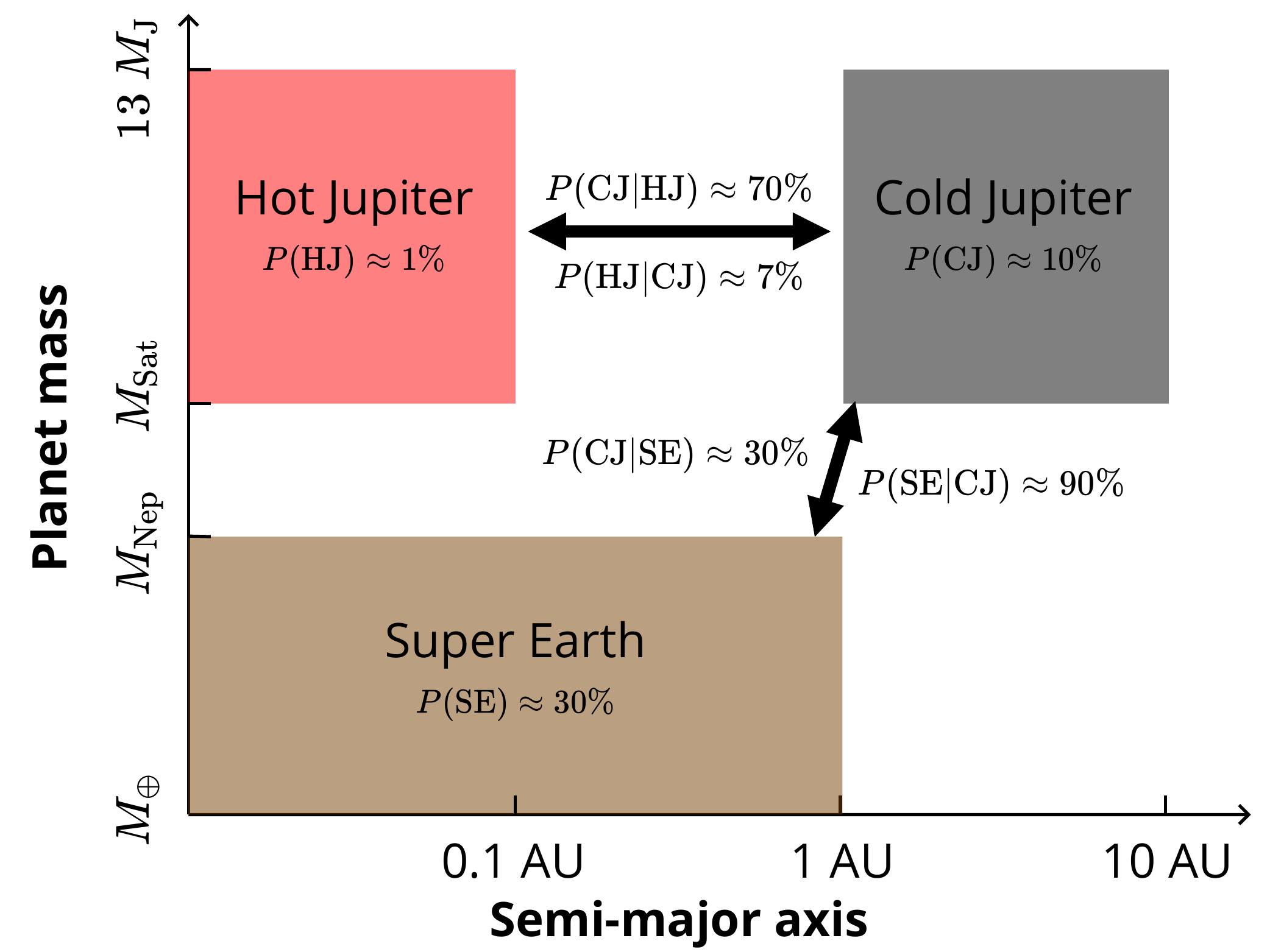}
\caption{Correlations between inner planets and outer cold Jupiters. Although only two types of inner planets are highlighted here, they are representative of the known inner planet population: The majority of hot Jupiters do not have close neighbors (see Section~\ref{sec:hot_Jupiters}), whereas super Earths usually reside in systems that include other types of inner planets (see Section~\ref{sec:multiplicity}). The unconditional probability shown here is the fraction of Sun-like stars with a specific type of planets, and the conditional probability is the fraction of Sun-like stars with a specific type of planets given that another type of planets is present in the system. 
\label{fig:correlation}}
\end{figure}

The term ``super Earth'' has different meanings in different studies. Here we call a planet super Earth if its mass (or radius) is between the masses (or radii) of Earth and Neptune, and the correlation under discussion applies specifically to the super Earths from the inner region. These super Earths dominate the known inner planet population, and they can co-exist with almost all types of inner planets except hot Jupiters (see Section~\ref{sec:kepler_planets}). For this reason, this super Earth population is representative of the inner planet population.

About $1/3$ of the inner super Earths have outer cold Jupiter companions, as studies have shown \citep{ZhuWu:2018,Bryan:2019}. The RV signal on the star induced by a super Earth is systematically smaller than the RV signal induced by a cold Jupiter. Making use of this point, \citet{ZhuWu:2018} constructed a sample of 54 super Earth systems around Sun-like hosts that received long-term RV observations, and they found that the fraction of super Earth hosts with cold Jupiter companions is $32\pm8\%$. This is $\sim3$ times higher than the frequency of cold Jupiters around field Sun-like stars. The fraction further rises to $\sim60\%$ for metal-rich systems (with [Fe/H]$>0.1$). These results were later confirmed by the independent study of \citet{Bryan:2019}. In that work the authors refit RV data sets of 65 super Earth hosts, some of which are M dwarfs, and reported an occurrence rate of $39\pm7\%$ for companions with masses in the range 0.5--20$\,M_{\rm J}$ and semi-major axes in the range 1--20$\,$AU. In this review, we take a rather conservative value of $P({\rm CJ}|{\rm SE})\approx30\%$, which is also shown in \textbf{Figure~\ref{fig:correlation}}.

The inversion of the above conditional probability reveals an even more interesting result. With $\sim30\%$ of Sun-like stars hosting inner super Earths and $\sim10\%$ of Sun-like stars hosting cold Jupiters, one finds from the Bayes theorem that $P({\rm SE}|{\rm CJ})\approx90\%$, suggesting that nearly all of the cold Jupiters should have inner small planets \citep{ZhuWu:2018,Bryan:2019}. Together with the fraction of cold Jupiter hosts with hot Jupiters, $P({\rm HJ}|{\rm CJ})\approx7\%$, outer giant planets almost all have inner companions.\footnote{It is possible that hot Jupiters were born cold and that their later evolution had cleared out the small planets originally present in the inner region. This would mean that essentially all cold Jupiters were born with inner small planets.}
We illustrate in \textbf{Figure~\ref{fig:correlation}} the connections between the outer giant planets and the two representative types of inner planets.

The above strong correlations are also confirmed by studies that utilized the rare but valuable long-period \emph{Kepler} transiting planets \citep{Uehara:2016,Herman:2019,Masuda:2020}. These studies find that the fraction of long-period ($P\gtrsim2\,$years) transiting planets with inner transiting companions is so high that it can only be explained by a strong inner--outer correlation. They also reported evidence that the dynamical hotness of inner and outer planets may also be correlated. Specifically, a dynamically hot outer Jupiter is likely associated with a dynamically hot inner planetary system. This provides a plausible explanation for the surprisingly large eccentricities and mutual inclinations of the inner systems with low multiplicities (\citealt{Masuda:2020}; see Section~\ref{sec:eccentricity}). It may also help explain the reduced super Earth multiplicities around metal-rich stars (\citealt{ZhuWu:2018}; see Section~\ref{sec:metallicity}).

The strong inner--outer correlation has implications on the frequency of planetary systems similar to our own. On the one hand, this correlation suggests that the general solar system-like architecture---with the inner region containing small planets and the outer region containing giant planets---is probably common among other planetary systems. On the other hand, planetary systems with properties very similar to ours, namely a system with both outer Jupiter-like ($\sim M_{\rm J}$ at a few AU) and inner Earth-like ($\lesssim M_\oplus$ within 1$\,$AU) planets may be rare ($\lesssim1\%$, \citealt{ZhuWu:2018}).
A possible explanation could be that our Jupiter formed very early and hence prevented the growth of inner embryos into super Earths \citep{Morbidelli:2015,Izidoro:2015}. 
This early Jupiter formation scenario also explains the isotope measurements on solar system iron meteorites \citep{Kruijer:2017}, although the question remains why the majority of cold Jupiters in other systems do have inner super Earths.
We defer to Section~\ref{sec:constraints} for further discussions of theoretical implications.

Observationally, the strong inner--outer correlation implies interesting synergies between space-based transit missions and astrometric missions or ground-based long-term RV surveys. Indeed, at least two of the TESS transiting planets have been found around stars with known RV cold Jupiters \citep{Huang:2018,Teske:2020}. Future combined TESS and \emph{Gaia} planet catalogs should yield hundreds of similar systems that can enable detailed studies of the system architecture, as has been demonstrated in the pi Mensae system \citep{Xuan:2020,Damasso:2020,DeRosa:2020}.

\subsection{Mass-ratio function from microlensing} \label{sec:microlensing}

\begin{figure}
\includegraphics[width=\textwidth]{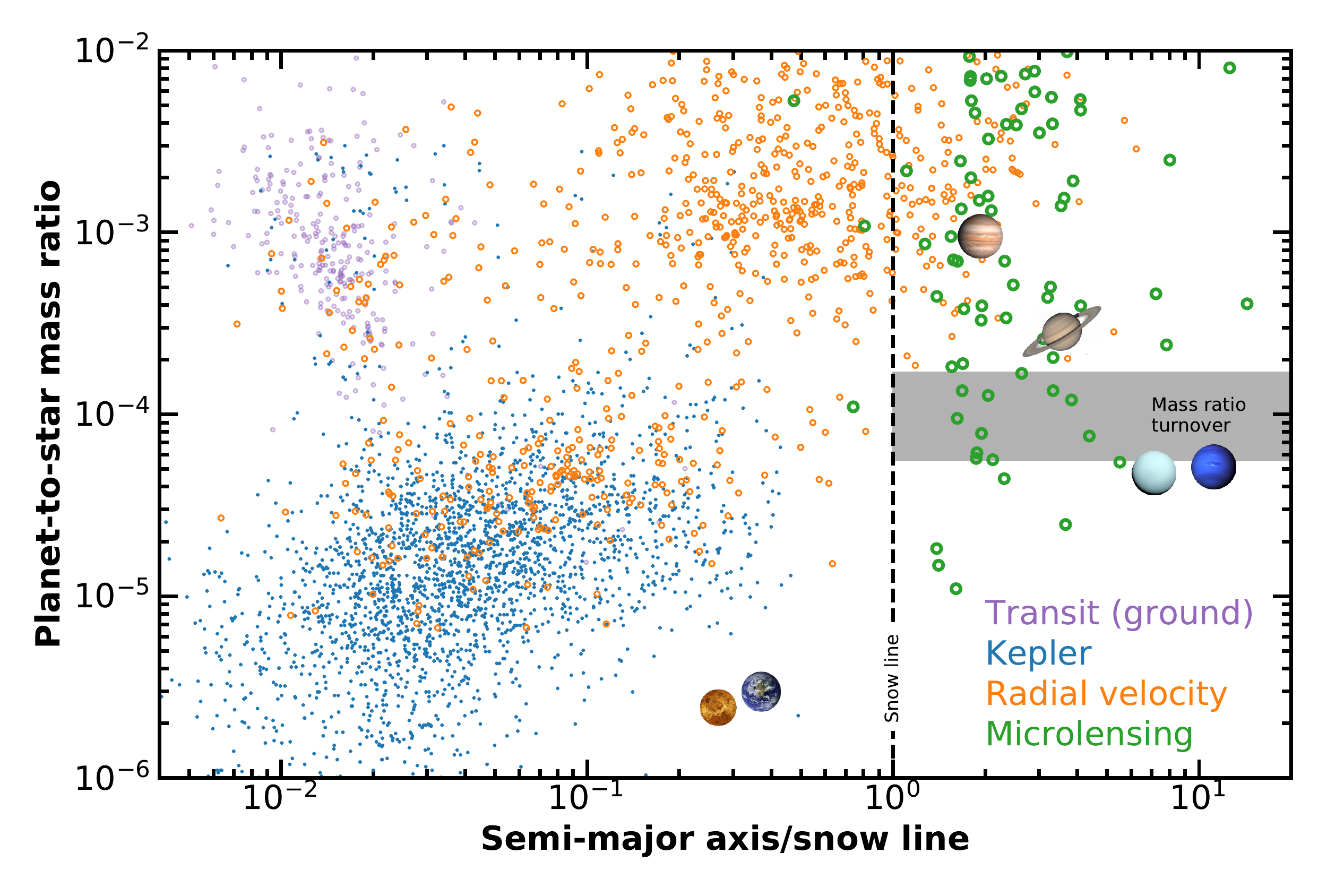}
\caption{Similar to \textbf{Figure~\ref{fig:overview}}, but here shows the planet-to-star mass ratio vs.\ the semi-major axis in units of the water snow line. The snow line is at $2.7\,$AU for $1\,M_\odot$ star and scales linearly with the host mass \citep{Kennedy:2008}. The location of the snow line is indicated with a black dashed line (note that this is only for illustrative purposes as the snow line should be determined in the protoplanetary disk), and the possible turnover in the mass ratio distribution found from microlensing, $0.55$--$1.7\times10^{-4}$ \citep{Suzuki:2016,Jung:2019}, is marked with the gray region. Six solar system planets are shown, with Mercury and Mars too low in mass ratio to appear on this plot.
\label{fig:q_temperature}}
\end{figure}

Gravitational microlensing probes a largely uncharted planet discovery space of cold planets \citep{Mao:1991,Gould:1992}, where $>99\%$ of the planetary mass of the solar system resides. Ground-based microlensing surveys are sensitive to planets down to Earth masses \citep[e.g.,][]{BennettRhie:1996, Dong:2006}, and a space-based survey will be capable of discovering all solar system planet analogs except Mercury (\citealt{Penny:2019}; see also \textbf{Figure~\ref{fig:overview}}). 
With the increasing number of discoveries, microlensing searches have been continuing to unveil the distribution of planets in this under-explored parameter space and offer insights into the planet formation outside the water snow line. 
We refer interested readers to \citet{Gaudi:2012} for an overview of the microlensing technique and its application in exoplanet discoveries \citep[see also][]{Mao:2012}. Below we focus on the important progress made since the review of \citet{Gaudi:2012}.

Several recent studies found evidence for a possible turnover in the planet-to-star mass ratio function for planets beyond the water snow line 
(see \textbf{Figure~\ref{fig:q_temperature}}). In the early era, a key microlensing finding was that cold Neptunes (with planet-to-star mass ratio $q\sim10^{-4}$) are a factor of a few more common than cold Jupiters (with $q\sim10^{-3}$, \citealt{Gould:2006}), and \citet{Sumi:2010} found that the distribution of the mass ratio $q$ could be described by a power law ${dN}/d\log_{10}{q} \propto q^{\nu}$, where $\nu=-0.7\pm0.2$. Using the planet sample from the MOA-II microlensing survey, \citet{Suzuki:2016} found that a single power law of the mass ratio function does not extend to very low mass ratios. Specifically, these authors reported a break in the mass ratio function at $q\sim10^{-4}$, corresponding to the mass of Neptune for a typical host star mass of $0.5\,M_\odot$. Based on a total sample of 30 planets that combines the MOA-II and previous statistical samples \citep{Gould:2010,Cassan:2012}, \citet{Suzuki:2016} reported a broken power-law mass ratio function with a break at $q_{\rm brk}= 1.7\times 10^{-4}$. The power-law indexes above and below the break are $\nu=-0.93\pm0.13$ and $0.6^{+0.5}_{-0.4}$, respectively. The normalization is such that $\bar{n}_{\rm p}=0.79$ for planets with mass ratio $q>5\times10^{-5}$ and projected separation $s$ in units of Einstein radius in the range 0.3--5. For typical microlenses with $0.5\,M_\odot$, these numbers correspond to the planetary mass $M_{\rm p}>8\,M_\oplus$ and the orbital separation between 1--15\,AU. The reported planet frequency is compatible with those from other detection techniques (i.e., RV and direct imaging) following a simple joint planet distribution function \citep{Clanton:2014,Clanton:2016}.

Further studies by \citet{Udalski:2018}, who studied an ensemble of seven (as compared to four in \citealt{Suzuki:2016}) planets with $q<10^{-4}$, and \citet{Jung:2019}, who analyzed a sample of 15 planets with $q<3\times10^{-4}$, investigated the possible turnover in the mass ratio function. 
Adopting a power-law form of the detection efficiency, \citet{Jung:2019} modeled the intrinsic mass-ratio distribution with a broken power law and revised the break to $q_{\rm brk}\approx 5.5\times10^{-5}$, which is a factor of three below the value found by \citet{Suzuki:2016}, but their low-mass planet sample was too small to distinguish a pile-up at that mass ratio from broken power law. Nevertheless, a break or pile-up in the planet-to-star mass ratio function could have important theoretical implications \citep[e.g.,][]{Pascucci2018, Wu:2019}, and further probing the distribution of sub-Neptune microlensing planets will be a research focus in the near future. Observations from high-cadence and nearly continuous microlensing surveys such as KMTNet \citep{Kim:2016} are pushing toward detecting more planets at low mass ratios (e.g., $q=1.8\times10^{-5}$ from \citealt{Gould:2020}, $q\approx1.4\times10^{-5}$ from \citealt{Yee:2021}, $q\approx1.1\times10^{-5}$ from \citealt{Zang:2021}), so a large enough sample is expected to be available soon to improve the determination of the mass ratio function at the low end (see \citealt{Zang:2021} and discussions therein).

Another interesting feature of the mass ratio function of \citet{Suzuki:2016} is its apparent smoothness between Neptune and Jupiter masses. Intriguingly, the derived radius distribution of cold planets from the single transit events in \emph{Kepler} also appears similarly continuous between Neptune and Saturn \citep{Herman:2019}. These results are surprising in view of the standard core accretion theory \citep{Pollack:1996}, which builds on the solar system and predicts a deficit of planets at such intermediate masses/radii \citep{IdaLin:2004a,Mordasini:2009}. This tension may suggest that the giant planet formation involves physical processes that have been overlooked in the standard models \citep{Suzuki:2018}.
Alternatively, it could be due to the limited sample sizes of cold intermediate-mass planets (\citealt{Suzuki:2016} sample contains nine detections in the range $10^{-4}<q<5\times10^{-4}$ and \citealt{Herman:2019} has four in their intermediate radius bin of $0.67$--$1.00\,R_{\rm J}$).
Therefore, increasing the sample size of cold intermediate-mass planets will clarify the degree of tension between observation and theory. Furthermore, physical mass (rather than mass ratio) determinations of a large sample of microlensing planets through measurements of the microlensing parallax or the lens flux \citep[e.g.,][]{Dong:2009} are needed to enable a more direct comparison with theories. This will be possible for essentially all microlensing planets detected to date at first light of adaptive optics on 30m class  telescopes \citep[e.g.,][]{Skidmore:2015} or for a significant fraction of planet hosts in a space-based microlensing survey such as the \emph{Roman} microlensing survey \citep{Penny:2019}.

\subsection{Free-floating planets}

The prevalence of eccentric and/or inclined planetary orbits suggest likely histories of violent dynamical interactions in the planetary systems, such as planet-planet scatterings, which naturally eject a significant fraction of the planets from the system and form unbound planets with no hosts \citep[e.g.,][]{RasioFord:1996,JuricTremaine:2008,Chatterjee:2008}. The distributions of free-floating planets bear important signatures of not only the initial configurations of the planetary systems at birth, but also their subsequent dynamical evolution. 

While it is possible to directly image young sub-stellar objects down to a few Jupiter masses \citep[see, e.g.,][]{Osorio:2000}, gravitational microlensing is the only known method in probing the lower-mass objects, which are believed to dominate the dynamically ejected FFP population. Low-mass objects produce relatively short-timescale microlensing light curves as the Einstein radius crossing time $t_{\rm E}\propto \sqrt{M}$. For typical stellar-mass microlenses, the timescale is $\sim20\,$days, whereas for planetary-mass objects it is $\lesssim1\,$day. The detection of such short and rare events thus demands wide-field high-cadence surveys that have only been available since the past decade.

\citet{Mroz:2017} analysed a sample of 2,617 microlensing events from the OGLE-IV survey and concluded that the frequency of Jupiter-mass free-floating (or wide-orbit) planets should be no more than $0.25$ planets per main-sequence star at the 95\% confidence level. This result is broadly compatible with the inferred occurrence rate of bound giant planets from RV surveys \citep[e.g.,][]{Cumming:2008}, microlensing searches \citep[e.g.,][]{Gould:2010} or direct imaging \citep[e.g.,][]{Bowler:2016}, and contradicts a previous claim that free-floating Jupiter-mass planets are more abundant than stars \citep{Sumi:2011}. The sample of \citet{Mroz:2017} also includes six short events with timescales in the range $0.1\,{\rm days}<t_{\rm E}<0.4\,{\rm days}$. Assuming the microlensing nature of these events and given the low detection efficiency at such ultrashort timescales, their sample suggests that there may be up to a few free-floating planets in the Earth-mass to super-Earth-mass range per main-sequence star. The results of \citet{Mroz:2017} about the absence of free-floating Jupiter-mass planets and the potential existence of free-floating Earth-mass and super-Earth-mass planets are generally consistent with theoretical expectations \citep[e.g.,][]{Ida:2013, Ma:2016}.

The existence of such ultrashort-timescale events was soon confirmed thanks to the coordinated observations of multiple microlensing survey telescopes around the globe. \citet{Mroz:2018} reported the first convincing example of a microlensing event with timescale $t_{\rm E}=0.32\,$days, and subsequent dedicated searches led to the discovery of a few more similar events \citep{Mroz:2019,Mroz:2020,Mroz:2021,Kim:2020,Ryu:2020}. These events all show strong finite-source effects that arise from the lenses \textit{transiting} distant giant sources, yielding the immediate measurement of the angular Einstein radius $\theta_{\rm E}$.  The lens mass scales as
\begin{equation} \label{eqn:ffp-mass_bulge}
M ({\rm bulge}) = \frac{\theta_{\rm E}^2}{\kappa \pi_{\rm rel}} = 250\,M_\oplus \left(\frac{\theta_{\rm E}}{10\,\mu{\rm as}}\right)^2 \left(\frac{\pi_{\rm rel}}{16\,\mu{\rm as}}\right)^{-1} ,
\end{equation}
with the normalization of the lens-source relative parallax $\pi_{\rm rel}$ chosen such that the lens and source are both in the Galactic bulge and separated by about $1\,{\rm kpc}$. Here the constant $\kappa \approx 8.14\,{\rm mas}\,M_\odot^{-1}$. For lens in the Galactic disk ($\pi_{\rm rel}\approx 125\,\mu{\rm as}$), the mass scales as
\begin{equation} \label{eqn:ffp-mass_disk}
M ({\rm disk}) = \frac{\theta_{\rm E}^2}{\kappa \pi_{\rm rel}} =32\,M_\oplus \left(\frac{\theta_{\rm E}}{10\,\mu{\rm as}}\right)^2 \left(\frac{\pi_{\rm rel}}{125\,\mu{\rm as}}\right)^{-1} .
\end{equation} 
\citet{Kim:2020} and \citet{Ryu:2020} argue that $\theta_{\rm E}$ is a better discriminator than $t_{\rm E}$ for selecting FFPs. In fact, from a small number of events with finite-source effects, there is a possible gap between $\sim 10\,\mu{\rm as}$ and $\sim 30\,\mu{\rm as}$ in the $\theta_{\rm E}$ distribution, and this ``Einstein desert'' may separate brown dwarfs from free-floating super Earths (and terrestrial planets) in the disk \citep{Ryu:2020}. We list in \textbf{Table~\ref{tab:ffp}} the relevant parameters and inferred masses of the FFP candidate events with $\theta_{\rm E}<10\,\mu{\rm as}$. The preliminary analyses by \citet{Mroz:2019} and \citet{Ryu:2020} suggest that low-mass unbound (or wide-orbit) planets may be more common than stars in the Galaxy. Future space-based microlensing surveys can assemble a large sample for quantitative assessments of the FFP population \citep[e.g.,][]{Johnson:2020}, and a satellite augmented with microlensing parallax measurements can directly measure the masses and distances of such FFP events \citep[e.g.,][and references therein]{Gould:2021}.

\begin{table}
\caption{Published microlensing free-floating planet candidates ($\theta_{\rm E}<10\mu{\rm as}$), sorted by the inferred lens mass (see Equation~\ref{eqn:ffp-mass_bulge} and Equation~\ref{eqn:ffp-mass_disk} for typical estimates in the bulge and disk, respectively).
\label{tab:ffp}}
\centering
\begin{tabular}{rccccl}
\hline
Event name & $\theta_{\rm E}/\mu{\rm as}$ & $t_{\rm E}/{\rm d}$ & $M({\rm disk})$ & $M({\rm bulge})$ &  Reference \\
\hline
OGLE-2016-BLG-1928 & 0.84 & 0.029 & $0.2\,M_\oplus$ & $1.8\,M_\oplus$ & \citet{Mroz:2021} \\
OGLE-2012-BLG-1323 & 2.4 & 0.16 & $1.8\,M_\oplus$ & $14\,M_\oplus$ & \citet{Mroz:2019} \\
OGLE-2019-BLG-0551 & 4.4 & 0.38 & $6.1\,M_\oplus$ & $48\,M_\oplus$ & \citet{Mroz:2020} \\
KMT-2019-BLG-2073 & 4.8 & 0.27 & $7.6\,M_\oplus$ & $59\,M_\oplus$ & \citet{Kim:2020} \\
KMT-2017-BLG-2820 & 5.9 & 0.29 & $11\,M_\oplus$ & $87\,M_\oplus$ & \citet{Ryu:2020} \\
OGLE-2016-BLG-1540 & 9.2 & 0.32 & $28\,M_\oplus$ & $217\,M_\oplus$ & \citet{Mroz:2018} \\
\hline
\end{tabular}
\end{table}

While the events listed in \textbf{Table~\ref{tab:ffp}} are promising candidates for FFPs, 
it is also plausible that these objects are actually in such wide orbits that no microlensing signatures from their hosts were detected. Light curve analyses can exclude the existence of any massive companions (i.e., hosts) out to a few Einstein radii, corresponding to $\sim15$--$20\,$AU away \citep[e.g.,][]{Mroz:2018,Kim:2020}. In other words, these FFP candidates could well be planets at Uranus-like or Neptune-like orbits \citep[e.g.,][]{Poleski:2014}.
Future high-resolution imaging observations that can resolve the hosts for wide-separation planets will be able to tell whether these objects are truly free-floating or loosely bound to some unidentified stellar hosts \citep{Han:2005,Gould:2016,Ryu:2020}.

%% file: sec4.tex
\section{THEORETICAL IMPLICATIONS} \label{sec:theory}

The observed distribution of planets and the architecture of planetary systems, as reviewed in Sections~\ref{sec:inner_planets} and \ref{sec:outer_planets}, are the consequence of $\sim$10--100$\,$Myr formation and later $\sim$Gyr evolution. 
In this section, we discuss the constraints from these observations on theoretical models.
A comprehensive overview on the formation and evolution theories is beyond the scope of the current review. Instead, we focus on the key physical processes that lead to observational signatures. To reduce the complexity, we restrict to planetary systems around Sun-like hosts.

\subsection{A brief overview of theories} \label{sec:theory_overview}

The generally accepted picture of planet formation can be traced back to the nebular model originally proposed by Immanuel Kant and Pierre Laplace in the 1700s. Modern theorists generally believe that planets were formed out of the gas and the dust in the protoplanetary disk. Small solid particles first accumulate to form asteroid-sized ($\sim$1--100$\,$km) planetesimals, and the collisions between planetesimals eventually lead to the formation of planet-sized objects \citep{Chamberlin:1916,Safronov:1972}. See \citet{Woolfson:1993} for a historical overview on the planet formation theories.

In the core accretion theory that explains the solar system formation\citep[e.g.,][]{Lissauer:1993,Pollack:1996}, the primary building blocks for planet formation are planetesimals. The growth of planetesimals is first divergent (i.e., the run-away phase) and then convergent (i.e., the oligarchic phase), until they have cleared nearly all planetesimals in their feeding zone ($\sim5\,R_{\rm H}$). These so-called protoplanets (or embryos) are now $\gtrsim1000\,$km in size and around Mars-mass \citep[e.g.,][]{IdaMakino:1993,Kokubo:1998}. The further growth of the protoplanets involves planetesimal accretion as well as dynamical interactions between protoplanets. 
At a few AU separation, the growth of protoplanets is sufficient and allows the formation of giant planets \citep{Mizuno:1980,Pollack:1996}. In the classical picture, the giant planet formation has three phases: core formation, hydrostatic gas accretion, and run-away gas accretion \citep{Pollack:1996}. The hydrostatic gas accretion phase starts when the embryo reaches a critical core mass ($\sim10\,M_\oplus$, \citealt{Mizuno:1980,Stevenson:1982}). This phase can take up to $\sim10\,$Myr and is the most time-consuming step in this classical core accretion model. The run-away gas accretion is triggered once the envelope and the core have comparable masses, and it sufficiently pushes the total mass to the giant planet regime ($\gtrsim100\,M_\oplus$).
In the inner region, embryos grow slowly and never reach the critical core mass before the gaseous disk is depleted. The later evolution involves collisions between these embryos in the gas-free environment. This so-called giant impact phase lasts $\sim100\,$Myr and eventually forms the terrestrial planets \citep[e.g.,][]{Chambers:2001}.

A new paradigm that has attracted much attention recent years is pebble accretion. In the astrophysical context, pebbles are dust particles that are weakly coupled to the gas and thus drift in the disk. The inclusion of pebbles in the formation diagram provides a plausible scenario for the formation of planetesimals via streaming instability \citep{Youdin:2005,Johansen:2007,Chiang:2010}. Unlike planetesimals that are decoupled from the gas, pebbles ``feel'' the aerodynamic drag from the gas and drift inward toward the star \citep{Nakagawa:1986}. This means that the ``food'' supply to a protoplanet is not limited to the local material. Additionally, the cross-section for protoplanets (or planetesimals) to accrete pebbles is larger than the cross-section for the same objects to accrete planetesimals \citep{Ormel:2010,Lambrechts:2012}. These two factors together make pebble accretion more efficient in building up cores of protoplanets. 
When the protoplanet becomes massive enough, it starts to carve a gap in the pebble disk, and the subsequent pressure bump outside of the orbit stops the inward drifting pebbles. The corresponding mass of the protoplanet is called the pebble isolation mass 
\begin{equation}
M_{\rm iso} \approx 10 \left(\frac{h/r}{0.04}\right)^3 \, M_\oplus ~,
\end{equation}
where $h/r$ is the disk aspect ratio at location $r$ and the prefactor is determined numerically and depends on disk properties \citep{Lambrechts:2014}. The above relation assumes a solar-mass host star. For other stellar masses, the pebble isolation mass scales linearly with the mass of the host star \citep{Liu:2019}. Once a protoplanet reaches the pebble isolation mass, it effectively cuts off the pebble flux and ``starves'' the protoplanetary core and all embryos interior to its orbit. With the halted pebble accretion, the critical core mass required to trigger the rapid gas accretion is reduced and thus giant planets can form more efficiently \citep{Lambrechts:2014}. Furthermore, pebbles can easily vaporize in the hot envelopes before they can reach the cores \citep{Brouwers:2018}. The enriched envelopes also speed up the formation of giant planets (\citealt{Venturini:2016}; see also \citealt{Stevenson:1982} and \citealt{Hori:2011} for a similar mechanism in the planetesimal accretion scenario). We refer interested readers to the reviews by \citet{Johansen:2017} and \citet{Ormel:2017} for more details about the pebble accretion model.

Planets may undergo disk-driven migration while accreting pebbles, planetesimals, and/or gas (e.g., \citealt{Kley:2012} and references therein). Migration can substantially change the architecture of the planetary system, such as locking planets into mean motion resonances \citep{Goldreich:1980,Lee:2002}. However, such features are not prominent in \emph{Kepler} systems (see Section~\ref{sec:spacings}). This may suggest that most \emph{Kepler} planets have not undergone significant disk-driven migrations (see also Section~\ref{sec:constraints}). Alternatively, the \emph{Kepler} planets may have never entered into resonances during the migration \citep[e.g.,][]{Goldreich:2014}, or the long-term dynamical evolution after the disk dispersal has effectively removed most of these features \citep[e.g.,][]{Izidoro:2017,Izidoro:2019}.

\subsection{Constraints from observations} \label{sec:constraints}

Given the substantial uncertainties in theories and in some parts of observations, we think that it is premature to provide detailed and quantitative comparisons between theories and observations (but see attempts by, e.g., \citealt{Hansen:2013,Izidoro:2017,Mulders:2019,Bern_2:2020}). We therefore choose to focus on some selected constraints that are considered relatively robust and discuss their implications to the formation of \emph{Kepler}-like planets:
\begin{itemize}
\item \textbf{Prevalence and multiplicity.} Inner super Earth-like planets are known to exist around $\sim 30\%$ Sun-like stars, and they typically reside in multi-planet systems (Section~\ref{sec:multiplicity}). Additionally, they preferentially have outer cold Jupiter companions (Section~\ref{sec:correlation}), suggesting that the two types of planets do not inhibit, but perhaps promote, the formation of each other. Unlike giant planets, super Earths show a much weaker dependence on the host metallicity (Section~\ref{sec:metallicity}).
\item \textbf{Composition.} As inferred from population-level studies of the radius valley (Section~\ref{sec:valley}) as well as mass and radii measurements of individual \emph{Kepler} planets \citep[e.g.,][]{Wu:2013,Hadden:2014,Hadden:2017}, some inner small planets likely have Earth-like (i.e., rocky and ice-poor) cores, and these cores have acquired gaseous envelopes that weigh up to a few percent of the total mass while the disk is still present.
\end{itemize}

The prevalence and the early formation of super Earth-like planets suggest that the planet formation process is more efficient than what had been expected from solar system formation models \citep[e.g.,][]{IdaLin:2004a, Mordasini:2009}.
This alone may not be an issue to the pebble accretion scenario (see Section~\ref{sec:theory_overview}).\footnote{Pebble accretion is efficient in growing planet embryos into larger bodies. However, pebble accretion is also lossy, as $\gtrsim90\%$ of the planet-forming material falls onto the host star rather than being accreted onto the growing planets \citep{Liu:2018,Lin:2018}}
In fact, pebble accretion can be so efficient that preventing super Earth-mass planets from undergoing run-away gas accretion places another challenge, a possible solution to which could be a delayed formation near the end of the disk phase \citep[e.g.,][]{Lee:2014}. For the planetesimal accretion scenario, a very massive disk is typically required to form super Earths efficiently and early \citep[e.g.,][]{Bern_2:2020}. The rocky composition suggests that the cores are formed in the ice-poor environment, likely inside the water ice line. In order for embryos or protoplanets from outside of the ice line to not largely contaminate the inner region, the disk-driven migration is probably suppressed.

The strong correlation between inner super Earths and outer cold Jupiters is a bit challenging to both accretion scenarios under the typical protoplanetary disk conditions. The planetesimal accretion scenario usually requires relatively efficient disk-driven migrations to explain the presence of abundant super Earths around metal-poor hosts, but the same migration efficiency turns out an overkill in reproducing the inner--outer strong correlation (\citealt{Bern_3:2020}, see also \citealt{IdaLin:2010}). For the pebble accretion scenario, because the solid supply to protoplanets is not limited locally, there is potentially a direct competition between different embryos. Furthermore, once the core of the outer giant planet first reaches the pebble isolation mass, the further growth of the inner planets is significantly limited, and the giant planet also acts as a barrier to the inward migrating embryos from outside of its orbit. Therefore, the pebble accretion scenario typically expects an anti-correlation between inner and outer planets \citep{Morbidelli:2015,Izidoro:2015,Lambrechts:2019}. Alternatively, the cores of both inner super Earths and outer cold Jupiters could be formed at such large separations (tens of AU) that enough material is available to the inner cores \citep{Bitsch:2015,Bitsch:2019}, although it is unclear whether such an approach can reproduce quantitatively the observed correlation and forms rocky core planets.
The difficulty in reproducing the inner--outer correlation may suggest that many protoplanetary disks start heavier than what has been typically assumed. Indeed, if \emph{Kepler} planets are formed \textit{in situ} based on the local material (i.e., not the inward-drifting pebbles), the required surface density is much higher than the minimum-mass solar nebula model \citep{Weidenschilling:1977,Hayashi:1981} and almost reaches the gravitational instability limit \citep{Chiang:2013, Schlichting2014}.

%% file: sec5.tex
\section{SUMMARY AND DISCUSSION} \label{sec:summary}

The discovery of thousands of exoplanets from the combination of multiple detection techniques have substantially advanced our understanding of the distribution of planets and the architecture of planetary systems. This review aims to update our knowledge of exoplanet statistics since the \citet{WinnFabrycky:2015} review. In Section~\ref{sec:inner_planets}, we described the distribution and properties of planets in the inner region, based mostly on discoveries from the \emph{Kepler} mission. In Section~\ref{sec:outer_planets}, we reviewed the recent progress on the cold planet population, with an emphasis on their connections to the close-in companions. Section~\ref{sec:theory} briefly described the theoretical models and the key constraints from observations. We summarize the key results below:

\begin{summary}[SUMMARY POINTS]
\begin{enumerate}
\item In the inner region ($\lesssim1\,$AU), about $30\%$ of Sun-like stars host planets with masses/radii down to Earth mass/radii, and each planetary system on average has about three such planets. These suggest that the planet formation process is more efficient than what had been expected from solar system formation models.
\item Planetary systems with more planets appear colder dynamically, with smaller orbital eccentricities, mutual inclinations, and orbital spacings. For systems with few planets in the inner region, planets can have $\sim0.3$ orbital eccentricities and $\gtrsim 10^\circ$ mutual inclinations. These support the idea that dynamical evolution has played a significant role reshaping the system architecture.
\item There exists a ``radius valley'' at $R_{\rm p}\sim 2\,R_\oplus$ and $P\lesssim30\,$days. The valley was predicted by the photoevaporation theory, although alternative explanations have also been proposed. Population-level analyses of the radius valley suggest that these planets were probably born with rocky cores and gaseous atmospheres up to a few percent  of the core masses.
\item Cold Neptune-like planets are a few times more abundant than cold Jupiter-like ones in the outer region. The inner ($\lesssim1\,$AU) and the outer ($\sim$1--10\,AU) planetary systems appear strongly correlated such that inner small planets preferentially have cold Jupiter-like companions and that outer cold Jupiters almost always have inner planetary companions.
\end{enumerate}
\end{summary}

With the ongoing and upcoming missions that have better capabilities and/or open up new observational channels, our understanding of exoplanets and planetary systems will continue to be improved. We outline below several promising directions that may see substantial advancement in the near future:
\begin{issues}[FUTURE PROSPECTS]
\begin{enumerate}
\item Exoplanet atmosphere and mass--radius relation: Space-based all-sky transit surveys like the TESS mission \citep{Ricker:2015} have been finding many bright targets, enabling detailed characterizations of more close-in planets \citep[e.g.,][]{Huang:2018,Armstrong:2020}. An improved mass--radius relation and better atmospheric characterizations will help to understand the composition and potentially the past evolution of the planet (see the recent review by \citealt{Madhusudhan:2019})
\item Planetary system architecture: The joint coverage of different surveys will potentially open up a larger parameter space (see \textbf{Figure~\ref{fig:overview}}) and reveal more interesting features about the planetary system architecture. The \emph{Gaia} mission alone is expected to detect at least thousands of giant planets around nearby stars \citep{Perryman:2014}, and its synergy with other surveys/missions will also open up new channels into the architecture study \citep[e.g.,][]{Xuan:2020,Damasso:2020,DeRosa:2020}. 
RV follow-ups of systems detected by other methods could also play increasingly indispensable roles in this aspect, in particular with the extreme RV instruments with capability down to $\sim0.3\,{\rm m\,s^{-1}}$ now coming online \citep[e.g.,][]{Fischer:2016}.
\item Planet across the HR diagram: Rapid advances of large-scale spectroscopic surveys and the {\it Gaia} satellite are continuing to revolutionize stellar astrophysics and Galactic astronomy. Further synergies of large samples of exoplanets with detailed stellar chemical compositions, kinematics and/or ages measurements are expected to place planet formation in the rich context of stellar populations and evolutions.
\item Planets around young stars: While the present review focused on the planetary systems around $\gtrsim$Gyr old stars, the demographics of planets around young ($\lesssim100\,$Myr) stars is a crucial link toward a more direct comparison with both planet formation theories and ALMA observations of protoplanetary disks (see the recent review by \citealt{Andrews:2020}). The detection and characterization of more planets around young stars (e.g., PDS 70b,\,c, \citealt{PDS70} and AU Microscopii~b, \citealt{AUMIc}) will be valuable.
\end{enumerate}
\end{issues}

%% file: planet_araa.bbl
\begin{thebibliography}{}
\expandafter\ifx\csname natexlab\endcsname\relax\def\natexlab#1{#1}\fi

\bibitem[{{Adibekyan} et~al.(2012){Adibekyan}, {Sousa}, {Santos}, {Delgado
  Mena}, {Gonz{\'a}lez Hern{\'a}ndez} et~al.}]{Adibekyan:2012}
{Adibekyan} VZ, {Sousa} SG, {Santos} NC, {Delgado Mena} E, {Gonz{\'a}lez
  Hern{\'a}ndez} JI, et~al. 2012.
\textit{\aap} 545:A32

\bibitem[{{Agol} et~al.(2005){Agol}, {Steffen}, {Sari} \&
  {Clarkson}}]{Agol:2005}
{Agol} E, {Steffen} J, {Sari} R, {Clarkson} W. 2005.
\textit{\mnras} 359:567--579

\bibitem[{{Akeson} et~al.(2013){Akeson}, {Chen}, {Ciardi}, {Crane}, {Good}
  et~al.}]{Akeson:2013}
{Akeson} RL, {Chen} X, {Ciardi} D, {Crane} M, {Good} J, et~al. 2013.
\textit{\pasp} 125:989

\bibitem[{{Andrews}(2020)}]{Andrews:2020}
{Andrews} SM. 2020.
\textit{\araa} 58:483--528

\bibitem[{{Andrews} et~al.(2013){Andrews}, {Rosenfeld}, {Kraus} \&
  {Wilner}}]{Andrews:2013}
{Andrews} SM, {Rosenfeld} KA, {Kraus} AL, {Wilner} DJ. 2013.
\textit{\apj} 771:129

\bibitem[{{Ansdell} et~al.(2016){Ansdell}, {Williams}, {van der Marel},
  {Carpenter}, {Guidi} et~al.}]{Ansdell:2016}
{Ansdell} M, {Williams} JP, {van der Marel} N, {Carpenter} JM, {Guidi} G,
  et~al. 2016.
\textit{\apj} 828:46

\bibitem[{{Armstrong} et~al.(2020){Armstrong}, {Lopez}, {Adibekyan}, {Booth},
  {Bryant} et~al.}]{Armstrong:2020}
{Armstrong} DJ, {Lopez} TA, {Adibekyan} V, {Booth} RA, {Bryant} EM, et~al.
  2020.
\textit{\nat} 583:39--42

\bibitem[{{Artymowicz} \& {Lubow}(1994)}]{Artymowicz:1994}
{Artymowicz} P, {Lubow} SH. 1994.
\textit{\apj} 421:651

\bibitem[{{Bailey} \& {Batygin}(2018)}]{Bailey:2018}
{Bailey} E, {Batygin} K. 2018.
\textit{\apjl} 866:L2

\bibitem[{{Ballard} \& {Johnson}(2016)}]{Ballard:2016}
{Ballard} S, {Johnson} JA. 2016.
\textit{\apj} 816:66

\bibitem[{{Beaug{\'e}} \& {Nesvorn{\'y}}(2013)}]{Beauge:2013}
{Beaug{\'e}} C, {Nesvorn{\'y}} D. 2013.
\textit{\apj} 763:12

\bibitem[{{Becker} et~al.(2015){Becker}, {Vanderburg}, {Adams}, {Rappaport} \&
  {Schwengeler}}]{Becker:2015}
{Becker} JC, {Vanderburg} A, {Adams} FC, {Rappaport} SA, {Schwengeler} HM.
  2015.
\textit{\apjl} 812:L18

\bibitem[{{Bennett} \& {Rhie}(1996)}]{BennettRhie:1996}
{Bennett} DP, {Rhie} SH. 1996.
\textit{\apj} 472:660

\bibitem[{{Berger} et~al.(2018){Berger}, {Huber}, {Gaidos} \& {van
  Saders}}]{Berger:2018}
{Berger} TA, {Huber} D, {Gaidos} E, {van Saders} JL. 2018.
\textit{\apj} 866:99

\bibitem[{{Berger} et~al.(2020{\natexlab{a}}){Berger}, {Huber}, {Gaidos}, {van
  Saders} \& {Weiss}}]{Berger:2020b}
{Berger} TA, {Huber} D, {Gaidos} E, {van Saders} JL, {Weiss} LM.
  2020{\natexlab{a}}.
\textit{\aj} 160:108

\bibitem[{{Berger} et~al.(2020{\natexlab{b}}){Berger}, {Huber}, {van Saders},
  {Gaidos}, {Tayar} \& {Kraus}}]{Berger:2020a}
{Berger} TA, {Huber} D, {van Saders} JL, {Gaidos} E, {Tayar} J, {Kraus} AL.
  2020{\natexlab{b}}.
\textit{\aj} 159:280

\bibitem[{{Bitsch} et~al.(2019){Bitsch}, {Izidoro}, {Johansen}, {Raymond},
  {Morbidelli} et~al.}]{Bitsch:2019}
{Bitsch} B, {Izidoro} A, {Johansen} A, {Raymond} SN, {Morbidelli} A, et~al.
  2019.
\textit{\aap} 623:A88

\bibitem[{{Bitsch} et~al.(2015){Bitsch}, {Lambrechts} \&
  {Johansen}}]{Bitsch:2015}
{Bitsch} B, {Lambrechts} M, {Johansen} A. 2015.
\textit{\aap} 582:A112

\bibitem[{{Borucki} et~al.(2010){Borucki}, {Koch}, {Basri}, {Batalha}, {Brown}
  et~al.}]{Borucki:2010}
{Borucki} WJ, {Koch} D, {Basri} G, {Batalha} N, {Brown} T, et~al. 2010.
\textit{Science} 327:977

\bibitem[{{Bouma} et~al.(2018){Bouma}, {Masuda} \& {Winn}}]{Bouma:2018}
{Bouma} LG, {Masuda} K, {Winn} JN. 2018.
\textit{\aj} 155:244

\bibitem[{{Bowler}(2016)}]{Bowler:2016}
{Bowler} BP. 2016.
\textit{\pasp} 128:102001

\bibitem[{{Brewer} et~al.(2018){Brewer}, {Wang}, {Fischer} \&
  {Foreman-Mackey}}]{Brewer:2018}
{Brewer} JM, {Wang} S, {Fischer} DA, {Foreman-Mackey} D. 2018.
\textit{\apjl} 867:L3

\bibitem[{{Brouwers} et~al.(2018){Brouwers}, {Vazan} \&
  {Ormel}}]{Brouwers:2018}
{Brouwers} MG, {Vazan} A, {Ormel} CW. 2018.
\textit{\aap} 611:A65

\bibitem[{{Bryan} et~al.(2016){Bryan}, {Knutson}, {Howard}, {Ngo}, {Batygin}
  et~al.}]{Bryan:2016}
{Bryan} ML, {Knutson} HA, {Howard} AW, {Ngo} H, {Batygin} K, et~al. 2016.
\textit{\apj} 821:89

\bibitem[{{Bryan} et~al.(2019){Bryan}, {Knutson}, {Lee}, {Fulton}, {Batygin}
  et~al.}]{Bryan:2019}
{Bryan} ML, {Knutson} HA, {Lee} EJ, {Fulton} BJ, {Batygin} K, et~al. 2019.
\textit{\aj} 157:52

\bibitem[{{Buchhave} et~al.(2018){Buchhave}, {Bitsch}, {Johansen}, {Latham},
  {Bizzarro} et~al.}]{Buchhave:2018}
{Buchhave} LA, {Bitsch} B, {Johansen} A, {Latham} DW, {Bizzarro} M, et~al.
  2018.
\textit{\apj} 856:37

\bibitem[{{Buchhave} et~al.(2012){Buchhave}, {Latham}, {Johansen}, {Bizzarro},
  {Torres} et~al.}]{Buchhave:2012}
{Buchhave} LA, {Latham} DW, {Johansen} A, {Bizzarro} M, {Torres} G, et~al.
  2012.
\textit{\nat} 486:375--377

\bibitem[{{Burke} \& {Catanzarite}(2017{\natexlab{a}})}]{Burke:2017}
{Burke} CJ, {Catanzarite} J. 2017{\natexlab{a}}.
{Planet Detection Metrics: Per-Target Detection Contours for Data Release 25}.
Kepler Science Document KSCI-19111-002

\bibitem[{{Burke} \& {Catanzarite}(2017{\natexlab{b}})}]{Burke:2017b}
{Burke} CJ, {Catanzarite} J. 2017{\natexlab{b}}.
{Planet Detection Metrics: Per-Target Flux-Level Transit Injection Tests of TPS
  for Data Release 25}.
Kepler Science Document KSCI-19109-002

\bibitem[{{Burke} et~al.(2015){Burke}, {Christiansen}, {Mullally}, {Seader},
  {Huber} et~al.}]{Burke:2015}
{Burke} CJ, {Christiansen} JL, {Mullally} F, {Seader} S, {Huber} D, et~al.
  2015.
\textit{\apj} 809:8

\bibitem[{{Burt} et~al.(2020){Burt}, {Nielsen}, {Quinn}, {Mamajek}, {Matthews}
  et~al.}]{Burt:2020}
{Burt} JA, {Nielsen} LD, {Quinn} SN, {Mamajek} EE, {Matthews} EC, et~al. 2020.
\textit{\aj} 160:153

\bibitem[{{Ca{\~n}as} et~al.(2019){Ca{\~n}as}, {Wang}, {Mahadevan}, {Bender},
  {De Lee} et~al.}]{Canas:2019}
{Ca{\~n}as} CI, {Wang} S, {Mahadevan} S, {Bender} CF, {De Lee} N, et~al. 2019.
\textit{\apjl} 870:L17

\bibitem[{{Cassan} et~al.(2012){Cassan}, {Kubas}, {Beaulieu}, {Dominik},
  {Horne} et~al.}]{Cassan:2012}
{Cassan} A, {Kubas} D, {Beaulieu} JP, {Dominik} M, {Horne} K, et~al. 2012.
\textit{\nat} 481:167--169

\bibitem[{{Chamberlin}(1916)}]{Chamberlin:1916}
{Chamberlin} TC. 1916.
\textit{\jrasc} 10:473

\bibitem[{{Chambers}(2001)}]{Chambers:2001}
{Chambers} JE. 2001.
\textit{\icarus} 152:205--224

\bibitem[{{Chambers} et~al.(1996){Chambers}, {Wetherill} \&
  {Boss}}]{Chambers:1996}
{Chambers} JE, {Wetherill} GW, {Boss} AP. 1996.
\textit{\icarus} 119:261--268

\bibitem[{{Chaplin} \& {Miglio}(2013)}]{Chaplin:2013}
{Chaplin} WJ, {Miglio} A. 2013.
\textit{\araa} 51:353--392

\bibitem[{{Chatterjee} et~al.(2008){Chatterjee}, {Ford}, {Matsumura} \&
  {Rasio}}]{Chatterjee:2008}
{Chatterjee} S, {Ford} EB, {Matsumura} S, {Rasio} FA. 2008.
\textit{\apj} 686:580--602

\bibitem[{{Chen} \& {Kipping}(2017)}]{ChenKipping:2017}
{Chen} J, {Kipping} D. 2017.
\textit{\apj} 834:17

\bibitem[{{Chiang} \& {Laughlin}(2013)}]{Chiang:2013}
{Chiang} E, {Laughlin} G. 2013.
\textit{\mnras} 431:3444--3455

\bibitem[{{Chiang} \& {Youdin}(2010)}]{Chiang:2010}
{Chiang} E, {Youdin} AN. 2010.
\textit{Annual Review of Earth and Planetary Sciences} 38:493--522

\bibitem[{Christiansen et~al.(2020)Christiansen, Clarke, Burke, Jenkins, Bryson
  et~al.}]{Christiansen:2020}
Christiansen JL, Clarke BD, Burke CJ, Jenkins JM, Bryson ST, et~al. 2020.
\textit{\aj} 160:159

\bibitem[{{Christiansen} et~al.(2015){Christiansen}, {Clarke}, {Burke},
  {Seader}, {Jenkins} et~al.}]{Christiansen:2015}
{Christiansen} JL, {Clarke} BD, {Burke} CJ, {Seader} S, {Jenkins} JM, et~al.
  2015.
\textit{\apj} 810:95

\bibitem[{{Ciardi} et~al.(2015){Ciardi}, {Beichman}, {Horch} \&
  {Howell}}]{Ciardi:2015}
{Ciardi} DR, {Beichman} CA, {Horch} EP, {Howell} SB. 2015.
\textit{\apj} 805:16

\bibitem[{{Ciardi} et~al.(2013){Ciardi}, {Fabrycky}, {Ford}, {Gautier},
  {Howell} et~al.}]{Ciardi:2013}
{Ciardi} DR, {Fabrycky} DC, {Ford} EB, {Gautier} T.~N. I, {Howell} SB, et~al.
  2013.
\textit{\apj} 763:41

\bibitem[{{Clanton} \& {Gaudi}(2014)}]{Clanton:2014}
{Clanton} C, {Gaudi} BS. 2014.
\textit{\apj} 791:91

\bibitem[{{Clanton} \& {Gaudi}(2016)}]{Clanton:2016}
{Clanton} C, {Gaudi} BS. 2016.
\textit{\apj} 819:125

\bibitem[{{Cumming} et~al.(2008){Cumming}, {Butler}, {Marcy}, {Vogt}, {Wright}
  \& {Fischer}}]{Cumming:2008}
{Cumming} A, {Butler} RP, {Marcy} GW, {Vogt} SS, {Wright} JT, {Fischer} DA.
  2008.
\textit{\pasp} 120:531

\bibitem[{{Dai} et~al.(2018){Dai}, {Masuda} \& {Winn}}]{Dai:2018}
{Dai} F, {Masuda} K, {Winn} JN. 2018.
\textit{\apjl} 864:L38

\bibitem[{{Damasso} et~al.(2020){Damasso}, {Sozzetti}, {Lovis}, {Barros},
  {Sousa} et~al.}]{Damasso:2020}
{Damasso} M, {Sozzetti} A, {Lovis} C, {Barros} SCC, {Sousa} SG, et~al. 2020.
\textit{\aap} 642:A31

\bibitem[{{Dawson} \& {Chiang}(2014)}]{Dawson:2014}
{Dawson} RI, {Chiang} E. 2014.
\textit{Science} 346:212--216

\bibitem[{{Dawson} \& {Johnson}(2018)}]{Dawson:2018}
{Dawson} RI, {Johnson} JA. 2018.
\textit{\araa} 56:175--221

\bibitem[{{Dawson} \& {Murray-Clay}(2013)}]{Dawson:2013}
{Dawson} RI, {Murray-Clay} RA. 2013.
\textit{\apjl} 767:L24

\bibitem[{{De Cat} et~al.(2015){De Cat}, {Fu}, {Ren}, {Yang}, {Shi}
  et~al.}]{DeCat:2015}
{De Cat} P, {Fu} JN, {Ren} AB, {Yang} XH, {Shi} JR, et~al. 2015.
\textit{\apjs} 220:19

\bibitem[{{De Rosa} et~al.(2020){De Rosa}, {Dawson} \& {Nielsen}}]{DeRosa:2020}
{De Rosa} RJ, {Dawson} R, {Nielsen} EL. 2020.
\textit{\aap} 640:A73

\bibitem[{{Deck} et~al.(2013){Deck}, {Payne} \& {Holman}}]{Deck:2013}
{Deck} KM, {Payne} M, {Holman} MJ. 2013.
\textit{\apj} 774:129

\bibitem[{{Dong} et~al.(2006){Dong}, {DePoy}, {Gaudi}, {Gould}, {Han}
  et~al.}]{Dong:2006}
{Dong} S, {DePoy} DL, {Gaudi} BS, {Gould} A, {Han} C, et~al. 2006.
\textit{\apj} 642:842--860

\bibitem[{{Dong} et~al.(2009){Dong}, {Gould}, {Udalski}, {Anderson}, {Christie}
  et~al.}]{Dong:2009}
{Dong} S, {Gould} A, {Udalski} A, {Anderson} J, {Christie} GW, et~al. 2009.
\textit{\apj} 695:970--987

\bibitem[{{Dong} et~al.(2014{\natexlab{a}}){Dong}, {Katz} \&
  {Socrates}}]{Dong:2014b}
{Dong} S, {Katz} B, {Socrates} A. 2014{\natexlab{a}}.
\textit{\apjl} 781:L5

\bibitem[{{Dong} et~al.(2018){Dong}, {Xie}, {Zhou}, {Zheng} \&
  {Luo}}]{Dong:2018}
{Dong} S, {Xie} JW, {Zhou} JL, {Zheng} Z, {Luo} A. 2018.
\textit{Proceedings of the National Academy of Science} 115:266--271

\bibitem[{{Dong} et~al.(2014{\natexlab{b}}){Dong}, {Zheng}, {Zhu}, {De Cat},
  {Fu} et~al.}]{Dong:2014a}
{Dong} S, {Zheng} Z, {Zhu} Z, {De Cat} P, {Fu} JN, et~al. 2014{\natexlab{b}}.
\textit{\apjl} 789:L3

\bibitem[{{Dong} \& {Zhu}(2013)}]{Dong:2013}
{Dong} S, {Zhu} Z. 2013.
\textit{\apj} 778:53

\bibitem[{{Dressing} \& {Charbonneau}(2013)}]{Dressing:2013}
{Dressing} CD, {Charbonneau} D. 2013.
\textit{\apj} 767:95

\bibitem[{{Dressing} \& {Charbonneau}(2015)}]{Dressing:2015}
{Dressing} CD, {Charbonneau} D. 2015.
\textit{\apj} 807:45

\bibitem[{{Duch{\^e}ne} \& {Kraus}(2013)}]{Duchene:2013}
{Duch{\^e}ne} G, {Kraus} A. 2013.
\textit{\araa} 51:269--310

\bibitem[{{Emsenhuber} et~al.(2020){Emsenhuber}, {Mordasini}, {Burn},
  {Alibert}, {Benz} \& {Asphaug}}]{Bern_2:2020}
{Emsenhuber} A, {Mordasini} C, {Burn} R, {Alibert} Y, {Benz} W, {Asphaug} E.
  2020.
\textit{arXiv e-prints} :arXiv:2007.05562

\bibitem[{{Fabrycky} et~al.(2014){Fabrycky}, {Lissauer}, {Ragozzine}, {Rowe},
  {Steffen} et~al.}]{Fabrycky:2014}
{Fabrycky} DC, {Lissauer} JJ, {Ragozzine} D, {Rowe} JF, {Steffen} JH, et~al.
  2014.
\textit{\apj} 790:146

\bibitem[{{Fang} \& {Margot}(2012)}]{FangMargot:2012}
{Fang} J, {Margot} JL. 2012.
\textit{\apj} 761:92

\bibitem[{{Fang} \& {Margot}(2013)}]{FangMargot:2013}
{Fang} J, {Margot} JL. 2013.
\textit{\apj} 767:115

\bibitem[{{Fernandes} et~al.(2019){Fernandes}, {Mulders}, {Pascucci},
  {Mordasini} \& {Emsenhuber}}]{Fernandes:2019}
{Fernandes} RB, {Mulders} GD, {Pascucci} I, {Mordasini} C, {Emsenhuber} A.
  2019.
\textit{\apj} 874:81

\bibitem[{{Figueira} et~al.(2012){Figueira}, {Marmier}, {Bou{\'e}}, {Lovis},
  {Santos} et~al.}]{Figueira:2012}
{Figueira} P, {Marmier} M, {Bou{\'e}} G, {Lovis} C, {Santos} NC, et~al. 2012.
\textit{\aap} 541:A139

\bibitem[{{Fischer} et~al.(2016){Fischer}, {Anglada-Escude}, {Arriagada},
  {Baluev}, {Bean} et~al.}]{Fischer:2016}
{Fischer} DA, {Anglada-Escude} G, {Arriagada} P, {Baluev} RV, {Bean} JL, et~al.
  2016.
\textit{\pasp} 128:066001

\bibitem[{{Fischer} \& {Valenti}(2005)}]{Fischer:2005}
{Fischer} DA, {Valenti} J. 2005.
\textit{\apj} 622:1102--1117

\bibitem[{{Ford} et~al.(2008){Ford}, {Quinn} \& {Veras}}]{Ford:2008}
{Ford} EB, {Quinn} SN, {Veras} D. 2008.
\textit{\apj} 678:1407--1418

\bibitem[{{Ford} et~al.(2011){Ford}, {Rowe}, {Fabrycky}, {Carter}, {Holman}
  et~al.}]{Ford:2011}
{Ford} EB, {Rowe} JF, {Fabrycky} DC, {Carter} JA, {Holman} MJ, et~al. 2011.
\textit{\apjs} 197:2

\bibitem[{{Foreman-Mackey} et~al.(2014){Foreman-Mackey}, {Hogg} \&
  {Morton}}]{ForemanMackey:2014}
{Foreman-Mackey} D, {Hogg} DW, {Morton} TD. 2014.
\textit{\apj} 795:64

\bibitem[{{Foreman-Mackey} et~al.(2016){Foreman-Mackey}, {Morton}, {Hogg},
  {Agol} \& {Sch{\"o}lkopf}}]{ForemanMackey:2016}
{Foreman-Mackey} D, {Morton} TD, {Hogg} DW, {Agol} E, {Sch{\"o}lkopf} B. 2016.
\textit{\aj} 152:206

\bibitem[{{Fressin} et~al.(2013){Fressin}, {Torres}, {Charbonneau}, {Bryson},
  {Christiansen} et~al.}]{Fressin:2013}
{Fressin} F, {Torres} G, {Charbonneau} D, {Bryson} ST, {Christiansen} J, et~al.
  2013.
\textit{\apj} 766:81

\bibitem[{{Fulton} \& {Petigura}(2018)}]{Fulton:2018}
{Fulton} BJ, {Petigura} EA. 2018.
\textit{\aj} 156:264

\bibitem[{{Fulton} et~al.(2017){Fulton}, {Petigura}, {Howard}, {Isaacson},
  {Marcy} et~al.}]{Fulton:2017}
{Fulton} BJ, {Petigura} EA, {Howard} AW, {Isaacson} H, {Marcy} GW, et~al. 2017.
\textit{\aj} 154:109

\bibitem[{{Funk} et~al.(2010){Funk}, {Wuchterl}, {Schwarz}, {Pilat-Lohinger} \&
  {Eggl}}]{Funk:2010}
{Funk} B, {Wuchterl} G, {Schwarz} R, {Pilat-Lohinger} E, {Eggl} S. 2010.
\textit{\aap} 516:A82

\bibitem[{{Furlan} et~al.(2017){Furlan}, {Ciardi}, {Everett}, {Saylors},
  {Teske} et~al.}]{Furlan:2017}
{Furlan} E, {Ciardi} DR, {Everett} ME, {Saylors} M, {Teske} JK, et~al. 2017.
\textit{\aj} 153:71

\bibitem[{{Gaia Collaboration} et~al.(2018){Gaia Collaboration}, {Brown},
  {Vallenari}, {Prusti}, {de Bruijne} et~al.}]{Gaia_dr2}
{Gaia Collaboration}, {Brown} AGA, {Vallenari} A, {Prusti} T, {de Bruijne} JHJ,
  et~al. 2018.
\textit{\aap} 616:A1

\bibitem[{{Gaia Collaboration} et~al.(2016){Gaia Collaboration}, {Prusti}, {de
  Bruijne}, {Brown}, {Vallenari} et~al.}]{Gaia}
{Gaia Collaboration}, {Prusti} T, {de Bruijne} JHJ, {Brown} AGA, {Vallenari} A,
  et~al. 2016.
\textit{\aap} 595:A1

\bibitem[{{Gandolfi} et~al.(2018){Gandolfi}, {Barrag{\'a}n}, {Livingston},
  {Fridlund}, {Justesen} et~al.}]{Gandolfi:2018}
{Gandolfi} D, {Barrag{\'a}n} O, {Livingston} JH, {Fridlund} M, {Justesen} AB,
  et~al. 2018.
\textit{\aap} 619:L10

\bibitem[{{Gaudi}(2012)}]{Gaudi:2012}
{Gaudi} BS. 2012.
\textit{\araa} 50:411--453

\bibitem[{{Gilbert} \& {Fabrycky}(2020)}]{Gilbert:2020}
{Gilbert} GJ, {Fabrycky} DC. 2020.
\textit{\aj} 159:281

\bibitem[{{Ginzburg} et~al.(2018){Ginzburg}, {Schlichting} \&
  {Sari}}]{Ginzburg:2018}
{Ginzburg} S, {Schlichting} HE, {Sari} R. 2018.
\textit{\mnras} 476:759--765

\bibitem[{{Goldreich} \& {Schlichting}(2014)}]{Goldreich:2014}
{Goldreich} P, {Schlichting} HE. 2014.
\textit{\aj} 147:32

\bibitem[{{Goldreich} \& {Tremaine}(1980)}]{Goldreich:1980}
{Goldreich} P, {Tremaine} S. 1980.
\textit{\apj} 241:425--441

\bibitem[{{Gould}(2016)}]{Gould:2016}
{Gould} A. 2016.
\textit{Journal of Korean Astronomical Society} 49:123--126

\bibitem[{{Gould} et~al.(2010){Gould}, {Dong}, {Gaudi}, {Udalski}, {Bond}
  et~al.}]{Gould:2010}
{Gould} A, {Dong} S, {Gaudi} BS, {Udalski} A, {Bond} IA, et~al. 2010.
\textit{\apj} 720:1073--1089

\bibitem[{{Gould} et~al.(2006{\natexlab{a}}){Gould}, {Dorsher}, {Gaudi} \&
  {Udalski}}]{Gould:2006a}
{Gould} A, {Dorsher} S, {Gaudi} BS, {Udalski} A. 2006{\natexlab{a}}.
\textit{\actaa} 56:1--50

\bibitem[{{Gould} \& {Loeb}(1992)}]{Gould:1992}
{Gould} A, {Loeb} A. 1992.
\textit{\apj} 396:104

\bibitem[{{Gould} et~al.(2020{\natexlab{a}}){Gould}, {Ryu}, {Calchi Novati},
  {Zang}, {Albrow} et~al.}]{Gould:2020}
{Gould} A, {Ryu} YH, {Calchi Novati} S, {Zang} W, {Albrow} MD, et~al.
  2020{\natexlab{a}}.
\textit{Journal of Korean Astronomical Society} 53:9--26

\bibitem[{{Gould} et~al.(2006{\natexlab{b}}){Gould}, {Udalski}, {An},
  {Bennett}, {Zhou} et~al.}]{Gould:2006}
{Gould} A, {Udalski} A, {An} D, {Bennett} DP, {Zhou} AY, et~al.
  2006{\natexlab{b}}.
\textit{\apjl} 644:L37--L40

\bibitem[{{Gould} et~al.(2020{\natexlab{b}}){Gould}, {Zang}, {Mao} \&
  {Dong}}]{Gould:2021}
{Gould} A, {Zang} W, {Mao} S, {Dong} S. 2020{\natexlab{b}}.
\textit{arXiv e-prints} :arXiv:2010.09671

\bibitem[{{Guo} et~al.(2017){Guo}, {Johnson}, {Mann}, {Kraus}, {Curtis} \&
  {Latham}}]{Guo:2017}
{Guo} X, {Johnson} JA, {Mann} AW, {Kraus} AL, {Curtis} JL, {Latham} DW. 2017.
\textit{\apj} 838:25

\bibitem[{{Gupta} \& {Schlichting}(2019)}]{Gupta:2019}
{Gupta} A, {Schlichting} HE. 2019.
\textit{\mnras} 487:24--33

\bibitem[{{Gupta} \& {Schlichting}(2020)}]{Gupta:2020}
{Gupta} A, {Schlichting} HE. 2020.
\textit{\mnras} 493:792--806

\bibitem[{{Hadden} \& {Lithwick}(2014)}]{Hadden:2014}
{Hadden} S, {Lithwick} Y. 2014.
\textit{\apj} 787:80

\bibitem[{{Hadden} \& {Lithwick}(2017)}]{Hadden:2017}
{Hadden} S, {Lithwick} Y. 2017.
\textit{\aj} 154:5

\bibitem[{{Hadden} \& {Lithwick}(2018)}]{Hadden:2018}
{Hadden} S, {Lithwick} Y. 2018.
\textit{\aj} 156:95

\bibitem[{{Haffert} et~al.(2019){Haffert}, {Bohn}, {de Boer}, {Snellen},
  {Brinchmann} et~al.}]{PDS70}
{Haffert} SY, {Bohn} AJ, {de Boer} J, {Snellen} IAG, {Brinchmann} J, et~al.
  2019.
\textit{Nature Astronomy} 3:749--754

\bibitem[{{Hallatt} \& {Lee}(2020)}]{Hallatt:2020}
{Hallatt} T, {Lee} EJ. 2020.
\textit{\apj} 904:134

\bibitem[{{Hamer} \& {Schlaufman}(2020)}]{Hamer:2020}
{Hamer} JH, {Schlaufman} KC. 2020.
\textit{\aj} 160:138

\bibitem[{{Han} et~al.(2005){Han}, {Gaudi}, {An} \& {Gould}}]{Han:2005}
{Han} C, {Gaudi} BS, {An} JH, {Gould} A. 2005.
\textit{\apj} 618:962--972

\bibitem[{{Hansen} \& {Murray}(2013)}]{Hansen:2013}
{Hansen} BMS, {Murray} N. 2013.
\textit{\apj} 775:53

\bibitem[{{Hayashi}(1981)}]{Hayashi:1981}
{Hayashi} C. 1981.
\textit{Progress of Theoretical Physics Supplement} 70:35--53

\bibitem[{{He} et~al.(2019){He}, {Ford} \& {Ragozzine}}]{He:2019}
{He} MY, {Ford} EB, {Ragozzine} D. 2019.
\textit{\mnras} 490:4575--4605

\bibitem[{{He} et~al.(2020){He}, {Ford}, {Ragozzine} \& {Carrera}}]{He:2020}
{He} MY, {Ford} EB, {Ragozzine} D, {Carrera} D. 2020.
\textit{\aj} 160:276

\bibitem[{{Herman} et~al.(2019){Herman}, {Zhu} \& {Wu}}]{Herman:2019}
{Herman} MK, {Zhu} W, {Wu} Y. 2019.
\textit{\aj} 157:248

\bibitem[{{Hirano} et~al.(2018){Hirano}, {Dai}, {Gandolfi}, {Fukui},
  {Livingston} et~al.}]{Hirano:2018}
{Hirano} T, {Dai} F, {Gandolfi} D, {Fukui} A, {Livingston} JH, et~al. 2018.
\textit{\aj} 155:127

\bibitem[{{Hoffman} \& {Rowe}(2017)}]{Hoffman:2017}
{Hoffman} Kelsey L, {Rowe} JF. 2017.
{Uniform Modeling of KOIs: MCMC Notes for Data Release 25}.
Kepler Science Document KSCI-19113-001

\bibitem[{{Holczer} et~al.(2016){Holczer}, {Mazeh}, {Nachmani},
  {Jontof-Hutter}, {Ford} et~al.}]{Holczer:2016}
{Holczer} T, {Mazeh} T, {Nachmani} G, {Jontof-Hutter} D, {Ford} EB, et~al.
  2016.
\textit{\apjs} 225:9

\bibitem[{{Holman} \& {Murray}(2005)}]{Holman:2005}
{Holman} MJ, {Murray} NW. 2005.
\textit{Science} 307:1288--1291

\bibitem[{{Holman} \& {Wiegert}(1999)}]{Holman:1999}
{Holman} MJ, {Wiegert} PA. 1999.
\textit{\aj} 117:621--628

\bibitem[{{Hori} \& {Ikoma}(2011)}]{Hori:2011}
{Hori} Y, {Ikoma} M. 2011.
\textit{\mnras} 416:1419--1429

\bibitem[{{Howard} et~al.(2012){Howard}, {Marcy}, {Bryson}, {Jenkins}, {Rowe}
  et~al.}]{Howard:2012}
{Howard} AW, {Marcy} GW, {Bryson} ST, {Jenkins} JM, {Rowe} JF, et~al. 2012.
\textit{\apjs} 201:15

\bibitem[{{Hsu} et~al.(2019){Hsu}, {Ford}, {Ragozzine} \& {Ashby}}]{Hsu:2019}
{Hsu} DC, {Ford} EB, {Ragozzine} D, {Ashby} K. 2019.
\textit{\aj} 158:109

\bibitem[{{Hsu} et~al.(2018){Hsu}, {Ford}, {Ragozzine} \&
  {Morehead}}]{Hsu:2018}
{Hsu} DC, {Ford} EB, {Ragozzine} D, {Morehead} RC. 2018.
\textit{\aj} 155:205

\bibitem[{{Huang} et~al.(2016){Huang}, {Wu} \& {Triaud}}]{Huang:2016}
{Huang} C, {Wu} Y, {Triaud} AHMJ. 2016.
\textit{\apj} 825:98

\bibitem[{{Huang} et~al.(2018){Huang}, {Burt}, {Vanderburg}, {G{\"u}nther},
  {Shporer} et~al.}]{Huang:2018}
{Huang} CX, {Burt} J, {Vanderburg} A, {G{\"u}nther} MN, {Shporer} A, et~al.
  2018.
\textit{\apjl} 868:L39

\bibitem[{{Huang} et~al.(2017){Huang}, {Petrovich} \& {Deibert}}]{Huang:2017}
{Huang} CX, {Petrovich} C, {Deibert} E. 2017.
\textit{\aj} 153:210

\bibitem[{{Huang} et~al.(2020){Huang}, {Quinn}, {Vanderburg}, {Becker},
  {Rodriguez} et~al.}]{Huang:2020}
{Huang} CX, {Quinn} SN, {Vanderburg} A, {Becker} J, {Rodriguez} JE, et~al.
  2020.
\textit{\apjl} 892:L7

\bibitem[{{Ida} et~al.(1993){Ida}, {Kokubo} \& {Makino}}]{Ida:1993}
{Ida} S, {Kokubo} E, {Makino} J. 1993.
\textit{\mnras} 263:875

\bibitem[{{Ida} \& {Lin}(2004{\natexlab{a}})}]{IdaLin:2004a}
{Ida} S, {Lin} DNC. 2004{\natexlab{a}}.
\textit{\apj} 604:388--413

\bibitem[{{Ida} \& {Lin}(2004{\natexlab{b}})}]{IdaLin:2004}
{Ida} S, {Lin} DNC. 2004{\natexlab{b}}.
\textit{\apj} 616:567--572

\bibitem[{{Ida} \& {Lin}(2010)}]{IdaLin:2010}
{Ida} S, {Lin} DNC. 2010.
\textit{\apj} 719:810--830

\bibitem[{{Ida} et~al.(2013){Ida}, {Lin} \& {Nagasawa}}]{Ida:2013}
{Ida} S, {Lin} DNC, {Nagasawa} M. 2013.
\textit{\apj} 775:42

\bibitem[{{Ida} \& {Makino}(1993)}]{IdaMakino:1993}
{Ida} S, {Makino} J. 1993.
\textit{\icarus} 106:210--227

\bibitem[{{Izidoro} et~al.(2019){Izidoro}, {Bitsch}, {Raymond}, {Johansen},
  {Morbidelli} et~al.}]{Izidoro:2019}
{Izidoro} A, {Bitsch} B, {Raymond} SN, {Johansen} A, {Morbidelli} A, et~al.
  2019.
\textit{arXiv e-prints} :arXiv:1902.08772

\bibitem[{{Izidoro} et~al.(2017){Izidoro}, {Ogihara}, {Raymond}, {Morbidelli},
  {Pierens} et~al.}]{Izidoro:2017}
{Izidoro} A, {Ogihara} M, {Raymond} SN, {Morbidelli} A, {Pierens} A, et~al.
  2017.
\textit{\mnras} 470:1750--1770

\bibitem[{{Izidoro} et~al.(2015){Izidoro}, {Raymond}, {Morbidelli}, {Hersant}
  \& {Pierens}}]{Izidoro:2015}
{Izidoro} A, {Raymond} SN, {Morbidelli} Ar, {Hersant} F, {Pierens} A. 2015.
\textit{\apjl} 800:L22

\bibitem[{{Jackson} et~al.(2012){Jackson}, {Davis} \&
  {Wheatley}}]{Jackson:2012}
{Jackson} AP, {Davis} TA, {Wheatley} PJ. 2012.
\textit{\mnras} 422:2024--2043

\bibitem[{{Jackson} et~al.(2013){Jackson}, {Stark}, {Adams}, {Chambers} \&
  {Deming}}]{Jackson:2013}
{Jackson} B, {Stark} CC, {Adams} ER, {Chambers} J, {Deming} D. 2013.
\textit{\apj} 779:165

\bibitem[{{Jiang} et~al.(2020){Jiang}, {Xie} \& {Zhou}}]{Jiang:2020}
{Jiang} CF, {Xie} JW, {Zhou} JL. 2020.
\textit{\aj} 160:180

\bibitem[{{Jin} \& {Mordasini}(2018)}]{Jin:2018}
{Jin} S, {Mordasini} C. 2018.
\textit{\apj} 853:163

\bibitem[{{Johansen} et~al.(2012){Johansen}, {Davies}, {Church} \&
  {Holmelin}}]{Johansen:2012}
{Johansen} A, {Davies} MB, {Church} RP, {Holmelin} V. 2012.
\textit{\apj} 758:39

\bibitem[{{Johansen} \& {Lambrechts}(2017)}]{Johansen:2017}
{Johansen} A, {Lambrechts} M. 2017.
\textit{Annual Review of Earth and Planetary Sciences} 45:359--387

\bibitem[{{Johansen} et~al.(2007){Johansen}, {Oishi}, {Mac Low}, {Klahr},
  {Henning} \& {Youdin}}]{Johansen:2007}
{Johansen} A, {Oishi} JS, {Mac Low} MM, {Klahr} H, {Henning} T, {Youdin} A.
  2007.
\textit{\nat} 448:1022--1025

\bibitem[{{Johnson} et~al.(2010){Johnson}, {Aller}, {Howard} \&
  {Crepp}}]{Johnson:2010}
{Johnson} JA, {Aller} KM, {Howard} AW, {Crepp} JR. 2010.
\textit{\pasp} 122:905

\bibitem[{{Johnson} et~al.(2007){Johnson}, {Butler}, {Marcy}, {Fischer}, {Vogt}
  et~al.}]{Johnson:2007}
{Johnson} JA, {Butler} RP, {Marcy} GW, {Fischer} DA, {Vogt} SS, et~al. 2007.
\textit{\apj} 670:833--840

\bibitem[{{Johnson} et~al.(2017){Johnson}, {Petigura}, {Fulton}, {Marcy},
  {Howard} et~al.}]{Johnson:2017}
{Johnson} JA, {Petigura} EA, {Fulton} BJ, {Marcy} GW, {Howard} AW, et~al. 2017.
\textit{\aj} 154:108

\bibitem[{{Johnson} et~al.(2020){Johnson}, {Penny}, {Gaudi}, {Kerins},
  {Rattenbury} et~al.}]{Johnson:2020}
{Johnson} SA, {Penny} M, {Gaudi} BS, {Kerins} E, {Rattenbury} NJ, et~al. 2020.
\textit{\aj} 160:123

\bibitem[{{Jung} et~al.(2019){Jung}, {Gould}, {Zang}, {Hwang}, {Ryu}
  et~al.}]{Jung:2019}
{Jung} YK, {Gould} A, {Zang} W, {Hwang} KH, {Ryu} YH, et~al. 2019.
\textit{\aj} 157:72

\bibitem[{{Juri{\'c}} \& {Tremaine}(2008)}]{JuricTremaine:2008}
{Juri{\'c}} M, {Tremaine} S. 2008.
\textit{\apj} 686:603--620

\bibitem[{{Kane} et~al.(2012){Kane}, {Ciardi}, {Gelino} \& {von
  Braun}}]{Kane:2012}
{Kane} SR, {Ciardi} DR, {Gelino} DM, {von Braun} K. 2012.
\textit{\mnras} 425:757--762

\bibitem[{{Kawahara} \& {Masuda}(2019)}]{Kawahara:2019}
{Kawahara} H, {Masuda} K. 2019.
\textit{\aj} 157:218

\bibitem[{{Kennedy} \& {Kenyon}(2008)}]{Kennedy:2008}
{Kennedy} GM, {Kenyon} SJ. 2008.
\textit{\apj} 673:502--512

\bibitem[{{Kim} et~al.(2020){Kim}, {Hwang}, {Gould}, {Yee}, {Ryu}
  et~al.}]{Kim:2020}
{Kim} HW, {Hwang} KH, {Gould} A, {Yee} JC, {Ryu} YH, et~al. 2020.
\textit{arXiv e-prints} :arXiv:2007.06870

\bibitem[{{Kim} et~al.(2016){Kim}, {Lee}, {Park}, {Kim}, {Cha}
  et~al.}]{Kim:2016}
{Kim} SL, {Lee} CU, {Park} BG, {Kim} DJ, {Cha} SM, et~al. 2016.
\textit{Journal of Korean Astronomical Society} 49:37--44

\bibitem[{{King} \& {Wheatley}(2021)}]{King:2020}
{King} GW, {Wheatley} PJ. 2021.
\textit{\mnras} 501:L28--L32

\bibitem[{{Kipping}(2018)}]{Kipping:2018}
{Kipping} D. 2018.
\textit{\mnras} 473:784--795

\bibitem[{{Kley} \& {Nelson}(2012)}]{Kley:2012}
{Kley} W, {Nelson} RP. 2012.
\textit{\araa} 50:211--249

\bibitem[{{Knutson} et~al.(2014){Knutson}, {Fulton}, {Montet}, {Kao}, {Ngo}
  et~al.}]{Knutson:2014}
{Knutson} HA, {Fulton} BJ, {Montet} BT, {Kao} M, {Ngo} H, et~al. 2014.
\textit{\apj} 785:126

\bibitem[{{Kokubo} \& {Ida}(1998)}]{Kokubo:1998}
{Kokubo} E, {Ida} S. 1998.
\textit{\icarus} 131:171--178

\bibitem[{{Kraus} et~al.(2016){Kraus}, {Ireland}, {Huber}, {Mann} \&
  {Dupuy}}]{Kraus:2016}
{Kraus} AL, {Ireland} MJ, {Huber} D, {Mann} AW, {Dupuy} TJ. 2016.
\textit{\aj} 152:8

\bibitem[{{Kruijer} et~al.(2017){Kruijer}, {Burkhardt}, {Budde} \&
  {Kleine}}]{Kruijer:2017}
{Kruijer} TS, {Burkhardt} C, {Budde} G, {Kleine} T. 2017.
\textit{Proceedings of the National Academy of Science} 114:6712--6716

\bibitem[{{Kurokawa} \& {Nakamoto}(2014)}]{Kurokawa:2014}
{Kurokawa} H, {Nakamoto} T. 2014.
\textit{\apj} 783:54

\bibitem[{{Kutra} \& {Wu}(2020)}]{Kutra:2020}
{Kutra} T, {Wu} Y. 2020.
\textit{arXiv e-prints} :arXiv:2003.08431

\bibitem[{{Lambrechts} \& {Johansen}(2012)}]{Lambrechts:2012}
{Lambrechts} M, {Johansen} A. 2012.
\textit{\aap} 544:A32

\bibitem[{{Lambrechts} et~al.(2014){Lambrechts}, {Johansen} \&
  {Morbidelli}}]{Lambrechts:2014}
{Lambrechts} M, {Johansen} A, {Morbidelli} A. 2014.
\textit{\aap} 572:A35

\bibitem[{{Lambrechts} et~al.(2019){Lambrechts}, {Morbidelli}, {Jacobson},
  {Johansen}, {Bitsch} et~al.}]{Lambrechts:2019}
{Lambrechts} M, {Morbidelli} A, {Jacobson} SA, {Johansen} A, {Bitsch} B, et~al.
  2019.
\textit{\aap} 627:A83

\bibitem[{{Laskar}(1997)}]{Laskar:1997}
{Laskar} J. 1997.
\textit{\aap} 317:L75--L78

\bibitem[{{Laskar} \& {Petit}(2017)}]{Laskar:2017}
{Laskar} J, {Petit} AC. 2017.
\textit{\aap} 605:A72

\bibitem[{{Lee}(2019)}]{Lee:2019}
{Lee} EJ. 2019.
\textit{\apj} 878:36

\bibitem[{{Lee} \& {Chiang}(2017)}]{Lee:2017}
{Lee} EJ, {Chiang} E. 2017.
\textit{\apj} 842:40

\bibitem[{{Lee} et~al.(2014){Lee}, {Chiang} \& {Ormel}}]{Lee:2014}
{Lee} EJ, {Chiang} E, {Ormel} CW. 2014.
\textit{\apj} 797:95

\bibitem[{{Lee} \& {Connors}(2020)}]{Lee:2020}
{Lee} EJ, {Connors} NJ. 2020.
\textit{arXiv e-prints} :arXiv:2008.01105

\bibitem[{{Lee} \& {Peale}(2002)}]{Lee:2002}
{Lee} MH, {Peale} SJ. 2002.
\textit{\apj} 567:596--609

\bibitem[{{Limbach} \& {Turner}(2015)}]{Limbach:2015}
{Limbach} MA, {Turner} EL. 2015.
\textit{Proceedings of the National Academy of Science} 112:20--24

\bibitem[{{Lin} et~al.(1996){Lin}, {Bodenheimer} \& {Richardson}}]{Lin:1996}
{Lin} DNC, {Bodenheimer} P, {Richardson} DC. 1996.
\textit{\nat} 380:606--607

\bibitem[{{Lin} et~al.(2018){Lin}, {Lee} \& {Chiang}}]{Lin:2018}
{Lin} JW, {Lee} EJ, {Chiang} E. 2018.
\textit{\mnras} 480:4338--4354

\bibitem[{{Lissauer}(1993)}]{Lissauer:1993}
{Lissauer} JJ. 1993.
\textit{\araa} 31:129--174

\bibitem[{{Lissauer} et~al.(2011){Lissauer}, {Ragozzine}, {Fabrycky},
  {Steffen}, {Ford} et~al.}]{Lissauer:2011}
{Lissauer} JJ, {Ragozzine} D, {Fabrycky} DC, {Steffen} JH, {Ford} EB, et~al.
  2011.
\textit{\apjs} 197:8

\bibitem[{{Lithwick} et~al.(2012){Lithwick}, {Xie} \& {Wu}}]{Lithwick:2012}
{Lithwick} Y, {Xie} J, {Wu} Y. 2012.
\textit{\apj} 761:122

\bibitem[{{Liu} et~al.(2019){Liu}, {Lambrechts}, {Johansen} \&
  {Liu}}]{Liu:2019}
{Liu} B, {Lambrechts} M, {Johansen} A, {Liu} F. 2019.
\textit{\aap} 632:A7

\bibitem[{{Liu} \& {Ormel}(2018)}]{Liu:2018}
{Liu} B, {Ormel} CW. 2018.
\textit{\aap} 615:A138

\bibitem[{{Liu} et~al.(2016){Liu}, {Yong}, {Asplund}, {Ram{\'\i}rez},
  {Mel{\'e}ndez} et~al.}]{Liu:2016}
{Liu} F, {Yong} D, {Asplund} M, {Ram{\'\i}rez} I, {Mel{\'e}ndez} J, et~al.
  2016.
\textit{\mnras} 456:2636--2646

\bibitem[{{Lloyd}(2011)}]{Lloyd:2011}
{Lloyd} JP. 2011.
\textit{\apjl} 739:L49

\bibitem[{{Lopez} \& {Fortney}(2013)}]{Lopez:2013}
{Lopez} ED, {Fortney} JJ. 2013.
\textit{\apj} 776:2

\bibitem[{{Lopez} \& {Rice}(2018)}]{Lopez:2018}
{Lopez} ED, {Rice} K. 2018.
\textit{\mnras} 479:5303--5311

\bibitem[{{Lundkvist} et~al.(2016){Lundkvist}, {Kjeldsen}, {Albrecht},
  {Davies}, {Basu} et~al.}]{Lundkvist:2016}
{Lundkvist} MS, {Kjeldsen} H, {Albrecht} S, {Davies} GR, {Basu} S, et~al. 2016.
\textit{Nature Communications} 7:11201

\bibitem[{{Ma} et~al.(2016){Ma}, {Mao}, {Ida}, {Zhu} \& {Lin}}]{Ma:2016}
{Ma} S, {Mao} S, {Ida} S, {Zhu} W, {Lin} DNC. 2016.
\textit{\mnras} 461:L107--L111

\bibitem[{{Madhusudhan}(2019)}]{Madhusudhan:2019}
{Madhusudhan} N. 2019.
\textit{\araa} 57:617--663

\bibitem[{{Malla} et~al.(2020){Malla}, {Stello}, {Huber}, {Montet}, {Bedding}
  et~al.}]{Malla:2020}
{Malla} SP, {Stello} D, {Huber} D, {Montet} BT, {Bedding} TR, et~al. 2020.
\textit{\mnras} 496:5423--5435

\bibitem[{{Mao}(2012)}]{Mao:2012}
{Mao} S. 2012.
\textit{Research in Astronomy and Astrophysics} 12:947--972

\bibitem[{{Mao} \& {Paczynski}(1991)}]{Mao:1991}
{Mao} S, {Paczynski} B. 1991.
\textit{\apjl} 374:L37

\bibitem[{{Masuda} et~al.(2020){Masuda}, {Winn} \& {Kawahara}}]{Masuda:2020}
{Masuda} K, {Winn} JN, {Kawahara} H. 2020.
\textit{\aj} 159:38

\bibitem[{{Matsakos} \& {K{\"o}nigl}(2016)}]{Matsakos:2016}
{Matsakos} T, {K{\"o}nigl} A. 2016.
\textit{\apjl} 820:L8

\bibitem[{{Mayor} et~al.(2011){Mayor}, {Marmier}, {Lovis}, {Udry},
  {S{\'e}gransan} et~al.}]{Mayor:2011}
{Mayor} M, {Marmier} M, {Lovis} C, {Udry} S, {S{\'e}gransan} D, et~al. 2011.
\textit{arXiv e-prints} :arXiv:1109.2497

\bibitem[{{Mayor} \& {Queloz}(1995)}]{Mayor:1995}
{Mayor} M, {Queloz} D. 1995.
\textit{\nat} 378:355--359

\bibitem[{{Mazeh} et~al.(2016){Mazeh}, {Holczer} \& {Faigler}}]{Mazeh:2016}
{Mazeh} T, {Holczer} T, {Faigler} S. 2016.
\textit{\aap} 589:A75

\bibitem[{{McArthur} et~al.(2010){McArthur}, {Benedict}, {Barnes}, {Martioli},
  {Korzennik} et~al.}]{McArthur:2010}
{McArthur} BE, {Benedict} GF, {Barnes} R, {Martioli} E, {Korzennik} S, et~al.
  2010.
\textit{\apj} 715:1203--1220

\bibitem[{{McDonald} et~al.(2019){McDonald}, {Kreidberg} \&
  {Lopez}}]{McDonald:2019}
{McDonald} GD, {Kreidberg} L, {Lopez} E. 2019.
\textit{\apj} 876:22

\bibitem[{{Millholland} \& {Laughlin}(2019)}]{Millholland:2019}
{Millholland} S, {Laughlin} G. 2019.
\textit{Nature Astronomy} 3:424--433

\bibitem[{{Millholland} et~al.(2017){Millholland}, {Wang} \&
  {Laughlin}}]{Millholland:2017}
{Millholland} S, {Wang} S, {Laughlin} G. 2017.
\textit{\apjl} 849:L33

\bibitem[{{Mills} \& {Fabrycky}(2017)}]{Mills:2017}
{Mills} SM, {Fabrycky} DC. 2017.
\textit{\aj} 153:45

\bibitem[{{Mills} et~al.(2019){Mills}, {Howard}, {Petigura}, {Fulton},
  {Isaacson} \& {Weiss}}]{Mills:2019}
{Mills} SM, {Howard} AW, {Petigura} EA, {Fulton} BJ, {Isaacson} H, {Weiss} LM.
  2019.
\textit{\aj} 157:198

\bibitem[{{Mizuno}(1980)}]{Mizuno:1980}
{Mizuno} H. 1980.
\textit{Progress of Theoretical Physics} 64:544--557

\bibitem[{{Moe} \& {Kratter}(2019)}]{Moe:2020}
{Moe} M, {Kratter} KM. 2019.
\textit{arXiv e-prints} :arXiv:1912.01699

\bibitem[{{Moe} et~al.(2019){Moe}, {Kratter} \& {Badenes}}]{Moe:2019}
{Moe} M, {Kratter} KM, {Badenes} C. 2019.
\textit{\apj} 875:61

\bibitem[{{Moorhead} et~al.(2011){Moorhead}, {Ford}, {Morehead}, {Rowe},
  {Borucki} et~al.}]{Moorhead:2011}
{Moorhead} AV, {Ford} EB, {Morehead} RC, {Rowe} J, {Borucki} WJ, et~al. 2011.
\textit{\apjs} 197:1

\bibitem[{{Morbidelli} et~al.(2015){Morbidelli}, {Lambrechts}, {Jacobson} \&
  {Bitsch}}]{Morbidelli:2015}
{Morbidelli} A, {Lambrechts} M, {Jacobson} S, {Bitsch} B. 2015.
\textit{\icarus} 258:418--429

\bibitem[{{Mordasini} et~al.(2009){Mordasini}, {Alibert} \&
  {Benz}}]{Mordasini:2009}
{Mordasini} C, {Alibert} Y, {Benz} W. 2009.
\textit{\aap} 501:1139--1160

\bibitem[{{Morton} et~al.(2016){Morton}, {Bryson}, {Coughlin}, {Rowe},
  {Ravichandran} et~al.}]{Morton:2016}
{Morton} TD, {Bryson} ST, {Coughlin} JL, {Rowe} JF, {Ravichandran} G, et~al.
  2016.
\textit{\apj} 822:86

\bibitem[{{Mr{\'o}z} et~al.(2020{\natexlab{a}}){Mr{\'o}z}, {Poleski}, {Gould},
  {Udalski}, {Sumi} et~al.}]{Mroz:2021}
{Mr{\'o}z} P, {Poleski} R, {Gould} A, {Udalski} A, {Sumi} T, et~al.
  2020{\natexlab{a}}.
\textit{\apjl} 903:L11

\bibitem[{{Mr{\'o}z} et~al.(2020{\natexlab{b}}){Mr{\'o}z}, {Poleski}, {Han},
  {Udalski}, {Gould} et~al.}]{Mroz:2020}
{Mr{\'o}z} P, {Poleski} R, {Han} C, {Udalski} A, {Gould} A, et~al.
  2020{\natexlab{b}}.
\textit{\aj} 159:262

\bibitem[{{Mr{\'o}z} et~al.(2018){Mr{\'o}z}, {Ryu}, {Skowron}, {Udalski},
  {Gould} et~al.}]{Mroz:2018}
{Mr{\'o}z} P, {Ryu} YH, {Skowron} J, {Udalski} A, {Gould} A, et~al. 2018.
\textit{\aj} 155:121

\bibitem[{{Mr{\'o}z} et~al.(2019){Mr{\'o}z}, {Udalski}, {Bennett}, {Ryu},
  {Sumi} et~al.}]{Mroz:2019}
{Mr{\'o}z} P, {Udalski} A, {Bennett} DP, {Ryu} YH, {Sumi} T, et~al. 2019.
\textit{\aap} 622:A201

\bibitem[{{Mr{\'o}z} et~al.(2017){Mr{\'o}z}, {Udalski}, {Skowron}, {Poleski},
  {Koz{\l}owski} et~al.}]{Mroz:2017}
{Mr{\'o}z} P, {Udalski} A, {Skowron} J, {Poleski} R, {Koz{\l}owski} S, et~al.
  2017.
\textit{\nat} 548:183--186

\bibitem[{{Mulders} et~al.(2019){Mulders}, {Mordasini}, {Pascucci}, {Ciesla},
  {Emsenhuber} \& {Apai}}]{Mulders:2019}
{Mulders} GD, {Mordasini} C, {Pascucci} I, {Ciesla} FJ, {Emsenhuber} A, {Apai}
  D. 2019.
\textit{\apj} 887:157

\bibitem[{{Mulders} et~al.(2015{\natexlab{a}}){Mulders}, {Pascucci} \&
  {Apai}}]{Mulders:2015a}
{Mulders} GD, {Pascucci} I, {Apai} D. 2015{\natexlab{a}}.
\textit{\apj} 798:112

\bibitem[{{Mulders} et~al.(2015{\natexlab{b}}){Mulders}, {Pascucci} \&
  {Apai}}]{Mulders:2015b}
{Mulders} GD, {Pascucci} I, {Apai} D. 2015{\natexlab{b}}.
\textit{\apj} 814:130

\bibitem[{{Mulders} et~al.(2018){Mulders}, {Pascucci}, {Apai} \&
  {Ciesla}}]{Mulders:2018}
{Mulders} GD, {Pascucci} I, {Apai} D, {Ciesla} FJ. 2018.
\textit{\aj} 156:24

\bibitem[{{Mulders} et~al.(2016){Mulders}, {Pascucci}, {Apai}, {Frasca} \&
  {Molenda-{\.Z}akowicz}}]{Mulders:2016}
{Mulders} GD, {Pascucci} I, {Apai} D, {Frasca} A, {Molenda-{\.Z}akowicz} J.
  2016.
\textit{\aj} 152:187

\bibitem[{{Munoz Romero} \& {Kempton}(2018)}]{MunozRomero:2018}
{Munoz Romero} CE, {Kempton} EMR. 2018.
\textit{\aj} 155:134

\bibitem[{{Murchikova} \& {Tremaine}(2020)}]{Murchikova:2020}
{Murchikova} L, {Tremaine} S. 2020.
\textit{\aj} 160:160

\bibitem[{{Nakagawa} et~al.(1986){Nakagawa}, {Sekiya} \&
  {Hayashi}}]{Nakagawa:1986}
{Nakagawa} Y, {Sekiya} M, {Hayashi} C. 1986.
\textit{\icarus} 67:375--390

\bibitem[{{Ngo} et~al.(2016){Ngo}, {Knutson}, {Hinkley}, {Bryan}, {Crepp}
  et~al.}]{Ngo:2016}
{Ngo} H, {Knutson} HA, {Hinkley} S, {Bryan} M, {Crepp} JR, et~al. 2016.
\textit{\apj} 827:8

\bibitem[{{Nielsen} et~al.(2019){Nielsen}, {De Rosa}, {Macintosh}, {Wang},
  {Ruffio} et~al.}]{Nielsen:2019}
{Nielsen} EL, {De Rosa} RJ, {Macintosh} B, {Wang} JJ, {Ruffio} JB, et~al. 2019.
\textit{\aj} 158:13

\bibitem[{{Ofir} et~al.(2018){Ofir}, {Xie}, {Jiang}, {Sari} \&
  {Aharonson}}]{Ofir:2018}
{Ofir} A, {Xie} JW, {Jiang} CF, {Sari} R, {Aharonson} O. 2018.
\textit{\apjs} 234:9

\bibitem[{{Ormel}(2017)}]{Ormel:2017}
{Ormel} CW. 2017.
\textit{{The Emerging Paradigm of Pebble Accretion}}, vol. 445 of
  \textit{Astrophysics and Space Science Library}.
 197

\bibitem[{{Ormel} \& {Klahr}(2010)}]{Ormel:2010}
{Ormel} CW, {Klahr} HH. 2010.
\textit{\aap} 520:A43

\bibitem[{{Owen}(2019)}]{Owen:2019}
{Owen} JE. 2019.
\textit{Annual Review of Earth and Planetary Sciences} 47:67--90

\bibitem[{{Owen} \& {Lai}(2018)}]{OwenLai:2018}
{Owen} JE, {Lai} D. 2018.
\textit{\mnras} 479:5012--5021

\bibitem[{{Owen} \& {Murray-Clay}(2018)}]{OwenMurrayClay:2018}
{Owen} JE, {Murray-Clay} R. 2018.
\textit{\mnras} 480:2206--2216

\bibitem[{{Owen} \& {Wu}(2013)}]{OwenWu:2013}
{Owen} JE, {Wu} Y. 2013.
\textit{\apj} 775:105

\bibitem[{{Owen} \& {Wu}(2017)}]{OwenWu:2017}
{Owen} JE, {Wu} Y. 2017.
\textit{\apj} 847:29

\bibitem[{{Pascucci} et~al.(2018){Pascucci}, {Mulders}, {Gould} \& {Fernand
  es}}]{Pascucci2018}
{Pascucci} I, {Mulders} GD, {Gould} A, {Fernand es} R. 2018.
\textit{\apjl} 856:L28

\bibitem[{{Penny} et~al.(2019){Penny}, {Gaudi}, {Kerins}, {Rattenbury}, {Mao}
  et~al.}]{Penny:2019}
{Penny} MT, {Gaudi} BS, {Kerins} E, {Rattenbury} NJ, {Mao} S, et~al. 2019.
\textit{\apjs} 241:3

\bibitem[{{Perryman} et~al.(2014){Perryman}, {Hartman}, {Bakos} \&
  {Lindegren}}]{Perryman:2014}
{Perryman} M, {Hartman} J, {Bakos} G{\'A}, {Lindegren} L. 2014.
\textit{\apj} 797:14

\bibitem[{{Petigura} et~al.(2013){Petigura}, {Howard} \&
  {Marcy}}]{Petigura:2013}
{Petigura} EA, {Howard} AW, {Marcy} GW. 2013.
\textit{Proceedings of the National Academy of Science} 110:19273--19278

\bibitem[{{Petigura} et~al.(2017){Petigura}, {Howard}, {Marcy}, {Johnson},
  {Isaacson} et~al.}]{Petigura:2017}
{Petigura} EA, {Howard} AW, {Marcy} GW, {Johnson} JA, {Isaacson} H, et~al.
  2017.
\textit{\aj} 154:107

\bibitem[{{Petigura} et~al.(2018){Petigura}, {Marcy}, {Winn}, {Weiss}, {Fulton}
  et~al.}]{Petigura:2018}
{Petigura} EA, {Marcy} GW, {Winn} JN, {Weiss} LM, {Fulton} BJ, et~al. 2018.
\textit{\aj} 155:89

\bibitem[{{Petrovich} et~al.(2019){Petrovich}, {Deibert} \&
  {Wu}}]{Petrovich:2019}
{Petrovich} C, {Deibert} E, {Wu} Y. 2019.
\textit{\aj} 157:180

\bibitem[{{Petrovich} \& {Tremaine}(2016)}]{Petrovich:2016}
{Petrovich} C, {Tremaine} S. 2016.
\textit{\apj} 829:132

\bibitem[{{Plavchan} et~al.(2020){Plavchan}, {Barclay}, {Gagn{\'e}}, {Gao},
  {Cale} et~al.}]{AUMIc}
{Plavchan} P, {Barclay} T, {Gagn{\'e}} J, {Gao} P, {Cale} B, et~al. 2020.
\textit{\nat} 582:497--500

\bibitem[{{Plavchan} et~al.(2014){Plavchan}, {Bilinski} \&
  {Currie}}]{Plavchan:2014}
{Plavchan} P, {Bilinski} C, {Currie} T. 2014.
\textit{\pasp} 126:34

\bibitem[{{Poleski} et~al.(2014){Poleski}, {Skowron}, {Udalski}, {Han},
  {Koz{\l}owski} et~al.}]{Poleski:2014}
{Poleski} R, {Skowron} J, {Udalski} A, {Han} C, {Koz{\l}owski} S, et~al. 2014.
\textit{\apj} 795:42

\bibitem[{{Pollack} et~al.(1996){Pollack}, {Hubickyj}, {Bodenheimer},
  {Lissauer}, {Podolak} \& {Greenzweig}}]{Pollack:1996}
{Pollack} JB, {Hubickyj} O, {Bodenheimer} P, {Lissauer} JJ, {Podolak} M,
  {Greenzweig} Y. 1996.
\textit{\icarus} 124:62--85

\bibitem[{{Pu} \& {Lai}(2019)}]{PuLai:2019}
{Pu} B, {Lai} D. 2019.
\textit{\mnras} 488:3568--3587

\bibitem[{{Pu} \& {Lai}(2020)}]{PuLai:2020}
{Pu} B, {Lai} D. 2020.
\textit{arXiv e-prints} :arXiv:2008.05698

\bibitem[{{Pu} \& {Wu}(2015)}]{PuWu:2015}
{Pu} B, {Wu} Y. 2015.
\textit{\apj} 807:44

\bibitem[{{Quillen}(2011)}]{Quillen:2011}
{Quillen} AC. 2011.
\textit{\mnras} 418:1043--1054

\bibitem[{{Rasio} \& {Ford}(1996{\natexlab{a}})}]{Rasio:1996}
{Rasio} FA, {Ford} EB. 1996{\natexlab{a}}.
\textit{Science} 274:954--956

\bibitem[{{Rasio} \& {Ford}(1996{\natexlab{b}})}]{RasioFord:1996}
{Rasio} FA, {Ford} EB. 1996{\natexlab{b}}.
\textit{Science} 274:954--956

\bibitem[{{Raymond} et~al.(2008){Raymond}, {Barnes} \&
  {Mandell}}]{Raymond:2008}
{Raymond} SN, {Barnes} R, {Mandell} AM. 2008.
\textit{\mnras} 384:663--674

\bibitem[{{Ricker} et~al.(2015){Ricker}, {Winn}, {Vanderspek}, {Latham},
  {Bakos} et~al.}]{Ricker:2015}
{Ricker} GR, {Winn} JN, {Vanderspek} R, {Latham} DW, {Bakos} G{\'A}, et~al.
  2015.
\textit{Journal of Astronomical Telescopes, Instruments, and Systems} 1:014003

\bibitem[{{Rogers} \& {Owen}(2020)}]{RogersOwen:2020}
{Rogers} JG, {Owen} JE. 2020.
\textit{arXiv e-prints} :arXiv:2007.11006

\bibitem[{{Ryu} et~al.(2020){Ryu}, {Mr{\'o}z}, {Gould}, {Hwang}, {Kim}
  et~al.}]{Ryu:2020}
{Ryu} YH, {Mr{\'o}z} P, {Gould} A, {Hwang} KH, {Kim} HW, et~al. 2020.
\textit{arXiv e-prints} :arXiv:2010.07527

\bibitem[{{Safronov}(1972)}]{Safronov:1972}
{Safronov} VS. 1972.
\textit{{Evolution of the protoplanetary cloud and formation of the earth and
  planets.}}

\bibitem[{{Sanchis-Ojeda} et~al.(2014){Sanchis-Ojeda}, {Rappaport}, {Winn},
  {Kotson}, {Levine} \& {El Mellah}}]{SanchisOjeda:2014}
{Sanchis-Ojeda} R, {Rappaport} S, {Winn} JN, {Kotson} MC, {Levine} A, {El
  Mellah} I. 2014.
\textit{\apj} 787:47

\bibitem[{{Sandford} et~al.(2019){Sandford}, {Kipping} \&
  {Collins}}]{Sandford:2019}
{Sandford} E, {Kipping} D, {Collins} M. 2019.
\textit{\mnras} 489:3162--3173

\bibitem[{{Santerne} et~al.(2016){Santerne}, {Moutou}, {Tsantaki}, {Bouchy},
  {H{\'e}brard} et~al.}]{Santerne:2016}
{Santerne} A, {Moutou} C, {Tsantaki} M, {Bouchy} F, {H{\'e}brard} G, et~al.
  2016.
\textit{\aap} 587:A64

\bibitem[{{Santos} et~al.(2001){Santos}, {Israelian} \& {Mayor}}]{Santos:2001}
{Santos} NC, {Israelian} G, {Mayor} M. 2001.
\textit{\aap} 373:1019--1031

\bibitem[{{Schlaufman} et~al.(2010){Schlaufman}, {Lin} \&
  {Ida}}]{Schlaufman:2010}
{Schlaufman} KC, {Lin} DNC, {Ida} S. 2010.
\textit{\apjl} 724:L53--L58

\bibitem[{{Schlaufman} \& {Winn}(2013)}]{Schlaufman:2013}
{Schlaufman} KC, {Winn} JN. 2013.
\textit{\apj} 772:143

\bibitem[{{Schlaufman} \& {Winn}(2016)}]{Schlaufman:2016}
{Schlaufman} KC, {Winn} JN. 2016.
\textit{\apj} 825:62

\bibitem[{{Schlecker} et~al.(2020){Schlecker}, {Mordasini}, {Emsenhuber},
  {Klahr}, {Henning} et~al.}]{Bern_3:2020}
{Schlecker} M, {Mordasini} C, {Emsenhuber} A, {Klahr} H, {Henning} T, et~al.
  2020.
\textit{arXiv e-prints} :arXiv:2007.05563

\bibitem[{{Schlichting}(2014)}]{Schlichting2014}
{Schlichting} HE. 2014.
\textit{\apjl} 795:L15

\bibitem[{{Seager} \& {Mall{\'e}n-Ornelas}(2003)}]{Seager:2003}
{Seager} S, {Mall{\'e}n-Ornelas} G. 2003.
\textit{\apj} 585:1038--1055

\bibitem[{{Shallue} \& {Vanderburg}(2018)}]{Shallue:2018}
{Shallue} CJ, {Vanderburg} A. 2018.
\textit{\aj} 155:94

\bibitem[{{Skidmore} et~al.(2015){Skidmore}, {TMT International Science
  Development Teams} \& {Science Advisory Committee}}]{Skidmore:2015}
{Skidmore} W, {TMT International Science Development Teams}, {Science Advisory
  Committee} T. 2015.
\textit{Research in Astronomy and Astrophysics} 15:1945

\bibitem[{{Sousa} et~al.(2008){Sousa}, {Santos}, {Mayor}, {Udry}, {Casagrande}
  et~al.}]{Sousa:2008}
{Sousa} SG, {Santos} NC, {Mayor} M, {Udry} S, {Casagrande} L, et~al. 2008.
\textit{\aap} 487:373--381

\bibitem[{{Steffen} et~al.(2010){Steffen}, {Batalha}, {Borucki}, {Buchhave},
  {Caldwell} et~al.}]{Steffen:2010}
{Steffen} JH, {Batalha} NM, {Borucki} WJ, {Buchhave} LA, {Caldwell} DA, et~al.
  2010.
\textit{\apj} 725:1226--1241

\bibitem[{{Steffen} \& {Farr}(2013)}]{Steffen:2013}
{Steffen} JH, {Farr} WM. 2013.
\textit{\apjl} 774:L12

\bibitem[{{Steffen} et~al.(2012){Steffen}, {Ragozzine}, {Fabrycky}, {Carter},
  {Ford} et~al.}]{Steffen:2012}
{Steffen} JH, {Ragozzine} D, {Fabrycky} DC, {Carter} JA, {Ford} EB, et~al.
  2012.
\textit{Proceedings of the National Academy of Science} 109:7982--7987

\bibitem[{{Stevenson}(1982)}]{Stevenson:1982}
{Stevenson} DJ. 1982.
\textit{\planss} 30:755--764

\bibitem[{{Sumi} et~al.(2010){Sumi}, {Bennett}, {Bond}, {Udalski}, {Batista}
  et~al.}]{Sumi:2010}
{Sumi} T, {Bennett} DP, {Bond} IA, {Udalski} A, {Batista} V, et~al. 2010.
\textit{\apj} 710:1641--1653

\bibitem[{{Sumi} et~al.(2011){Sumi}, {Kamiya}, {Bennett}, {Bond}, {Abe}
  et~al.}]{Sumi:2011}
{Sumi} T, {Kamiya} K, {Bennett} DP, {Bond} IA, {Abe} F, et~al. 2011.
\textit{\nat} 473:349--352

\bibitem[{{Suzuki} et~al.(2018){Suzuki}, {Bennett}, {Ida}, {Mordasini},
  {Bhattacharya} et~al.}]{Suzuki:2018}
{Suzuki} D, {Bennett} DP, {Ida} S, {Mordasini} C, {Bhattacharya} A, et~al.
  2018.
\textit{\apjl} 869:L34

\bibitem[{{Suzuki} et~al.(2016){Suzuki}, {Bennett}, {Sumi}, {Bond}, {Rogers}
  et~al.}]{Suzuki:2016}
{Suzuki} D, {Bennett} DP, {Sumi} T, {Bond} IA, {Rogers} LA, et~al. 2016.
\textit{\apj} 833:145

\bibitem[{{Szab{\'o}} \& {Kiss}(2011)}]{Szabo:2011}
{Szab{\'o}} GM, {Kiss} LL. 2011.
\textit{\apjl} 727:L44

\bibitem[{{Tabachnik} \& {Tremaine}(2002)}]{Tabachnik:2002}
{Tabachnik} S, {Tremaine} S. 2002.
\textit{\mnras} 335:151--158

\bibitem[{{Terquem} \& {Papaloizou}(2019)}]{Terquem:2019}
{Terquem} C, {Papaloizou} JCB. 2019.
\textit{\mnras} 482:530--549

\bibitem[{{Teske} et~al.(2020){Teske}, {D{\'\i}az}, {Luque}, {Mo{\v{c}}nik},
  {Seidel} et~al.}]{Teske:2020}
{Teske} J, {D{\'\i}az} MR, {Luque} R, {Mo{\v{c}}nik} T, {Seidel} JV, et~al.
  2020.
\textit{\aj} 160:96

\bibitem[{{Teske} et~al.(2019){Teske}, {Thorngren}, {Fortney}, {Hinkel} \&
  {Brewer}}]{Teske:2019}
{Teske} JK, {Thorngren} D, {Fortney} JJ, {Hinkel} N, {Brewer} JM. 2019.
\textit{\aj} 158:239

\bibitem[{{Thompson} et~al.(2018){Thompson}, {Coughlin}, {Hoffman}, {Mullally},
  {Christiansen} et~al.}]{Thompson:2018}
{Thompson} SE, {Coughlin} JL, {Hoffman} K, {Mullally} F, {Christiansen} JL,
  et~al. 2018.
\textit{\apjs} 235:38

\bibitem[{{Tremaine}(2015)}]{Tremaine:2015}
{Tremaine} S. 2015.
\textit{\apj} 807:157

\bibitem[{{Tremaine} \& {Dong}(2012)}]{Tremaine:2012}
{Tremaine} S, {Dong} S. 2012.
\textit{\aj} 143:94

\bibitem[{{Tu} et~al.(2015){Tu}, {Johnstone}, {G{\"u}del} \&
  {Lammer}}]{Tu:2015}
{Tu} L, {Johnstone} CP, {G{\"u}del} M, {Lammer} H. 2015.
\textit{\aap} 577:L3

\bibitem[{{Udalski} et~al.(2018){Udalski}, {Ryu}, {Sajadian}, {Gould},
  {Mr{\'o}z} et~al.}]{Udalski:2018}
{Udalski} A, {Ryu} YH, {Sajadian} S, {Gould} A, {Mr{\'o}z} P, et~al. 2018.
\textit{\actaa} 68:1--42

\bibitem[{{Uehara} et~al.(2016){Uehara}, {Kawahara}, {Masuda}, {Yamada} \&
  {Aizawa}}]{Uehara:2016}
{Uehara} S, {Kawahara} H, {Masuda} K, {Yamada} S, {Aizawa} M. 2016.
\textit{\apj} 822:2

\bibitem[{{Van Eylen} et~al.(2018){Van Eylen}, {Agentoft}, {Lundkvist},
  {Kjeldsen}, {Owen} et~al.}]{VanEylen:2018}
{Van Eylen} V, {Agentoft} C, {Lundkvist} MS, {Kjeldsen} H, {Owen} JE, et~al.
  2018.
\textit{\mnras} 479:4786--4795

\bibitem[{{Van Eylen} \& {Albrecht}(2015)}]{VanEylen:2015}
{Van Eylen} V, {Albrecht} S. 2015.
\textit{\apj} 808:126

\bibitem[{{Van Eylen} et~al.(2019){Van Eylen}, {Albrecht}, {Huang},
  {MacDonald}, {Dawson} et~al.}]{VanEylen:2019}
{Van Eylen} V, {Albrecht} S, {Huang} X, {MacDonald} MG, {Dawson} RI, et~al.
  2019.
\textit{\aj} 157:61

\bibitem[{{Venturini} et~al.(2016){Venturini}, {Alibert} \&
  {Benz}}]{Venturini:2016}
{Venturini} J, {Alibert} Y, {Benz} W. 2016.
\textit{\aap} 596:A90

\bibitem[{{Wang} \& {Fischer}(2015)}]{WangFischer:2015}
{Wang} J, {Fischer} DA. 2015.
\textit{\aj} 149:14

\bibitem[{{Wang} et~al.(2015){Wang}, {Fischer}, {Xie} \& {Ciardi}}]{Wang:2015}
{Wang} J, {Fischer} DA, {Xie} JW, {Ciardi} DR. 2015.
\textit{\apj} 813:130

\bibitem[{{Wang} et~al.(2014){Wang}, {Xie}, {Barclay} \& {Fischer}}]{Wang:2014}
{Wang} J, {Xie} JW, {Barclay} T, {Fischer} DA. 2014.
\textit{\apj} 783:4

\bibitem[{{Weidenschilling}(1977)}]{Weidenschilling:1977}
{Weidenschilling} SJ. 1977.
\textit{\apss} 51:153--158

\bibitem[{{Weidenschilling} \& {Marzari}(1996)}]{Weidenschilling:1996}
{Weidenschilling} SJ, {Marzari} F. 1996.
\textit{\nat} 384:619--621

\bibitem[{{Weiss} et~al.(2018{\natexlab{a}}){Weiss}, {Isaacson}, {Marcy},
  {Howard}, {Petigura} et~al.}]{Weiss:2018b}
{Weiss} LM, {Isaacson} HT, {Marcy} GW, {Howard} AW, {Petigura} EA, et~al.
  2018{\natexlab{a}}.
\textit{\aj} 156:254

\bibitem[{{Weiss} et~al.(2018{\natexlab{b}}){Weiss}, {Marcy}, {Petigura},
  {Fulton}, {Howard} et~al.}]{Weiss:2018a}
{Weiss} LM, {Marcy} GW, {Petigura} EA, {Fulton} BJ, {Howard} AW, et~al.
  2018{\natexlab{b}}.
\textit{\aj} 155:48

\bibitem[{{Weiss} \& {Petigura}(2020)}]{Weiss:2020}
{Weiss} LM, {Petigura} EA. 2020.
\textit{\apjl} 893:L1

\bibitem[{{Winn}(2010)}]{Winn:2010}
{Winn} JN. 2010.
\textit{arXiv e-prints} :arXiv:1001.2010

\bibitem[{{Winn} \& {Fabrycky}(2015)}]{WinnFabrycky:2015}
{Winn} JN, {Fabrycky} DC. 2015.
\textit{\araa} 53:409--447

\bibitem[{{Winn} et~al.(2018){Winn}, {Sanchis-Ojeda} \&
  {Rappaport}}]{Winn:2018}
{Winn} JN, {Sanchis-Ojeda} R, {Rappaport} S. 2018.
\textit{\nar} 83:37--48

\bibitem[{{Winn} et~al.(2017){Winn}, {Sanchis-Ojeda}, {Rogers}, {Petigura},
  {Howard} et~al.}]{Winn:2017}
{Winn} JN, {Sanchis-Ojeda} R, {Rogers} L, {Petigura} EA, {Howard} AW, et~al.
  2017.
\textit{\aj} 154:60

\bibitem[{{Wisdom}(1980)}]{Wisdom:1980}
{Wisdom} J. 1980.
\textit{\aj} 85:1122--1133

\bibitem[{{Wittenmyer} et~al.(2016){Wittenmyer}, {Butler}, {Tinney}, {Horner},
  {Carter} et~al.}]{Wittenmyer:2016}
{Wittenmyer} RA, {Butler} RP, {Tinney} CG, {Horner} J, {Carter} BD, et~al.
  2016.
\textit{\apj} 819:28

\bibitem[{{Woolfson}(1993)}]{Woolfson:1993}
{Woolfson} MM. 1993.
\textit{\qjras} 34:1--20

\bibitem[{{Wright} et~al.(2012){Wright}, {Marcy}, {Howard}, {Johnson}, {Morton}
  \& {Fischer}}]{Wright:2012}
{Wright} JT, {Marcy} GW, {Howard} AW, {Johnson} JA, {Morton} TD, {Fischer} DA.
  2012.
\textit{\apj} 753:160

\bibitem[{{Wright} et~al.(2009){Wright}, {Upadhyay}, {Marcy}, {Fischer}, {Ford}
  \& {Johnson}}]{Wright:2009}
{Wright} JT, {Upadhyay} S, {Marcy} GW, {Fischer} DA, {Ford} EB, {Johnson} JA.
  2009.
\textit{\apj} 693:1084--1099

\bibitem[{{Wright} et~al.(2011){Wright}, {Veras}, {Ford}, {Johnson}, {Marcy}
  et~al.}]{Wright:2011}
{Wright} JT, {Veras} D, {Ford} EB, {Johnson} JA, {Marcy} GW, et~al. 2011.
\textit{\apj} 730:93

\bibitem[{{Wu}(2019)}]{Wu:2019}
{Wu} Y. 2019.
\textit{\apj} 874:91

\bibitem[{{Wu} \& {Lithwick}(2013)}]{Wu:2013}
{Wu} Y, {Lithwick} Y. 2013.
\textit{\apj} 772:74

\bibitem[{{Xie} et~al.(2016){Xie}, {Dong}, {Zhu}, {Huber}, {Zheng}
  et~al.}]{Xie:2016}
{Xie} JW, {Dong} S, {Zhu} Z, {Huber} D, {Zheng} Z, et~al. 2016.
\textit{Proceedings of the National Academy of Science} 113:11431--11435

\bibitem[{{Xie} et~al.(2014){Xie}, {Wu} \& {Lithwick}}]{Xie:2014}
{Xie} JW, {Wu} Y, {Lithwick} Y. 2014.
\textit{\apj} 789:165

\bibitem[{{Xuan} \& {Wyatt}(2020)}]{Xuan:2020}
{Xuan} JW, {Wyatt} MC. 2020.
\textit{\mnras}

\bibitem[{{Yalinewich} \& {Petrovich}(2020)}]{Yalinewich:2020}
{Yalinewich} A, {Petrovich} C. 2020.
\textit{\apjl} 892:L11

\bibitem[{{Yang} et~al.(2020){Yang}, {Xie} \& {Zhou}}]{Yang:2020}
{Yang} JY, {Xie} JW, {Zhou} JL. 2020.
\textit{\aj} 159:164

\bibitem[{{Yee} et~al.(2021){Yee}, {Zang}, {Udalski}, {Ryu}, {Green}
  et~al.}]{Yee:2021}
{Yee} JC, {Zang} W, {Udalski} A, {Ryu} YH, {Green} J, et~al. 2021.
\textit{arXiv e-prints} :arXiv:2101.04696

\bibitem[{{Youdin}(2011)}]{Youdin:2011}
{Youdin} AN. 2011.
\textit{\apj} 742:38

\bibitem[{{Youdin} \& {Goodman}(2005)}]{Youdin:2005}
{Youdin} AN, {Goodman} J. 2005.
\textit{\apj} 620:459--469

\bibitem[{{Zang} et~al.(2021){Zang}, {Han}, {Kondo}, {Yee}, {Lee}
  et~al.}]{Zang:2021}
{Zang} W, {Han} C, {Kondo} I, {Yee} JC, {Lee} CU, et~al. 2021.
\textit{arXiv e-prints} :arXiv:2103.01896

\bibitem[{{Zapatero Osorio} et~al.(2000){Zapatero Osorio}, {B{\'e}jar},
  {Mart{\'\i}n}, {Rebolo}, {Barrado y Navascu{\'e}s} et~al.}]{Osorio:2000}
{Zapatero Osorio} MR, {B{\'e}jar} VJS, {Mart{\'\i}n} EL, {Rebolo} R, {Barrado y
  Navascu{\'e}s} D, et~al. 2000.
\textit{Science} 290:103--107

\bibitem[{{Zhou} et~al.(2007){Zhou}, {Lin} \& {Sun}}]{Zhou:2007}
{Zhou} JL, {Lin} DNC, {Sun} YS. 2007.
\textit{\apj} 666:423--435

\bibitem[{{Zhu}(2019)}]{Zhu:2019}
{Zhu} W. 2019.
\textit{\apj} 873:8

\bibitem[{{Zhu}(2020)}]{Zhu:2020}
{Zhu} W. 2020.
\textit{\aj} 159:188

\bibitem[{{Zhu} et~al.(2018{\natexlab{a}}){Zhu}, {Dai} \& {Masuda}}]{Zhu_note}
{Zhu} W, {Dai} F, {Masuda} K. 2018{\natexlab{a}}.
\textit{Research Notes of the American Astronomical Society} 2:160

\bibitem[{{Zhu} et~al.(2018{\natexlab{b}}){Zhu}, {Petrovich}, {Wu}, {Dong} \&
  {Xie}}]{Zhu:2018}
{Zhu} W, {Petrovich} C, {Wu} Y, {Dong} S, {Xie} J. 2018{\natexlab{b}}.
\textit{\apj} 860:101

\bibitem[{{Zhu} et~al.(2016){Zhu}, {Wang} \& {Huang}}]{Zhu:2016}
{Zhu} W, {Wang} J, {Huang} C. 2016.
\textit{\apj} 832:196

\bibitem[{{Zhu} \& {Wu}(2018)}]{ZhuWu:2018}
{Zhu} W, {Wu} Y. 2018.
\textit{\aj} 156:92

\bibitem[{{Ziegler} et~al.(2018){Ziegler}, {Law}, {Baranec}, {Riddle}, {Duev}
  et~al.}]{Ziegler:2018}
{Ziegler} C, {Law} NM, {Baranec} C, {Riddle} R, {Duev} DA, et~al. 2018.
\textit{\aj} 155:161

\bibitem[{{Zink} et~al.(2019){Zink}, {Christiansen} \& {Hansen}}]{Zink:2019}
{Zink} JK, {Christiansen} JL, {Hansen} BMS. 2019.
\textit{\mnras} 483:4479--4494

\bibitem[{{Zinzi} \& {Turrini}(2017)}]{Zinzi:2017}
{Zinzi} A, {Turrini} D. 2017.
\textit{\aap} 605:L4

\bibitem[{{Zong} et~al.(2018){Zong}, {Fu}, {De Cat}, {Shi}, {Luo}
  et~al.}]{Zong:2018}
{Zong} W, {Fu} JN, {De Cat} P, {Shi} J, {Luo} A, et~al. 2018.
\textit{\apjs} 238:30

\end{thebibliography}
